\documentclass[11pt, english]{article}
\usepackage{amsfonts}
\usepackage[T1]{fontenc}
\usepackage{mathpazo}
\usepackage{amssymb,amsmath,amsthm,bm,mathtools, commath}
\usepackage{thmtools}
\usepackage{enumitem}
\usepackage[margin=1in,bottom=1in,top=1in]{geometry}
\usepackage{comment}
\usepackage{float}
\usepackage{natbib}
\usepackage{booktabs}
\usepackage{setspace}
\usepackage{multirow}
\usepackage[flushleft]{threeparttable}
\usepackage{tikz}
\usepackage{caption}
\usepackage{xcolor}
\usepackage{hyperref}
\usepackage{url}
\usepackage{pifont}
\usepackage{verbatim}
\usepackage{booktabs}
\usepackage{soul}
\usepackage{authblk}

﻿
﻿
﻿
\setlist[enumerate,1]{label=(\roman*),ref=\roman*}
\hypersetup{
colorlinks = true,
linkcolor = blue,
citecolor = blue,
}

\newcommand{\RN}[1]{  \textup{\uppercase\expandafter{\romannumeral#1}}}
\renewcommand\thmcontinues[1]{Continued}

\newtheorem{theorem}{Theorem}

\newtheorem{assumption}{Assumption}

\input{tcilatex}
\begin{document}

\title{Principal Component Analysis for a Mix of Stationary and
Nonstationary Variables\thanks{%
An earlier version of this paper was circulated under the title, "Principal
Component Analysis for Nonstationary Time Series." We thank Serena Ng, Mark
Watson, Ken West, and anonymous referees and editors for helpful comments.
Xi gratefully acknowledges financial
support from the National Natural Science Foundation of China (No. 72503234).
Authors are listed in alphabetical order; all contributed equally to this
work.}}
\author[]{James D. Hamilton\thanks{%
Corresponding author. Email: \texttt{jhamilton@ucsd.edu.}} \and Xinwei Ma\thanks{%
Email: \texttt{x1ma@ucsd.edu.}} \and Jin Xi\thanks{%
Email: \texttt{xijin@amss.ac.cn.%
}}}
\date{May 28, 2022 \\
Revised: August 17, 2026}
\maketitle

\begin{abstract}
\doublespacing This paper develops a procedure for uncovering the common
cyclical factors that drive a mix of stationary and nonstationary variables.
The method does not require knowing which variables are nonstationary or the
nature of the nonstationarity. An application to the FRED-MD macroeconomic
dataset demonstrates that the approach offers similar benefits to those of
traditional principal component analysis with some added advantages.
\end{abstract}


\affil[1]{University of California at San Diego} 
\affil[2]{University of
California at San Diego} \affil[3]{Center for Forecasting Science, Academy
of Mathematics and Systems Science, Chinese Academy of Sciences}

\bigskip \thispagestyle{empty} \vspace{2em}

 \bigskip

\clearpage
\pagenumbering{arabic}

\section{Introduction}

Principal component analysis (PCA) has become a key tool for building
dynamic models of vector time series with a large cross-sectional dimension.
A typical application first subtracts the sample mean from each variable and
divides the demeaned variables by their sample standard deviations. PCA
finds linear combinations of the standardized variables that have maximum
sample variance. These principal components are then used to build dynamic
models for each individual series. For surveys of PCA and its usefulness in
economics see \cite{BaiNg2008} and \cite{StockWatson2016Handbook}.

One difficulty with PCA is that many of the time series encountered in
economics and finance are nonstationary. For a nonstationary variable, the
population mean is undefined and the sample standard deviation diverges to
infinity as the number of time-series observations gets large. \cite%
{onatski2021spurious} detailed some of the problems that can arise from
trying to apply PCA to nonstationary data. The typical solution to this
problem is for researchers to examine each series individually by hand to
determine the transformation that needs to be applied to that series before
calculating principal components of the set of variables.

This approach has two shortcomings. First, while for some variables it may
be fairly clear what transformation is necessary to achieve stationarity,
for others it is far from obvious. For example, interest rates have a strong
downward trend since 1980. Should yields be treated as stationary? If
nonstationary, should we take their first differences or deviations from a
time trend before performing PCA? Some authors work with first differences
of inflation and the unemployment rate while others leave those variables as
is. Many decisions like these have to be made before applying PCA to large
data sets. In this paper, we propose an automatic procedure that allows the
researcher to avoid these judgment calls.

A second fundamental problem is the appropriateness of the methodology
itself. Suppose we somehow overcame the first problem and knew for certain
the true nature of the trend in each individual series. Suppose for
illustration we knew correctly that the first variable $y_{1t}$ is a
stationary $AR(1)$ process with autoregressive coefficient $\rho =0.99$
while the second variable $y_{2t}$ is a random walk. The currently
prescribed procedure would instruct the researcher to use the first variable
as is and the second variable in first differences. But while the levels of $%
y_{1t}$ and $y_{2t}$ exhibit very similar properties, the level of $y_{1t}$
and first difference of $y_{2t}$ are radically different. Should we expect
that there is some linear combination of $y_{1t}$ and $\Delta y_{2t}$ that
can summarize the common economic drivers behind the two variables? If
differencing is the appropriate transformation for a random walk, it seems
we should be using some similar transformation for an $AR(1)$ process whose
root is close but not quite equal to unity.

This paper proposes an approach to PCA that solves these problems. We first
note as in \cite{hamilton2018you} that it is possible to use OLS to estimate
an $h$-period-ahead forecast of the level of any variable as a linear
function of its own lags without knowing the nature of the nonstationarity.
Moreover, the errors from these linear forecasts are stationary for a broad
class of underlying nonstationary processes. Our proposal is to use PCA to
identify common factors behind the forecast errors. Specifically, we
estimate an OLS regression of $y_{it}$ on $(1,y_{i,t-h},y_{i,t-h-1},\dots
,y_{i,t-h-p+1})^{\prime }$, where $i=1,\dots ,N$ represents the index of the
variable, $h$ is the forecasting horizon, and $p$ is the number of lags used
for the forecast. We then calculate principal components of the residuals.
This approach solves both of the problems identified above. The procedure is
fully automatic and treats every variable in the same way. The
transformation of $y_{it}$ is a continuous function of the estimated
autoregressive coefficients and implies no discontinuity in the way that
persistent stationary series are treated as the largest autoregressive root
tends toward unity.

\cite{hamilton2018you} suggested that a two-year-ahead forecast error could
be interpreted as the stationary cyclical component of the series. Our
empirical application follows this suggestion and uses\thinspace $h=24$ for
monthly data to uncover the common cyclical factors behind a range of
economic and financial indicators. We demonstrate that using $h=24$ also
offers a practical solution to the problem of outliers which has posed a
severe challenge to using PCA on macroeconomic observations in 2020.

\cite{BaiNg2004} and \cite{barigozzi2021large} developed methods that are
suitable when the observed data are an unknown mix of $I(0)$ and $I(1)$
variables. Their approaches start by applying PCA to first-differences of
all of the variables whether stationary or not. \cite{BaiNg2004} estimated
factors for the original data by accumulating factors estimated from the
differenced data. \cite{barigozzi2021large} estimated factors by applying
the loadings estimated from differenced data to the detrended levels of the
original data. \cite{barigozzi2021large} showed that their procedure can
estimate the set of factors that account for common stochastic trends in the
original data. In contrast to these approaches, the goal of our analysis is
to uncover the cyclical components that are driving both the stationary and
nonstationary variables.

Our results are also useful for estimation of dynamic factor models. One
common approach to estimating dynamic factor models is to apply a
transformation that is believed to make the individual series stationary,
calculate principal components of the transformed data, and then fit a
vector autoregression to the estimated factors (\cite%
{stock2002forecasting,StockWatson2016Handbook}; \cite{BaiNg2002,
BaiNgEcta2006}). In contrast, our approach allows the researcher to
implement the initial transformations without the need for subjective
judgments about each individual series.

Section \ref{sec_detrend} discusses the use of forecasting regressions to
isolate a stationary component of a possibly nonstationary time series.
Section \ref{sec_c_obs} describes how PCA could be used to uncover the
common factors behind the forecast errors if we somehow could know the true
process followed by each variable. Section \ref{sec_c_not_observed} analyzes
the case when PCA is applied to the residuals from estimated OLS regressions
and establishes consistency of the method. Simulations in Section \ref%
{sec_simulations} report results for a variety of possible data-generating
processes. We find that using two-year-ahead forecast errors works well when
we have 50 years of data and that $h=1$ performs quite well even in much
shorter samples. Section \ref{sec_fred_md} illustrates the promise of our
approach in an empirical analysis using the FRED-MD large macroeconomics
data set. Proofs of theorems and additional simulations and empirical
results are provided in an online supplement.

\section{Isolating a stationary component from a nonstationary series}

\label{sec_detrend} Let $y_{it}$ denote the level of a possibly
nonstationary variable at time $t$. Collect its $p$ most recent values as of
date $t-h$ along with a constant term in a vector {${z}%
_{i,t-h}=(1,y_{i,t-h},y_{i,t-h-1},...,y_{i,t-h-p+1})^{\prime }.$} The
population linear projection of $y_{it}$ on {${z}_{i,t-h}$} is given by {$%
\mathbb{P}(y_{it}|{z}_{i,t-h})=\alpha _{i0}^{\prime }{z}_{i,t-h}$} where the
coefficient $\alpha _{i0}$ is defined as the vector that minimizes the
expected squared error of a forecast of $y_{it}$ based on a linear function
of {${z}_{i,t-h}$}:%
\begin{equation}
\alpha _{i0}=\arg \underset{\alpha }{\min }E(y_{it}-\alpha ^{\prime }{z}%
_{i,t-h})^{2}.  \label{eqn_pop_lin_proj_definition}
\end{equation}%
For example, if $y_{it}$ is covariance stationary with {$E({z}_{i,t-h}{z}%
_{i,t-h}^{\prime })$} nonsingular, the population linear projection
coefficient is given by%
\begin{equation}
\alpha _{i0}=\left[ E({z}_{i,t-h}{z}_{i,t-h}^{\prime })\right] ^{-1}E({z}%
_{i,t-h}y_{it}).  \label{eqn_pop_lin_proj_stationary}
\end{equation}%
If $y_{it}$ is ergodic for second moments, the population parameter $\alpha
_{i0}$ can be consistently estimated by an OLS regression of $y_{it}$ on {${z%
}_{i,t-h}$}:%
\begin{equation*}
\hat{\alpha}_{i}=\left[ \tsum\nolimits_{t=1}^{T}{z}_{i,t-h}{z}%
_{i,t-h}^{\prime }\right] ^{-1}\left[ \tsum\nolimits_{t=1}^{T}{z}%
_{i,t-h}y_{it}\right] =\left[ T^{-1}\tsum\nolimits_{t=1}^{T}{z}_{i,t-h}{z}%
_{i,t-h}^{\prime }\right] ^{-1}\left[ T^{-1}\tsum\nolimits_{t=1}^{T}{z}%
_{i,t-h}y_{it}\right] \overset{p}{\rightarrow }\alpha _{i0}.
\end{equation*}%
Note that there is no assumption in the definition (\ref%
{eqn_pop_lin_proj_definition}) nor in its stationary solution (\ref%
{eqn_pop_lin_proj_stationary}) that the process is linear or that it can be
characterized by an ARMA representation of known order. There may be better
forecasts of $y_{it}$ that could be obtained using a nonlinear function or
more lags like $y_{i,t-h-p},y_{i,t-h-p-1},...$ But there is an optimal
forecast within the class of linear forecasts that use only $p$ lags. For a
covariance-stationary process, the optimal forecast within that class is
characterized by (\ref{eqn_pop_lin_proj_stationary}).

A solution to (\ref{eqn_pop_lin_proj_definition}) also exists if $y_{it}$ is
nonstationary but its first difference $\Delta y_{it}$ is covariance
stationary. This can be seen from the accounting identity that holds for any
time series:%
\begin{equation}
y_{it}=y_{i,t-h}+\Delta y_{i,t-h+1}+\Delta y_{i,t-h+2}+\cdots +\Delta y_{it}.
\label{eqn_I_1_identity}
\end{equation}%
For an\thinspace $I(1)$ process,\footnote{%
Note that the space spanned by $z_{i,t-h}$ is the same as the space spanned
by $(z_{i,t-h}^{\dag },y_{i,t-h})$ which is bigger than the space spanned by 
$z_{i,t-h}^{\dag }.$ However, the coefficient on $y_{i,t-h}$ in the
projection $\mathbb{P}(\Delta y_{it}|{z}_{i,t-h}^{\dag },y_{i,t-h})$ is
zero, because multiplying the $I(1)$ variable $y_{i,t-h}$ by any nonzero
value would produce a forecast of $y_{it}$ that has infinite MSE. For this
reason $\mathbb{P}(\Delta y_{it}|z_{i,t-h})=\mathbb{P}(\Delta y_{it}|{z}%
_{i,t-h}^{\dag }).$}%
\begin{equation}
\mathbb{P}(y_{it}|{z}_{i,t-h})=y_{i,t-h}+\mathbb{P}(\Delta y_{i,t-h+1}|{z}%
_{i,t-h}^{\dag })+\mathbb{P}(\Delta y_{i,t-h+2}|{z}_{i,t-h}^{\dag })+\cdots +%
\mathbb{P}(\Delta y_{it}|{z}_{i,t-h}^{\dag })  \label{eqn_lin_proj_I_1}
\end{equation}%
for ${z}_{i,t-h}=(1,y_{i,t-h},y_{i,t-h-1},...,y_{i,t-h-p+1})^{\prime },$ ${z}%
_{i,t-h}^{\dag }=(1,\Delta y_{i,t-h},\Delta y_{i,t-h-1},...,\Delta
y_{i,t-h-p+2})^{\prime }$ and \linebreak $\mathbb{P}(\Delta
y_{i,t-s}|z_{i,t-h}^{\dag })=\left[ E(\Delta y_{i,t-s}{z}_{i,t-h}^{\dag
\prime })\right] \left[ E({z}_{i,t-h}^{\dag }{z}_{i,t-h}^{\dag \prime })%
\right] ^{-1}{z}_{i,t-h}^{\dag }.$ Moreover, the population linear
projection coefficient $\alpha _{i0}$ can again be consistently estimated by
an OLS regression of the level of $y_{it}$ on $(1,y_{i,t-h},y_{i,t-h-1},%
\allowbreak ...,y_{i,t-h-p+1})^{\prime }.$ The intuition for this is that an
OLS levels regression minimizes the sample analog to the population
minimization in (\ref{eqn_pop_lin_proj_definition}):%
\begin{equation}
\hat{\alpha}_{i}=\arg \underset{\alpha }{\min }T^{-1}\tsum%
\nolimits_{t=1}^{T}(y_{it}-\alpha ^{\prime }{z}_{i,t-h})^{2}.
\label{eqn_OLS_objective_function}
\end{equation}%
If the OLS estimate $\hat{\alpha}_{i}$ picks up the unit root, the value of (%
\ref{eqn_OLS_objective_function}) converges to a finite number as $T$ gets
large. If it does not, the value of (\ref{eqn_OLS_objective_function}) would
diverge to infinity. Thus in large samples, OLS estimation is dominated by
the incentive inherent in the objective function of OLS to remove any
nonstationary features of the data. More formally, the fitted values from
the regression (\ref{eqn_OLS_objective_function}) are numerically identical
to the fitted values from an OLS regression of $y_{it}$ on {$(y_{i,t-h},{z}%
_{i,t-h}^{\dag ^{\prime }}).$} In the latter regression, the OLS coefficient
on the lagged level $y_{i,t-h}$ will converge in probability to one whenever
the dependent variable is an $I(1)$ process. In fact, the OLS estimate of
this parameter is superconsistent, converging at rate $T$ rather than at
rate $\sqrt{T}$, and the asymptotic distribution of the other coefficients
is identical to what would result if we regressed $\Delta y_{it}$ on {${z}%
_{i,t-h}^{\dag };$} see \cite{hamilton2018you}.

\cite{hamilton2018you} showed that these results generalize to any
nonstationary process that is covariance stationary around a deterministic
polynomial in time $t$ provided that the order of the polynomial is no
greater than the number of lags $p,$ or if the process is $I(d)$ with the $d$%
th difference of $y_{it}$ covariance stationary for some $d$ no greater than 
$p.$ One does not need to know the order of the polynomial or the value of $%
d $ in order to consistently estimate the population linear projection
coefficient $\alpha _{i0}$ using a regression of the level of $y_{it}$ on a
constant and its $p$ most recent levels as of date $t-h.$

\cite{hamilton2018you} also showed that the residual from the population
linear projection, {$c_{it}=y_{it}-\alpha _{i0}^{\prime }{z}_{i,t-h},$} is
covariance stationary for any of the above nonstationary processes. For
example, for an $I(1)$ process, we see from (\ref{eqn_I_1_identity}) and (%
\ref{eqn_lin_proj_I_1}) that

\begin{eqnarray*}
c_{it} &=&y_{it}-\mathbb{P}(y_{it}|{z}_{i,t-h}) \\
&=&\left[ \Delta y_{i,t-h+1}-\mathbb{P}(\Delta y_{i,t-h+1}|{z}_{i,t-h}^{\dag
})\right] +\left[ \Delta y_{i,t-h+2}-\mathbb{P}(\Delta y_{i,t-h+2}|{z}%
_{i,t-h}^{\dag })\right] + \\
&&\cdots +\left[ \Delta y_{it}-\mathbb{P}(\Delta y_{it}|{z}_{i,t-h}^{\dag })%
\right]
\end{eqnarray*}%
is covariance stationary for any finite $h.$ It is thus possible to isolate
a stationary component of a large range of nonstationary processes from the
residuals of an OLS regression of $y_{it}$ on $%
(1,y_{i,t-h},y_{i,t-h-1},...,y_{i,t-h-p+1})^{\prime }$ without knowing
whether the variable $y_{it}$ is stationary or nonstationary.\footnote{%
Specifically, if either: (i) $y_{it}$ is stationary around a deterministic
polynomial function of time of order $d_{i}\leq p$ satisfying%
\begin{equation*}
T^{-1/2}\sum_{s=1}^{[Tr]}(y_{it}-\delta _{i0}-\delta _{i1}t-\delta
_{i2}t^{2}-\cdots -\delta _{i}t^{d_{i}})\overset{d}{\rightarrow } \omega
_{i}W_{i}(r)
\end{equation*}%
where $[Tr]$ denotes the largest integer no greater than $Tr$, $W_{i}(r)$
denotes standard Brownian motion, and $\overset{d}{\rightarrow } $ denotes
convergence in distribution; or alternatively if (ii) $d_{i}$ differences of 
$y_{it}$ are stationary for some $d_{i}\leq p$ satisfying%
\begin{equation*}
T^{-1/2}\sum_{s=1}^{[Tr]}(\Delta ^{d_{i}}y_{it}-\mu _{i})\overset{d}{%
\rightarrow } \omega _{i}W_{i}(r);
\end{equation*}%
then \cite{hamilton2018you} showed that $c_{it}$ is stationary and that the
estimated coefficient $\hat{\alpha}_{i}$ from an OLS regression gives a
consistent estimate of the population parameter $\alpha _{i0}.$}

It is instructive to compare this approach with other common ways of
thinking about the trend in a nonstationary series. If the $i$th observed
variable is a deterministic function of time plus a zero-mean stationary
ergodic process, $y_{it}=\delta _{i}^{\det }(t)+c_{it}^{\det }$, we
typically could describe the deterministic time trend as the limit of a
forecast that would have been made in the arbitrarily distant past: 
\begin{equation*}
\delta _{i}^{\det }(t)=\underset{h\rightarrow \infty }{\,\lim }\underset{%
p\rightarrow \infty }{\,\lim }E(y_{it}\mid
y_{i,t-h},y_{i,t-h-1},...,y_{i,t-h-p+1}).
\end{equation*}
By contrast, if the first difference of the $i$th variable is a zero-mean
stationary process, \cite{beveridge1981new} suggested that we think of the
trend as the forecast of the variable in the arbitrarily distant future.
They decomposed $y_{it}=\delta _{it}^{BN}+c_{it}^{BN}$ where 
\begin{equation*}
\delta _{it}^{BN}=\underset{h\rightarrow \infty }{\,\lim }\underset{%
p\rightarrow \infty }{\,\lim }E(y_{i,t+h}\mid
y_{it},y_{i,t-1},...,y_{i,t-p+1}).
\end{equation*}
While these concepts of trend have some appeal, they have the significant
practical drawback that both are based on the properties of forecasts at
infinite horizons. They thus depend on conjectures of what happens at
infinity, conjectures that are impossible to verify on the basis of a finite
sample of observations.

By contrast, forecasts at a two-year horizon are something that can be
reasonably investigated without auxiliary assumptions in samples of typical
size. If we choose $h$ to correspond to a two-year horizon, \cite%
{hamilton2018you} argued that the decomposition%
\begin{equation}
y_{it}=\mathbb{P}(y_{it}|1,y_{i,t-h},y_{i,t-h-1},...,y_{i,t-h-p+1})+c_{it}
\label{eqn_hamilton_regression}
\end{equation}%
can be viewed as a way to implement traditional approaches to trend-cycle
decomposition that is practical and robust. Moreover, the primary reason we
would go wrong in making a two-year-ahead forecast of most economic time
series is due to unforeseen cyclical changes. For example, the value of $%
y_{it}$ will be significantly below {$\mathbb{P}(y_{it}|{z}_{i,t-h})$} if
the economy goes into a recession after $t-h$ and significantly above the
forecast if recovery from a downturn is more robust than expected. For this
reason, \cite{hamilton2018you} proposed to use $h$ corresponding to a
two-year horizon as a way to isolate the stationary cyclical component of
the $i$th variable, for example, $h=24$ for monthly data.

Another important benefit of setting $h=24$ is that it offers a solution to
the huge practical problem of dealing with outliers. To see why this is the
case, suppose for illustration that the $i$th variable is characterized by a
random walk: $y_{it}=y_{i,t-1}+\varepsilon _{it}.$ For a monthly random
walk, the residual from the two-year population linear projection is given
by $c_{it}=\tsum\nolimits_{s=0}^{23}\varepsilon _{i,t-s}.$ From the Central
Limit Theorem, $c_{it}$ typically exhibits much less kurtosis than the
one-period-ahead forecast error $\varepsilon _{it}.$\footnote{\label{fn_CLT}
For example, if $c_{it}^{(h)}= \varepsilon _{it}+\ldots + \varepsilon
_{i,t-h+1}$ and $\varepsilon _{it}$ is i.i.d. with $E(\varepsilon _{it})=0,$ 
$E(\varepsilon _{it}^{2})=\sigma ^{2},$ and $E(\varepsilon _{it}^{4})=\kappa
,$ then $E\left[ (c_{it}^{(h)})^{2}\right] =h\sigma ^{2}$ and $E\left[
(c_{it}^{(h)})^{4}\right] =h\kappa +3h(h-1)\sigma ^{4}.$ The excess kurtosis
of $c_{it}^{(h)}$ is then 
\begin{equation*}
\frac{E\left[ (c_{it}^{(h)})^{4}\right] }{\left\{ E\left[ (c_{it}^{(h)})^{2}%
\right] \right\} ^{2}}-3=\frac{h\kappa +3h(h-1)\sigma ^{4}}{h^{2}\sigma ^{4}}%
-3=\frac{\left[ (\kappa /\sigma ^{4})-3\right] }{h}.
\end{equation*}%
If $\varepsilon _{it}$ has positive excess kurtosis ($\kappa /\sigma
^{4}>3$), then the excess kurtosis of $c_{it}^{(h)}$ declines monotonically
to zero as $h$ increases. For a numerical illustration, suppose $\varepsilon
_{it}$ comes from a mixture of two Normals, where $\varepsilon _{it}\sim
N(0,\sigma _{1}^{2})$ with probability $p$ and $N(0,\sigma _{2}^{2})$ with
probability $1-p.$ Then $c_{it}^{(h)}=\tsum\nolimits_{s=0}^{h-1}\varepsilon
_{i,t-s}$ is a mixture of $h+1$ Normals, distributed 
\begin{equation*}
N(0,k\sigma _{1}^{2}+(h-k)\sigma _{2}^{2})\text{ with probability }\left( 
\begin{array}{l}
h \\ 
k%
\end{array}%
\right) p^{k}(1-p)^{h-k}
\end{equation*}%
and variance $E\left[ c_{it}^{(h)}\right] ^{2}=V^{(h)}=h[p\sigma
_{1}^{2}+(1-p)\sigma _{2}^{2}].$ For $\sigma _{1}=1,$ $\sigma _{2}=5,$ and $%
p=0.9,$ the probability that $c_{it}^{(h)}$ exceeds four of its standard
deviations, that is, Prob$\left( \left\vert c_{it}^{(h)}\right\vert >4\sqrt{%
V^{(h)}}\right) $, is 1.4\% for $h=1,$ 0.1\% for $h=12,$ and 0.005\% for $%
h=24.$} We find in our empirical application in Section \ref{sec_fred_md}
that this feature is extremely helpful in dealing with outliers such as
those seen during the COVID-19 pandemic in 2020.

Notwithstanding, these benefits do not come without a cost. The larger the
value of $h,$ the more serial correlation there will be in $c_{it},$ and the
more prone PCA can be in small samples to the spurious factor problem
identified by \cite{onatski2021spurious}. Our simulations in Section \ref%
{sec_simulations} suggest that if we have 50 years of data, using
two-year-ahead forecast errors works pretty well. For shorter samples, the
spurious factor problem can be more serious, and our recommendation for
those applications is to look for common factors in the one-year-ahead ($%
h=12)$ or one-month-ahead $(h=1)$ forecasts. In the latter case, additional
correction for outliers is likely needed.

To summarize, our procedure is to estimate the same regression for every
variable $y_{it},$ regardless of whether we think it is stationary and
without making any conjecture about the nature of any nonstationarity. To
allow for persistent seasonal components in $y_{it}$, we recommend choosing $%
p$ to be the number of observations in a year. Our procedure estimates the
following regression by OLS for every variable, 
\begin{equation}
y_{it}=k_{i}+\alpha _{i1}y_{i,t-h}+\alpha _{i2}y_{i,t-h-1}+\cdots +\alpha
_{ip}y_{i,t-h-p+1}+c_{it},  \label{eqn_recommended_regression}
\end{equation}%
with $h=8$ and $p=4$ for quarterly data and $h=24$ and $p=12$ for monthly
data. We will refer to the residual from the estimated regression $\hat{c}%
_{it}$ as the OLS residual for variable $y_{it}$ and the residual from the
population linear projection $c_{it}$ as the true cyclical component. The
value $\hat{c}_{it}$ is a consistent estimate of $c_{it},$ and the true
value $c_{it}$ is stationary as long as any nonstationarity in $y_{it}$ is
characterized by either a polynomial time trend of order $d_{i}$ or an $%
I(d_{i})$ process with $d_{i}\leq p.$ Our procedure is to perform PCA on the
regression residuals $\{\hat{c}_{1t},...,\hat{c}_{Nt}\}.$

One practical decision is whether a nonlinear transformation of the raw data
is necessary for $\Delta ^{d_{i}}y_{it}$ to be stationary for some $d_{i}$.
If taking first differences of the log is the correct way to produce a
stationary series, then taking the change in the level would not produce a
stationary series. We recommend using logs for variables like output or
prices which are usually described in terms of growth rates. \ For such
variables we use the log of the level of the variable, $y_{it}=\log Y_{it}$
as the variable in the regression (\ref{eqn_recommended_regression}). \ For
variables like interest rates or the unemployment rate that are already
quoted in percentage terms, we use the raw data $y_{it}=Y_{it}$ in the
regression.

The true cyclical component $c_{it}$ has mean zero and is stationary for a
wide range of processes. However, the population value of $\alpha _{i0}$ is
not known but must be estimated by regression. In Section \ref{sec_c_obs} we
characterize our assumptions about the factor structure that we hypothesize
describes the true values of $c_{it},$ and use standard results to establish
that these population factors could be consistently estimated if the true
values of $c_{it}$ were observed without error. Section \ref%
{sec_c_not_observed} considers the case when we do not know the value of $%
d_{i}$ for each series, do not know whether it is stationary or
characterized by a deterministic time trend or an $I(d_{i})$ process, and
the $c_{it}$ are not observed. In that section we analyze the consequences
of performing PCA on the estimated OLS residuals $\hat{c}_{it}.$

\section{Principal component analysis when the true cyclical component is
observed}

\label{sec_c_obs}In the previous section we defined the true cyclical
component $c_{it}$ to be the residual from a population linear projection of 
$y_{it}$ on $(1,y_{i,t-h},y_{i,t-h-1},...,y_{i,t-h-p+1})^{\prime }$, and
noted that $c_{it}$ is stationary for a broad class of possible processes.
In this section we provide sufficient conditions under which the true
cyclical components for a collection of $N$ different variables would have a
factor structure that could be consistently estimated using PCA if we
observed the true value of $c_{it}$ for each variable. The set-up and
results in this section closely follow \cite{stock2002forecasting}.

\subsection{Assumed factor structure of the true cyclical components}

Collect the true cyclical components for the $N$ different series at time $t$
in an $(N\times 1)$ vector $C_{t}=(c_{1t},...,c_{Nt})^{\prime }.$ We
postulate that these are characterized by a factor structure of the form 
\begin{equation}
\underset{(N\times 1)}{C_{t}}=\underset{(N\times r)}{\Lambda }\underset{%
(r\times 1)}{F_{t}}+\underset{(N\times 1)}{e_{t}}.
\label{eqn_factor_equation}
\end{equation}%
The number of latent factors $r$ is much less than the number of variables $%
N,$ but the $r$ factors are assumed to account for most of the variance of $%
C_{t}$ in a sense made formal below. Since the factors are unobserved, $%
C_{t}=\Lambda H^{-1}HF_{t}+e_{t}$ would imply the identical observable model
as (\ref{eqn_factor_equation}). Thus some normalizations are necessary in
order to talk about consistently estimating the $j$th factor $f_{jt}.$ In
empirical estimation, practitioners typically resolve this ambiguity by
estimating the $j$th column of $\Lambda $ by the eigenvector associated with
the $j$th largest eigenvalue of $T^{-1}\sum\nolimits_{t=1}^{T}C_{t}C_{t}^{%
\prime }.$ Note that such a procedure implies a normalization in which the
columns of $\Lambda $ are orthogonal to each other and the elements of $%
F_{t} $ are uncorrelated with each other and ordered by the size of their
variance. We follow \cite{stock2002forecasting} in how to characterize these
conventions as the cross-section dimension $N$ and time-series dimension $T$
get large.\footnote{%
See \cite{BaiNg2013} and \cite{StockWatson2016Handbook} for discussion of
alternative normalizations.}

\begin{assumption}[factor structure]
\label{ass_factor} \mbox{}

\begin{enumerate}
\item \label{ass_factor_a} $(\Lambda^{\prime }\Lambda/N) \rightarrow I_r$.

\item \label{ass_factor_b} $E[F_tF_t^{\prime }] = \Omega_{FF}$, where $%
\Omega_{FF}$ is a diagonal matrix with $\omega_{ii} > \omega_{jj} > 0 $ for $%
i < j$.

\item \label{ass_factor_c}$|\lambda _{ij}|\leq \bar{\lambda}<\infty $.

\item \label{ass_factor_d}$T^{-1}\sum_{t}F_{t}F_{t}^{\prime }\overset{p}{%
\rightarrow }\Omega _{FF}$.
\end{enumerate}
\end{assumption}

In addition to implementing the property that eigenvectors of a symmetric
matrix are orthogonal, Assumption \ref{ass_factor}(\textit{\ref{ass_factor_a}%
}) requires that each factor makes a nonnegligible contribution to the
average variance of $c_{it}$ across $i$. That is, if we were to imagine
adding more variables (increasing $N)$ with $\lambda _{ij}=0$ for all $i$
greater than some fixed $N_{0},$ then Assumption \ref{ass_factor}(\textit{%
\ref{ass_factor_a}}) could not hold. Likewise \ref{ass_factor}(\textit{\ref%
{ass_factor_b}}) and \ref{ass_factor}(\textit{\ref{ass_factor_d}}) require
that each factor continues to matter as the number of time-series
observations $T$ grows. These conditions are consistent with serial
dependence of the factors, but rely on the fact that $C_{t}$ is stationary.

Let $\gamma $ denote an $(N\times 1)$ vector and $\Gamma =\{\gamma :\gamma
^{\prime }\gamma /N=1\}.$ Note that if $\gamma $ were the $j$th column of $%
\Lambda ,$ the scalar $\gamma ^{\prime }\Lambda F_{t}/N$ would converge to $%
f_{jt}$ and $(N^{2}T)^{-1}\sum\nolimits_{t=1}^{T}\gamma ^{\prime }\Lambda
F_{t}F_{t}^{\prime }\Lambda ^{\prime }\gamma \overset{p}{\rightarrow }\omega
_{jj}.$ The assumption that the idiosyncratic elements $e_{t}$ do not have a
factor structure requires that there is no value of $\gamma $ for which the
analogous operation applied to $e_{t}$ would lead to anything other than
zero: $\underset{\gamma \in \Gamma }{\sup }(N^{2}T)^{-1}\sum%
\nolimits_{t=1}^{T}\gamma ^{\prime }e_{t}e_{t}^{\prime }\gamma \overset{p}{%
\rightarrow }0.$ \cite{stock2002forecasting} used the following assumptions
to guarantee the absence of a factor structure in $e_{t}.$

\begin{assumption}[moments of the errors]
\label{ass_error} \mbox{}

\begin{enumerate}
\item $\underset{N\rightarrow \infty}{\lim}\sup_t
\sum_{s=-\infty}^{\infty}|E[e_t^{\prime }e_{t+s}/N]|<\infty$.

\item $\underset{N\rightarrow \infty}{\lim}\sup_t
N^{-1}\sum_{i=1}^{N}\sum_{j=1}^{N}|E[e_{it}e_{jt}]|<\infty$, where $e_{it}$
denotes the $i$th element of $e_t$.

\item $\underset{N\rightarrow \infty }{\lim }\sup_{t,s}N^{-1}\sum_{i=1}^{N}%
\sum_{j=1}^{N}|cov[e_{is}e_{it},e_{js}e_{jt}]|<\infty $.
\end{enumerate}
\end{assumption}

Some might be concerned that we have simply postulated that the true
cyclical components are characterized by Assumptions \ref{ass_factor} and %
\ref{ass_error}. But something very similar is done in traditional
applications that assume conditions like these characterize specified
stationary transformations of the original data. Indeed, insofar as the
cyclical components have a common primitive definition in terms of $h$%
-period-ahead forecast errors, we would argue that these assumptions are
easier to defend in our application than in many others.

\subsection{Consequences of applying PCA to the true cyclical components}

Recall that the $(N\times 1)$ vector of true cyclical components $C_{t}$ is
stationary and has population mean zero. If $C_{t}$ was observed directly,
its estimated sample variance matrix would be $S=T^{-1}\tsum%
\nolimits_{t=1}^{T}C_{t}C_{t}^{\prime }$ and a linear combination $\gamma
^{\prime }C_{t}$ for any $(N\times 1)$ vector $\gamma $ would have sample
variance $\gamma ^{\prime }S\gamma .$ If $C_{t}$ was observed, the first
estimated principal component (denoted $\tilde{f}_{1t}=N^{-1}\tilde{\lambda}%
_{1}^{\prime }C_{t})$ would be defined as the linear combination that has
maximum sample variance subject to a normalization condition such as $\gamma
\in \Gamma =\{\gamma :\gamma ^{\prime }\gamma /N=1\}$:%
\begin{equation}
\tilde{\lambda}_{1}=\arg \underset{\gamma \in \Gamma }{\sup }\tilde{R}%
(\gamma )  \label{eqn def lambda1 tilde}
\end{equation}%
\begin{equation}
\tilde{R}(\gamma )=(N^{2}T)^{-1}\gamma ^{\prime
}\tsum\nolimits_{t=1}^{T}C_{t}C_{t}^{\prime }\gamma .  \label{eqn_Rtilde}
\end{equation}%
Note we are normalizing $\tilde{\lambda}_{1}^{\prime }\tilde{\lambda}%
_{1}/N=1 $ as we did asymptotically for the columns of $\Lambda $ in
Assumption \ref{ass_factor}(\ref{ass_factor_a})$.$ We also divide the sample
variance of $\gamma ^{\prime }C_{t}$ by $N^{2}$ in anticipation of the
result that $\tilde{R}(\tilde{\lambda}_{1})$ will converge to a fixed
constant as $N$ and $T$ grow. The solution to (\ref{eqn def lambda1 tilde})
is obtained by setting $\tilde{\lambda}_{1}$ proportional to the eigenvector
of $S=T^{-1}\sum\nolimits_{t=1}^{T}C_{t}C_{t}^{\prime }$ associated with the
largest eigenvalue. For example, if we calculated eigenvectors of this
matrix using code that normalizes eigenvectors to have unit length and
orders eigenvalues by decreasing size, $\tilde{\lambda}_{1}$ would be $\sqrt{%
N}$ times the first eigenvector. The largest eigenvalue of $S$ is equal to $%
T^{-1}\tsum\nolimits_{t=1}^{T}\tilde{f}_{1t}^{2},$ the sample variance of
the first principal component$.$ The $j$th principal component $N^{-1}\tilde{%
\lambda}_{j}^{\prime }C_{t}$ is found by maximizing $\tilde{R}(\gamma )$
subject to the constraint that $\gamma $ is orthogonal to $\tilde{\lambda}%
_{1},...,\tilde{\lambda}_{j-1}.$ The solution for $\tilde{\lambda}_{j}$ is
proportional to the eigenvector of $S$ associated with the $j$th largest
eigenvalue.

Alternatively, if we observed the true factors $F_{t}$ and loadings $\Lambda
,$ we could calculate the component of the variance of $\gamma ^{\prime
}C_{t}$ that is attributable to the $r$ factors alone:%
\begin{equation}
R^{\ast }(\gamma )=(N^{2}T)^{-1}\gamma ^{\prime
}\tsum\nolimits_{t=1}^{T}\Lambda F_{t}F_{t}^{\prime }\Lambda ^{\prime
}\gamma .  \label{eqn_R*}
\end{equation}%
\cite{stock2002forecasting} showed that under Assumptions \ref{ass_factor}
and \ref{ass_error}, the maximum value for (\ref{eqn_Rtilde}) (which is
given by the largest eigenvalue of $S)$ and the supremum of (\ref{eqn_R*})
over all $\gamma \in \Gamma $ converge in probability to the same number $%
\omega _{11},$ which is the population variance of the first factor, and
that $\tilde{\lambda}_{1}^{\prime }C_{t}/N$ gives a consistent estimate of $%
f_{1t}$ up to a sign. If we were to estimate $k>r$ principal components, the
first $r$ would consistently estimate $f_{jt}$ up to a sign normalization
and the last $k-r$ would asymptotically have zero variance. Their results
are a special case of Theorem \ref{thm_convergence_general_M} developed in
the following section.

\section{Principal component analysis when the cyclical component must be
estimated}

\label{sec_c_not_observed}

In this section we assume that we do not observe the true cyclical component 
$c_{it}$ of series $i$ but have an estimate $\hat{c}_{it}=c_{it}+\hat{v}%
_{it}.$ Let $\hat{C}_{t}=(\hat{c}_{1t},...,\hat{c}_{Nt})^{\prime }$ and $%
\hat{V}_{t}=(\hat{v}_{1t},...,\hat{v}_{Nt})^{\prime }.$ We investigate the
properties of principal components calculated from the estimated cyclical
components:%
\begin{equation}
\hat{f}_{jt}=N^{-1}\hat{\lambda}_{j}^{\prime }\hat{C}_{t}
\label{eqn_HX_f1_definition}
\end{equation}%
\begin{equation*}
\hat{\lambda}_{j}=\arg \underset{\{\gamma \in \Gamma ,\gamma ^{\prime }\hat{%
\lambda}_{1}=\cdots =\gamma ^{\prime }\hat{\lambda}_{j-1}=0\}}{\sup }\hat{R}%
(\gamma )
\end{equation*}%
\begin{equation*}
\hat{R}(\gamma )=(N^{2}T)^{-1}\gamma ^{\prime }\tsum\nolimits_{t=1}^{T}\hat{C%
}_{t}\hat{C}_{t}^{\prime }\gamma .
\end{equation*}%
We first state high-level sufficient conditions under which PCA applied to
the estimated cyclical components $\hat{C}_{t}$ gives consistent estimates
of the true factors $F_{t}.$ Let $\hat{v}_{it}=\hat{c}_{it}-c_{it}$ denote
the difference between the estimated and true cyclical component of series $%
i $ at date $t$. {\ The conditions require }$N$ to grow with $T$ so as to
ensure that $N^{-1}\tsum\nolimits_{i=1}^{N}\hat{v}_{it}^{2}$ and $%
T^{-1}\tsum\nolimits_{t=1}^{T}\hat{v}_{it}^{2}$ are $o_{p}(1).$

\begin{assumption}[high-level conditions on $\hat{v}_{it}$]
\label{ass_convergence_of_sum_vit2} The cross-section dimension $N$ grows
with the number of time series observations $T$ according to a function $%
N(T) $ such that for all $\delta ,\varepsilon >0$ there exists a $%
T_{0}(\delta ,\varepsilon )$ such that for all $T>T_{0}(\delta ,\varepsilon
) $:%
\begin{equation}
\mathnormal{Prob}\left\{ \frac{1}{N(T)}\sum_{i=1}^{N(T)}\hat{v}%
_{it}^{2}>\delta \right\} <\varepsilon  \tag{i}
\label{eqn_high_assumption_i}
\end{equation}%
\begin{equation}
\mathnormal{Prob}\left\{ \max_{1\leq i\leq N(T)}\frac{1}{T}\sum_{t=1}^{T}%
\hat{v}_{it}^{2}>\delta \right\} <\varepsilon .  \tag{ii}
\label{eqn_high_assumption_ii}
\end{equation}
\end{assumption}

The following result establishes that if the error in estimating the
cyclical component satisfies Assumption \ref{ass_convergence_of_sum_vit2},
then the results $(\hat{f}_{jt},\hat{\lambda}_{j},\hat{R}(\gamma ))$ of
applying PCA to the estimated cyclical components $\hat{C}_{t}$ give
consistent estimates of the magnitudes that characterize the true cyclical
components $C_{t}.$ For a proof see online Appendix A.

\begin{theorem}
\label{thm_convergence_general_M} Suppose that $C_{t}$ and $e_{t}$ in
equation (\ref{eqn_factor_equation}) satisfy Assumptions \ref{ass_factor}
and \ref{ass_error}. Let $\hat{C}_{t}=C_{t}+\hat{V}_{t}$ for $\hat{V}_{t}=(%
\hat{v}_{1t},...,\hat{v}_{Nt})^{\prime }$ where $\hat{v}_{it}$ satisfy
Assumption \ref{ass_convergence_of_sum_vit2}. Let $\hat{f}_{1t},...,\hat{f}%
_{kt}$ denote the first $k$ estimated principal components of $\hat{C}_{t}$ (%
$\hat{f}_{jt}=\hat{\lambda}_{j}^{\prime }\hat{C}_{t}/N)$ with $k\geq r$ and
let $\underset{(r\times 1)}{\hat{F}_{t}}=(\hat{f}_{1t},...,\hat{f}%
_{rt})^{\prime }$ and\ $\underset{(N\times r)}{\hat{\Lambda}}=\left[ 
\begin{array}{ccc}
\hat{\lambda}_{1} & \cdots & \hat{\lambda}_{r}%
\end{array}%
\right] .$ Then

\begin{enumerate}
\item $T^{-1}\tsum\nolimits_{t=1}^{T}\hat{f}_{jt}^{2}=\hat{R}(\hat{\lambda}%
_{j})\overset{p}{\rightarrow }\omega _{jj}$ for $j=1,...,r;$

\item $T^{-1}\sum\nolimits_{t=1}^{T}\hat{f}_{jt}^{2}\overset{p}{\rightarrow }%
0$ for $j=r+1,...,k;$

\item $\hat{\Xi}\hat{\Lambda}^{\prime }\Lambda /N\overset{p}{\rightarrow }%
I_{r}$ where $\hat{\Xi}$ is a diagonal matrix whose row $j$ column $j$
element is $+1$ if $\hat{\lambda}_{j}^{\prime }\lambda _{j}>0$ and $-1$ if $%
\hat{\lambda}_{j}^{\prime }\lambda _{j}<0;$

\item $\hat{\Xi}\hat{F}_{t}-F_{t}\overset{p}{\rightarrow }0$.
\end{enumerate}
\end{theorem}

Before presenting formal sufficient conditions for verifying Assumption \ref%
{ass_convergence_of_sum_vit2}, we first discuss the intuition for why we
might expect it to hold. For $z_{i,t-h}=(1,y_{i,t-h},y_{i,t-h-1},...,%
\allowbreak y_{i,t-h-p+1})^{\prime },$ the true cyclical component $c_{it}$
is the residual from a population linear projection of $y_{it}$ on $%
z_{i,t-h} $ and $\hat{c}_{it}$ is the residual from the corresponding
estimated regression:%
\begin{equation*}
c_{it}=y_{it}-\alpha _{i0}^{\prime }z_{i,t-h}
\end{equation*}%
\begin{equation*}
\hat{c}_{it}=y_{it}-\hat{\alpha}_{i}^{\prime }z_{i,t-h}
\end{equation*}%
\begin{equation*}
\hat{v}_{it}=\hat{c}_{it}-c_{it}=(\alpha _{i0}-\hat{\alpha}_{i})^{\prime
}z_{i,t-h}
\end{equation*}%
\begin{equation}
(\alpha _{i0}-\hat{\alpha}_{i})=-\left(
\tsum\nolimits_{t=1}^{T}z_{i,t-h}z_{i,t-h}^{\prime }\right) ^{-1}\left(
\tsum\nolimits_{t=1}^{T}z_{i,t-h}c_{it}\right)
\label{eqn_alpha_minus_alphahat}
\end{equation}%
\begin{equation}
\hat{v}_{it}^{2}=(\alpha _{i0}-\hat{\alpha}_{i})^{\prime
}z_{i,t-h}z_{i,t-h}^{\prime }(\alpha _{i0}-\hat{\alpha}_{i})
\label{eqn_vit2}
\end{equation}%
\begin{equation}
\tsum\nolimits_{t=1}^{T}\hat{v}_{it}^{2}=(\alpha _{i0}-\hat{\alpha}%
_{i})^{\prime }\left( \tsum\nolimits_{t=1}^{T}z_{i,t-h}z_{i,t-h}^{\prime
}\right) (\alpha _{i0}-\hat{\alpha}_{i}).  \label{eqn_sum_vit2}
\end{equation}%
Expression (\ref{eqn_sum_vit2}) will be recognized as the OLS Wald statistic
for testing the true null hypothesis $H_{0}:\alpha _{i}=\alpha _{i0}$
multiplied by $\hat{\sigma}_{i}^{2}=(T-k)^{-1}\tsum\nolimits_{t=1}^{T}\hat{c}%
_{it}^{2},$ the average squared regression residual. As we now sketch, this
statistic would be expected to be $O_{p}(1)$ for a wide class of stationary
and nonstationary processes. This would lead us to expect that the
individual terms $\hat{v}_{it}^{2}$ should be $o_{p}(1),$ with some
uniformity conditions across $i$ then guaranteeing Assumption \ref%
{ass_convergence_of_sum_vit2}(\ref{eqn_high_assumption_i}). Likewise if we
divide (\ref{eqn_sum_vit2}) by $T$ we should again get an $o_{p}(1)$ random
variable, as required by Assumption \ref{ass_convergence_of_sum_vit2}(\ref%
{eqn_high_assumption_ii}).

We now explore the intuition for why we would typically expect (\ref%
{eqn_sum_vit2}) to be $O_{p}(1).$ Note that the expression can be written 
\begin{eqnarray}
\tsum\nolimits_{t=1}^{T}\hat{v}_{it}^{2} &=&\left(
\tsum\nolimits_{t=1}^{T}c_{it}z_{i,t-h}^{\prime }\right) \left(
\tsum\nolimits_{t=1}^{T}z_{i,t-h}z_{i,t-h}^{\prime }\right) ^{-1}\left(
\tsum\nolimits_{t=1}^{T}z_{i,t-h}c_{it}\right)  \notag \\
&=&\left( \tsum\nolimits_{t=1}^{T}c_{it}z_{i,t-h}^{\prime }\Upsilon
_{iT}^{-1}\right) \left( \Upsilon
_{iT}^{-1}\tsum\nolimits_{t=1}^{T}z_{i,t-h}z_{i,t-h}^{\prime }\Upsilon
_{iT}^{-1}\right) ^{-1}\left( \Upsilon
_{iT}^{-1}\tsum\nolimits_{t=1}^{T}z_{i,t-h}c_{it}\right)
\label{eqn_sum_vit2_scaled}
\end{eqnarray}%
where $\Upsilon _{iT}^{-1}$ could be any sequence of nonsingular matrices.
Note that the researcher does not need to know the value of $\Upsilon _{iT}.$
This matrix is just a device to calculate the asymptotic properties of the
left side of (\ref{eqn_sum_vit2_scaled}) (which does not depend on $\Upsilon
_{iT})$ under various possible forms of nonstationarity for $z_{it}.$

If $y_{it}$ were stationary, we would use $\Upsilon
_{iT}^{-1}=T^{-1/2}I_{p+1}$ to find the asymptotic distribution of $%
\tsum\nolimits_{t=1}^{T}\hat{v}_{it}^{2}.$ In this case, $z_{i,t-h}c_{it}$
is a stationary random variable which by the definition of the true cyclical
component $c_{it}$ has population mean zero. In this case, a central limit
theorem could be used to establish that $\Upsilon
_{iT}^{-1}\tsum\nolimits_{t=1}^{T}z_{i,t-h}c_{it}=T^{-1/2}\tsum%
\nolimits_{t=1}^{T}z_{i,t-h}c_{it}$ converges to a Normal distribution.
Likewise in the stationary case, 
\begin{equation*}
\Upsilon _{iT}^{-1}\tsum\nolimits_{t=1}^{T}z_{i,t-h}z_{i,t-h}^{\prime
}\Upsilon
_{iT}^{-1}=T^{-1}\tsum\nolimits_{t=1}^{T}z_{i,t-h}z_{i,t-h}^{\prime }\overset%
{p}{\rightarrow }E(z_{i,t-h}z_{i,t-h}^{\prime }).
\end{equation*}%
For this reason, $\tsum\nolimits_{t=1}^{T}\hat{v}_{it}^{2}$ would typically
be $O_{p}(1)$ when $y_{it}$ is stationary.

Consider next the case when $d_{i}=1$ and $\Delta y_{it}$ (denoted below by $%
u_{it})$ is a zero-mean stationary process. Taking a regression with $p=3$
lags for illustration, an OLS regression of $y_{it}$ on $%
z_{i,t-h}=(1,y_{i,t-h},y_{i,t-h-1},y_{i,t-h-2})^{\prime }$ has the identical
fitted values as an OLS regression of $y_{it}$ on $%
R_{i}^{-1}z_{i,t-h}=(u_{i,t-h},u_{i,t-h-1},1,y_{i,t-h})^{\prime }.$ In this
case to calculate the asymptotic distribution of (\ref{eqn_sum_vit2_scaled})
we would take%
\begin{equation*}
\Upsilon _{iT}^{-1}=\left[ 
\begin{array}{ll}
T^{-1/2}I_{3} & 0 \\ 
0 & T^{-1}%
\end{array}%
\right] R_{i}^{-1}
\end{equation*}%
\begin{eqnarray*}
&&\Upsilon _{iT}^{-1}\tsum\nolimits_{t=1}^{T}z_{i,t-h}z_{i,t-h}^{\prime
}\Upsilon _{iT}^{-1} \\
&=&\left[ 
\begin{array}{llll}
T^{-1}\tsum\nolimits_{t=1}^{T}u_{i,t-h}^{2} & T^{-1}\tsum%
\nolimits_{t=1}^{T}u_{i,t-h}u_{i,t-h-1} & T^{-1}\tsum%
\nolimits_{t=1}^{T}u_{i,t-h} & T^{-3/2}\tsum%
\nolimits_{t=1}^{T}u_{i,t-h}y_{i,t-h} \\ 
T^{-1}\tsum\nolimits_{t=1}^{T}u_{i,t-h-1}u_{i,t-h} & T^{-1}\tsum%
\nolimits_{t=1}^{T}u_{i,t-h-1}^{2} & T^{-1}\tsum%
\nolimits_{t=1}^{T}u_{i,t-h-1} & T^{-3/2}\tsum%
\nolimits_{t=1}^{T}u_{i,t-h-1}y_{i,t-h} \\ 
T^{-1}\tsum\nolimits_{t=1}^{T}u_{i,t-h} & T^{-1}\tsum%
\nolimits_{t=1}^{T}u_{i,t-h-1} & 1 & T^{-3/2}\tsum%
\nolimits_{t=1}^{T}y_{i,t-h} \\ 
T^{-3/2}\tsum\nolimits_{t=1}^{T}y_{i,t-h}u_{i,t-h} & T^{-3/2}\tsum%
\nolimits_{t=1}^{T}y_{i,t-h}u_{i,t-h-1} & T^{-3/2}\tsum%
\nolimits_{t=1}^{T}y_{i,t-h} & T^{-2}\tsum\nolimits_{t=1}^{T}y_{i,t-h}^{2}%
\end{array}%
\right] .
\end{eqnarray*}%
Under standard unit-root asymptotics (e.g., \citet[p. 506]{hamilton1994}),
we expect $\Upsilon
_{iT}^{-1}\tsum\nolimits_{t=1}^{T}z_{i,t-h}z_{i,t-h}^{\prime }\Upsilon
_{iT}^{-1}\overset{d}{\rightarrow }Q_{i}$ with%
\begin{equation}
Q_{i}=\left[ 
\begin{array}{llll}
\gamma _{i0} & \gamma _{i1} & 0 & 0 \\ 
\gamma _{i1} & \gamma _{i0} & 0 & 0 \\ 
0 & 0 & 1 & \omega _{i}\int_{0}^{1}W_{i}(r)dr \\ 
0 & 0 & \omega _{i}\int_{0}^{1}W_{i}(r)dr & \omega _{i}^{2}\int_{0}^{1}\left[
W_{i}(r)\right] ^{2}dr%
\end{array}%
\right]  \label{eqn_Q_example}
\end{equation}%
for $\gamma _{ij}=E(u_{it}u_{i,t-j}),$ $\omega
_{i}^{2}=\tsum\nolimits_{j=-\infty }^{\infty }\gamma _{ij},$ and $W_{i}(r)$
standard Brownian motion. Similar tools can be used to establish that $%
\Upsilon _{iT}^{-1}\tsum\nolimits_{t=1}^{T}z_{i,t-h}c_{it}\overset{d}{%
\rightarrow }q_{i}$ and $\tsum\nolimits_{t=1}^{T}\hat{v}_{it}^{2}\overset{d}{%
\rightarrow }q_{i}^{\prime }Q_{i}^{-1}q_{i}\sim O_{p}(1).$

If instead $E(\Delta y_{it})=\mu _{i}\neq 0$, the level $y_{it}$ would be
dominated asymptotically by a deterministic time trend $\mu _{i}t.$\footnote{%
\label{fn_trend} That is, for any nonnegative integer $\nu $,%
\begin{equation*}
T^{-(\nu +1)}\tsum\nolimits_{t=1}^{T}y_{it}^{\nu }-T^{-(\nu
+1)}\tsum\nolimits_{t=1}^{T}(\mu _{i}t)^{\nu }\overset{p}{\rightarrow }0
\end{equation*}%
as $T\rightarrow \infty $ and $T^{-(\nu +1)}\tsum\nolimits_{t=1}^{T}(\mu
_{i}t)^{\nu }\rightarrow \mu _{i}^{\nu }/(\nu +1).$} In this case (and for
general $p\geq 2)$ we would use\ $R_{i}^{-1}z_{i,t-h}=(\Delta y_{i,t-h}-\mu
_{i},\Delta y_{i,t-h-1}-\mu _{i},...,\Delta y_{i,t-h+2}-\mu
_{i},1,y_{i,t-h})^{\prime }$ and 
\begin{equation*}
\Upsilon _{iT}^{-1}=\left[ 
\begin{array}{ll}
T^{-1/2}I_{p} & 0 \\ 
0 & T^{-3/2}%
\end{array}%
\right] R_{i}^{-1}
\end{equation*}%
to establish that $\tsum\nolimits_{t=1}^{T}\hat{v}_{it}^{2}\sim O_{p}(1).$
General cases for $d_{i}\leq p$ are examined in \cite{hamilton2018you}. Note
that we do not need to know the value of $d_{i}$ or $\Upsilon _{iT}$ to
estimate any magnitudes -- in every case we are talking about a regression
of $y_{it}$ on the lagged levels $%
z_{i,t-h}=(1,y_{i,t-h},y_{i,t-h-1},...,y_{i,t-h-p+1})^{\prime }.$ Instead, $%
\Upsilon _{iT}$ is just a device to analyze the properties if the levels
regression is applied to a variety of different stationary and nonstationary
processes.

We are now in a position to state formally sufficient conditions under which
Assumption \ref{ass_convergence_of_sum_vit2} would be satisfied, for which
we let $\left\Vert X\right\Vert $ denote the Frobenius/Euclidean norm (the
square root of the sums of squares of the elements of a vector or matrix $X)$%
.\footnote{%
An earlier version of this paper avoided the requirement that $T$ grows
faster than $N$ by assuming that the true cyclical component $c_{it}$ is
uniformly bounded.}

\begin{assumption}[sufficient conditions for Assumption \protect\ref%
{ass_convergence_of_sum_vit2}]
\label{ass_sufficient} For each $i$, there exist some nonsingular matrix $%
\Upsilon_{iT}$ and a possibly random matrix $Q_i$, such that \mbox{}

\begin{enumerate}
\item \label{ass_sufficient_a} For some positive constant $\mathfrak{c}_{1}$%
, 
\begin{align*}
\max_{1\leq i\leq N}\max_{1\leq t\leq T}E( \Vert \sqrt{T}%
\Upsilon_{iT}^{-1}z_{i,t-h}\Vert^2 ) \leq \mathfrak{c}_1\quad \text{ and}%
\quad \max_{1\leq i\leq N}E( \Vert\sum_{t=1}^T
\Upsilon_{iT}^{-1}z_{i,t-h}c_{it}\Vert^2 ) \leq \mathfrak{c}_1.
\end{align*}

\item \label{ass_sufficient_b} For some positive constants $\mathfrak{c}_2$, 
$\mathfrak{c}_3$, $\mathfrak{c}_4$, and $\mathfrak{c}_5$, 
\begin{align*}
\max_{1\leq i\leq N}\mathnormal{Prob}\big(\varrho_{\min}(Q_i) < \delta\big) %
\leq \mathfrak{c}_2 \exp\big\{-\mathfrak{c}_3\big(\frac{1}{\delta}\big)^{1/%
\mathfrak{c}_4}\big\}\quad \text{ for all $\delta\leq \mathfrak{c}_5$},
\end{align*}
where $\varrho_{\min}$ denotes the minimum eigenvalue. In addition, $N\log^{2%
\mathfrak{c}_4}(N)/T \rightarrow 0$.

\item \label{ass_sufficient_c} The cross-section dimension $N$ grows with
the number of time series observations $T$ according to a function $N(T) $,
such that 
\begin{align*}
\mathnormal{Prob}\big(\max_{1\leq i\leq N} \big\Vert \sum_{t=1}^T
\Upsilon_{iT}^{-1}z_{i,t-h}z_{i,t-h}^{\prime }\Upsilon_{iT}^{-1} - Q_i %
\big\Vert > \mathfrak{c}_{6}(N)\big) \rightarrow 0
\end{align*}
for some positive sequence $\mathfrak{c}_{6}(N)= o(\log^{-\mathfrak{c}%
_4}(N)) $.
\end{enumerate}
\end{assumption}

{Condition \ref{ass_sufficient}(\ref{ass_sufficient_a}) is quite mild, as it
only requires finite moments on the transformed regressors and the cyclical
components after suitable normalization. For example, when }$\Delta y_{it}$
is a zero-mean stationary process, the first inequality requires $\Delta
y_{it}$ and $y_{it}/\sqrt{T}$ to have finite variance. {Conditions \ref%
{ass_sufficient}(\ref{ass_sufficient_b}--\ref{ass_sufficient_c}) are key to
our theoretical developments. Specifically, they posit that the matrix that
gets inverted in (\ref{eqn_sum_vit2_scaled}) can be uniformly approximated
by a possibly random limit $Q_{i}$, and that the minimum eigenvalue of the
limiting matrix exhibits an exponential-type tail at zero. }

\begin{theorem}
\label{thm_main_result} Assumption \ref{ass_sufficient} implies Assumption %
\ref{ass_convergence_of_sum_vit2}.
\end{theorem}

{Next we relate Assumption \ref{ass_sufficient} to conditions for general
linear processes.}

\begin{assumption}[linear process]
\label{ass_linear_process} There exist positive constants $\mathfrak{c}_{7}$%
--$\mathfrak{c}_{11}$, such that for each $i\in \{1,2,\dots ,N\}$ there
exist a $\mu _{i}$ and an integer $d_{i}$ for which $%
u_{it}=(1-L)^{d_{i}}y_{it}-\mu _{i}=\sum_{l=0}^{\infty }\psi _{il}\eta
_{i,t-l}$, and

\begin{enumerate}
\item \label{ass_linear_process_a} $\max_{1\leq i\leq N}|\psi _{il}|\leq 
\mathfrak{c}_{7}l^{-\mathfrak{c}_{8}}$ with $\mathfrak{c}_{8}>2$.

\item \label{ass_linear_process_a1} $\min_{1\leq i\leq N}|\sum_{l=0}^{\infty
}\psi_{il}|\geq \mathfrak{c}_{9}>0$.

\item \label{ass_linear_process_b} $\eta _{it}$ is an i.i.d. process with
zero mean, unit variance, and $\max_{1\leq i\leq N}E(\eta _{it}^{4})\leq 
\mathfrak{c}_{10}$.

\item \label{ass_linear_process_d} Let $\tilde{Q}_{i}$ be the covariance
matrix of $(u_{it},\ u_{i,t-1},\ \dots ,\ u_{i,t-p+d_{i}+1})$. Then $\varrho
_{\min }(\tilde{Q}_{i})\geq \mathfrak{c}_{11}>0$.
\end{enumerate}
\end{assumption}

Conditions \ref{ass_linear_process}(\ref{ass_linear_process_a}) and \ref%
{ass_linear_process}(\ref{ass_linear_process_a1}) would be satisfied if $%
u_{it}$ is any zero-mean stationary $AR(g)$ process for some $g$,%
\begin{equation*}
u_{it}=\phi _{i1}u_{i,t-1}+\cdots +\phi _{ig}u_{i,t-g}+\eta _{it}\text{ \ \
with }\left\Vert z\right\Vert >1\text{ for any }z\text{ satisfying\ }%
\left\vert 1-\phi _{i1}z-\cdots -\phi _{ig}z^{g}\right\vert =0.
\end{equation*}%
Alternatively, \ref{ass_linear_process}(\ref{ass_linear_process_a}) and \ref%
{ass_linear_process}(\ref{ass_linear_process_a1}) would also be satisfied
for any $MA(g)$ process that has a representation that is not arbitrarily
close to being noninvertible,%
\begin{equation*}
u_{it}=\eta _{it}+\theta _{i1}\eta _{i,t-1}+\cdots +\theta _{ig}\eta _{i,t-g}%
\text{ \ \ with }\left\vert 1+\theta _{i1}+\cdots +\theta _{ig}\right\vert >%
\text{$\mathfrak{c}$}_{9}.
\end{equation*}

Condition \ref{ass_linear_process} (\ref{ass_linear_process_a1}) defines the
necessary degree of differencing $d_{i}.$ For example, if $y_{it}=\psi
_{i}(L)\eta _{it}$ with $\psi _{i\ell }$ satisfying $\psi _{i}(1)>\mathfrak{c%
}_{9},$ then the $MA(\infty )$ coefficients $\tilde{\psi}_{i\ell }$ of the
first difference $(1-L)y_{it}=(1-L)\psi _{i}(L)\eta _{it}=\tilde{\psi}%
_{i}(L)\eta _{it}$ would not satisfy (\ref{ass_linear_process_a1}) because $%
\tilde{\psi}_{i}(1)=(1-1)\psi _{i}(1)=0.$

\begin{theorem}
\label{thm_main_result_linear_process} Suppose that $p\geq 1$ and Assumption %
\ref{ass_linear_process} is satisfied for every $i\in\{1,2,\dots,N\}$ for
some $d_i\in\{0,1\}$, and that the true cyclical components $c_{it}$ satisfy
Assumptions \ref{ass_factor} and \ref{ass_error}. If $N \log^{12}(N) / T
\rightarrow 0$, then results (i)--(iv) of Theorem \ref%
{thm_convergence_general_M} hold.
\end{theorem}

We now briefly outline our proof technique and main technical contribution.
It is helpful to compare the current setting with the case where $N$ is
fixed and $T$ grows. The fixed-$N$ case is straightforward, as one can
establish either convergence in probability to a nonrandom $Q_i$ (under weak
stationarity) or convergence in distribution to a random $Q_i$ (as
functionals of Brownian motion). This aligns with the reasoning provided
earlier (see \eqref{eqn_Q_example} and the related discussion). However,
when $N$ is allowed to grow, the theoretical analysis requires uniform
control of the estimation error across $i=1, \dots, N$, as implied by
Assumption \ref{ass_convergence_of_sum_vit2}. The core of our proof is to
decompose the quantity of interest into a sum of i.i.d. components plus
negligible remainder terms. We then apply the coupling method of \cite%
{komlos1976approximation} to establish a distributional approximation that
is uniformly valid for $i=1,\dots,N$.

While our proof of Theorem \ref{thm_main_result_linear_process} only covers
the case when $d_{i}\leq 1,$ we conjecture that a similar result could be
obtained for any $d_{i}\leq p.$ Extending the proof to this case requires
verifying the three conditions in Assumption \ref{ass_sufficient}.
Assumption \ref{ass_sufficient}(\ref{ass_sufficient_a}) requires only finite
moments of the regressors and the cyclical components after suitable
transformation and rescaling by the appropriate $\Upsilon _{iT}$. This is
true for processes satisfying Assumption \ref{ass_linear_process} with $%
d_{i}=2,$ although the calculations demonstrating this would be more tedious
than the $d_{i}=1$ case we review. It is also possible to establish \ref%
{ass_sufficient}(\ref{ass_sufficient_c}) for $d_{i}\geq 2$. The key step in
our proof builds on the coupling result of \cite{komlos1976approximation},
which provides a uniform (in both $1\leq i\leq N$ and $1\leq t\leq T$)
strong approximation of a random walk by sums of independent Gaussian
variables. This result can also be applied to provide a uniform (in $1\leq
i\leq N$) approximation of the $Q_{iT}$ matrix using $d_{i}$-fold integrals
of Brownian motion. As another extension, we show in Appendix A.4 that the
strong approximation result also holds for series that are local to unity.

The main technical challenge in extending Theorem \ref%
{thm_main_result_linear_process} to the case when $d_{i}\geq 2$ and the
local-to-unity scenario is verifying Assumption \ref{ass_sufficient}(\ref%
{ass_sufficient_b}). When $d_{i}\geq 2$ and $\mu _{i}=0,$ the limiting random
matrix $Q_{i}$ involves $n$-fold integrals of Brownian motion. Based on the
literature of small ball probabilities for $n$-fold integrals of Brownian
motion \citep{chen2003quadratic,gao2003integrated}, we conjecture that \ref%
{ass_sufficient}(\ref{ass_sufficient_b}) holds with $\mathfrak{c}%
_{4}=2\max_{1\leq i\leq N}d_{i}$, though we were unable to locate a concrete
demonstration of this in the literature. Regarding the local-to-unity case,
existing results on small ball probabilities for Ornstein-Uhlenbeck
processes \citep{ai2016note} suggest that \ref{ass_sufficient}(\ref%
{ass_sufficient_b}) holds with $\mathfrak{c}_{4}=2$. Thus, although our
current Theorem \ref{thm_main_result_linear_process} applies only to
stationary and $I(1)$ processes, we expect that it extends to series with
any order of integration and those that are local to unity, with the only
change that the exponent on the log of $N$ may differ.

Finally, we note that no additional proof is necessary to extend Theorem \ref%
{thm_main_result_linear_process} to the case when $2\leq d_{i}\leq p$ if the
number of series for which $d_{i}\geq 2$ does not increase with $N$ or $T$.

\section{Results from simulations}

\label{sec_simulations}

In this section we report results from applying our method in a variety of
different settings. In these simulations we take the number of cross-section
variables to be $N=100$ and vary the number of time-series observations $T$
from 100 to 1000. For each generated sample, we calculate the $(N\times N)$
correlation matrices of: (1) the raw data; (2) the OLS regression residuals
for that data set; and (3) the true cyclical components for that data set
implied by the particular process that was used to generate that sample. For
each of these correlation matrices, let $\hat{\xi}_{j}$ denote the $j$th
largest eigenvalue of the correlation matrix. The fraction of the variance
of the sample explained by the $j$th principal component is%
\begin{equation}
R_{j}^{2}=\hat{\xi}_{j}/N.  \label{eqn_R2}
\end{equation}%
We also calculated the number of factors that would be selected for that
sample based on the $IC_{p2}$ criterion of \cite{BaiNg2002} recommended by %
\citet[p. 436]{StockWatson2016Handbook},%
\begin{equation}
r^{\ast }=\arg \left\{ \underset{r\in \{0,1,...,r_{0}\}}{\min }\log \left( 1-%
\frac{\tsum\nolimits_{j=0}^{r}\hat{\xi}_{j}}{N}\right) +r\left( \frac{N+T}{NT%
}\right) \log(\min \{N,T\})\right\} ,  \label{eqn_bic}
\end{equation}%
with $\hat{\xi}_{0}$ defined to be 0. We took $r_{0}=10$ and for each of the
different cases generated 1,000 different samples.

\subsection{Mix of unrelated stationary and nonstationary variables}

For our first example, half the variables are random walks and the other
half are white noise,%
\begin{equation*}
y_{it}=\left\{ 
\begin{array}{cc}
y_{it-1}+\varepsilon _{it}=\varepsilon _{it}+\varepsilon _{i,t-1}+\cdots
+\varepsilon _{i1} & \text{for }i=1,2,...,N/2 \\ 
\varepsilon _{it} & \text{for }i=(N/2)+1,...,N%
\end{array}%
\right.
\end{equation*}%
for $t=1,...,T.$ The innovations $\varepsilon _{it}\sim N(0,1)$ are
independent across all $i$ and $t.$ Thus each of the $N$ variables is
completely independent of the others and there is no factor structure in the
true data-generating process.

The first two columns of Table \ref{tab_random_walk} report the results from
calculating principal components of the raw data. In a sample of $T=100$
observations, the first three principal components seem to account for 38\%
of the variance of the full set of $N=100$ variables. This result is
entirely spurious, and illustrates the cautions raised by \cite%
{onatski2021spurious} about using PCA when some of the variables are
nonstationary. The criterion (\ref{eqn_bic}) would always lead us
incorrectly to conclude that there is more than one factor in a sample of
size $T=100.$ This problem in the apparent number of factors gets even worse
when the sample size increases. The latter is the expected result, since (%
\ref{eqn_bic}) is in the class of criteria for which 
\citet[p.
602]{onatski2021spurious} demonstrated that the number of factors selected
diverges as $N$ and $T$ go to infinity.

\begin{table}[tbph]
\caption{Mixture of independent random walks and white noise}
\label{tab_random_walk}%
\begin{tabular}{c|cc|cc|cc|cc|cc|cc}
\hline
& \multicolumn{2}{c|}{Raw data} & \multicolumn{2}{c|}{$c_t$ ($h$=24)} & 
\multicolumn{2}{c|}{$\hat{c}_t$ ($h$=24)} & \multicolumn{2}{c|}{$\hat{c}_t$ (%
$h$=12)} & \multicolumn{2}{c|}{$\hat{c}_t$ ($h$=8)} & \multicolumn{2}{c|}{$%
\hat{c}_t$ ($h$=1)} \\ 
\cmidrule{2-3} \cmidrule{4-5} \cmidrule{6-7} \cmidrule{8-9} \cmidrule{10-11} %
\cmidrule{12-13} $j$ & $R^2$ & $r^*$ & $R^2$ & $r^*$ & $R^2$ & $r^*$ & $R^2$
& $r^*$ & $R^2$ & $r^*$ & $R^2$ & $r^*$ \\ 
& (1) & (2) & (3) & (4) & (5) & (6) & (7) & (8) & (9) & (10) & (11) & (12)
\\ \hline
$T=100$ &  &  &  &  &  &  &  &  &  &  &  &  \\ 
0 & --- & 0 & --- & 0 & --- & 1 & --- & 13 & --- & 96 & --- & 100 \\ 
1 & 22.6 & 0 & 16.0 & 0 & 11.9 & 26 & 10.3 & 48 & 8.4 & 4 & 4.1 & 0 \\ 
2 & 9.9 & 67 & 12.0 & 6 & 9.4 & 54 & 8.5 & 35 & 7.1 & 0 & 3.8 & 0 \\ 
3 & 5.6 & 33 & 8.3 & 94 & 7.3 & 19 & 6.9 & 4 & 6.1 & 0 & 3.6 & 0 \\ \hline
$T=200$ &  &  &  &  &  &  &  &  &  &  &  &  \\ 
0 & --- & 0 & --- & 0 & --- & 0 & --- & 78 & --- & 100 & --- & 100 \\ 
1 & 22.7 & 0 & 9.4 & 0 & 10.2 & 0 & 6.6 & 22 & 5.2 & 0 & 2.9 & 0 \\ 
2 & 9.4 & 13 & 7.8 & 3 & 8.2 & 8 & 5.7 & 0 & 4.6 & 0 & 2.7 & 0 \\ 
3 & 5.3 & 87 & 6.6 & 97 & 6.5 & 92 & 4.9 & 0 & 4.2 & 0 & 2.6 & 0 \\ \hline
$T=400$ &  &  &  &  &  &  &  &  &  &  &  &  \\ 
0 & --- & 0 & --- & 23 & --- & 3 & --- & 100 & --- & 100 & --- & 100 \\ 
1 & 22.5 & 0 & 5.9 & 51 & 6.5 & 34 & 4.3 & 0 & 3.5 & 0 & 2.2 & 0 \\ 
2 & 9.4 & 0 & 5.2 & 22 & 5.5 & 47 & 3.8 & 0 & 3.2 & 0 & 2.1 & 0 \\ 
3 & 5.1 & 100 & 4.5 & 4 & 4.7 & 16 & 3.5 & 0 & 3.0 & 0 & 2.1 & 0 \\ \hline
$T=600$ &  &  &  &  &  &  &  &  &  &  &  &  \\ 
0 & --- & 0 & --- & 99 & --- & 79 & --- & 100 & --- & 100 & --- & 100 \\ 
1 & 22.4 & 0 & 4.6 & 1 & 5.0 & 21 & 3.4 & 0 & 2.9 & 0 & 1.9 & 0 \\ 
2 & 9.3 & 0 & 4.1 & 0 & 4.4 & 0 & 3.1 & 0 & 2.7 & 0 & 1.9 & 0 \\ 
3 & 5.2 & 100 & 3.7 & 0 & 3.9 & 0 & 2.9 & 0 & 2.5 & 0 & 1.8 & 0 \\ \hline
$T=800$ &  &  &  &  &  &  &  &  &  &  &  &  \\ 
0 & --- & 0 & --- & 100 & --- & 100 & --- & 100 & --- & 100 & --- & 100 \\ 
1 & 22.7 & 0 & 4.0 & 0 & 4.2 & 0 & 3.0 & 0 & 2.6 & 0 & 1.8 & 0 \\ 
2 & 9.3 & 0 & 3.6 & 0 & 3.7 & 0 & 2.7 & 0 & 2.4 & 0 & 1.7 & 0 \\ 
3 & 5.0 & 100 & 3.3 & 0 & 3.4 & 0 & 2.6 & 0 & 2.3 & 0 & 1.7 & 0 \\ \hline
$T=1000$ &  &  &  &  &  &  &  &  &  &  &  &  \\ 
0 & --- & 0 & --- & 100 & --- & 100 & --- & 100 & --- & 100 & --- & 100 \\ 
1 & 22.6 & 0 & 3.5 & 0 & 3.7 & 0 & 2.7 & 0 & 2.4 & 0 & 1.7 & 0 \\ 
2 & 9.3 & 1 & 3.2 & 0 & 3.3 & 0 & 2.5 & 0 & 2.2 & 0 & 1.7 & 0 \\ 
3 & 5.1 & 99 & 3.0 & 0 & 3.0 & 0 & 2.4 & 0 & 2.1 & 0 & 1.6 & 0 \\ 
\hline\hline
\end{tabular}
\vspace{5pt} \newline
Notes to Table \ref{tab_random_walk}. $R^2$ indicates the percentage of
total variance accounted for by the $j$th principal component for $j=1,2$ or
3. $r^*$ indicates the percentage of samples for which the criterion (\ref%
{eqn_bic}) selects the number of factors to be $j=0,1,2,$ or $\ge 3$. In
every case, the true number of factors is $r=0$ and the cross-section
dimension is $N=100$.
\end{table}
The next two columns of Table \ref{tab_random_walk} report what the results
would be if we somehow knew the true cyclical component of each variable.
For this case, we applied PCA to a sample of $T-h$ observations for which
the $i$th observed variable for $t=h+1,h+2,...,T$ is given by%
\begin{equation}
c_{it}=\left\{ 
\begin{array}{cc}
y_{it}-y_{i,t-h}=\varepsilon _{it}+\varepsilon _{i,t-1}+\cdots +\varepsilon
_{i,t-h+1} & \text{for }i=1,2,...,N/2 \\ 
\varepsilon _{it} & \text{for }i=(N/2)+1,...,N%
\end{array}%
\right. .  \label{eqn_true_cycle_random_walk}
\end{equation}%
Note that the cyclical components in (\ref{eqn_true_cycle_random_walk}) can
be serially correlated, but this autocorrelation vanishes for observations
separated by more than $h$ periods. The cyclical components $c_{it}$ thus
satisfy by construction the conditions under which \cite{BaiNg2002}
demonstrated that (\ref{eqn_bic}) would asymptotically select the correct
number of factors. We find in our simulations that (\ref{eqn_bic}) does
indeed correctly conclude there is no factor structure for these data sets
provided the number of time-series observations is 600 or larger. For
smaller $T$ it is less reliable. The reason is that there is a small-sample
version of the \cite{onatski2021spurious} spurious factor problem that
arises from the serial correlation of some of the variables that is induced
by the definition of the true cyclical component in equation (\ref%
{eqn_true_cycle_random_walk}). If $T$ is large enough, this problem goes
away, but for smaller $T$ it can make a difference.

Columns (5) and (6) examine the case where the analysis is based on the
residuals from running an OLS regression on the raw data $Y_{it}$ for all
variables $i=1,...,N$ without making any judgments about which variables are
stationary and which are not. For large samples, the results are similar to
those that we would obtain if we somehow knew the exact correct
transformation to use for every variable.

The small-sample problem in columns (3)-(6) results from the serial
correlation that is a consequence defining the cyclical component to be the
error from a 24-period-ahead forecast. Columns (7) and (8) report results if
we instead were to look for common factors in the 12-period-ahead forecast
errors. This typically would reach the correct conclusion even in a sample
of only $T=200$ observations. Columns (9) and (10) consider 8-period-ahead
forecast errors, such as our suggested cyclical calculation would use for
quarterly data. The results indicate that if we have more than 50 years of
data ($T=600$ for monthly data or $T=200$ for quarterly data), conducting
PCA on the two-year-ahead OLS forecast residuals ($h=24$ for monthly data or 
$h=8$ for quarterly data) is reasonably reliable. With less than 50 years of
data, some researchers might want to use a smaller value for $h$ or place
less reliance on (\ref{eqn_bic}) as a criterion for selecting the number of
factors.

The last two columns of Table \ref{tab_random_walk} examine looking for
common factors in the one-period-ahead forecast errors. In these
simulations, this reaches the correct conclusion 100\% of the time that
there are zero factors in these data sets even for a sample of $T=100$
observations. Thus our proposed method appears to be quite reliable if the
interest is in identifying common factors behind one-period-ahead forecast
errors. However, one-period-ahead forecast errors are more sensitive to
outliers. This is an important consideration, as will be demonstrated in our
analysis of actual data in Section \ref{sec_fred_md}.

\subsection{Mix of unrelated stationary variables with differing persistence}

In our second example, for $i=1,2,...,N/2$ the variables are generated by a
stationary but persistent AR(1) process:%
\begin{equation*}
y_{i1}\sim N(0,1/(1-\rho ^{2}))
\end{equation*}%
\begin{equation}
y_{it}=\rho y_{i,t-1}+\varepsilon _{it}\text{ \ \ for }t=2,3,...,T.
\label{eq_simulation_AR1}
\end{equation}%
The remaining $N/2$ variables are white noise $(y_{it}=\varepsilon _{it}$
for $i=(N/2)+1,...,N).$ Our example uses $\rho =0.99,$ so all the variables
are stationary but half of them are highly persistent. The innovations $%
\varepsilon _{it}\sim N(0,1)$ are independent across all $i$ and $t,$ so
there is no factor structure in the true data-generating processes.

Columns (1) and (2) of Table \ref{tab_stationary_coint}\ report the results
from applying PCA to the raw data. Note that for this example, the raw data
themselves satisfy the \cite{BaiNg2002} conditions for asymptotic validity
of PCA. Nevertheless, even in a sample of size $T=1000,$ the first principal
component alone appears to explain a third of the data, and the criterion in
(\ref{eqn_bic}) would always conclude incorrectly that there is at least one
factor. This is a small-sample manifestation of the \cite%
{onatski2021spurious} spurious factor phenomenon. In a sufficiently large
sample, this problem would go away. But $T=1000$ is not large enough for
persistence characterized by $\rho =0.99.$

\begin{table}[tbph]
\caption{Monte Carlo results in other settings}
\label{tab_stationary_coint}%
\begin{tabular}{c|cc|cc|cc|cc|cc|cc|cc|}
\hline
& \multicolumn{6}{c|}{$\rho=0.99$} & \multicolumn{4}{c|}{Cointegrated} & 
\multicolumn{4}{c|}{Mixed} \\ 
\cmidrule{2-7} \cmidrule{8-11} \cmidrule{12-15} & \multicolumn{2}{c|}{Raw
data} & \multicolumn{2}{c|}{$\hat{c}_t$ ($h$=24)} & \multicolumn{2}{c|}{$%
\hat{c}_t$ ($h$=1)} & \multicolumn{2}{c|}{Raw data} & \multicolumn{2}{c|}{$%
\hat{c}_t$ ($h $=24)} & \multicolumn{2}{c|}{Raw data} & \multicolumn{2}{c|}{$%
\hat{c}_t$ ($h $=24)} \\ 
\cmidrule{2-3} \cmidrule{4-5} \cmidrule{6-7} \cmidrule{8-9} \cmidrule{10-11} %
\cmidrule{12-13} \cmidrule{14-15} $j$ & $R^2$ & $r^*$ & $R^2$ & $r^*$ & $R^2$
& $r^*$ & $R^2$ & $r^*$ & $R^2$ & $r^*$ & $R^2$ & $r^*$ & $R^2$ & $r^*$ \\ 
& (1) & (2) & (3) & (4) & (5) & (6) & (7) & (8) & (9) & (10) & (11) & (12) & 
(13) & (14) \\ \hline
$T=100$ &  &  &  &  &  &  &  &  &  &  &  &  &  &  \\ 
0 & --- & 0 & --- & 0 & --- & 100 & --- & 0 & --- & 0 & --- & 0 & --- & 0 \\ 
1 & 43.0 & 98 & 15.8 & 3 & 4.1 & 0 & 46.1 & 100 & 40.3 & 100 & 27.5 & 0 & 
23.8 & 0 \\ 
2 & 3.7 & 2 & 10.6 & 97 & 3.8 & 0 & 2.8 & 0 & 3.4 & 0 & 18.1 & 0 & 11.2 & 1
\\ 
3 & 2.8 & 0 & 5.5 & 0 & 3.6 & 0 & 2.5 & 0 & 3.1 & 0 & 8.5 & 100 & 8.7 & 99
\\ \hline
$T=200$ &  &  &  &  &  &  &  &  &  &  &  &  &  &  \\ 
0 & --- & 0 & --- & 0 & --- & 100 & --- & 0 & --- & 0 & --- & 0 & --- & 0 \\ 
1 & 42.5 & 83 & 10.8 & 0 & 2.9 & 0 & 48.1 & 100 & 45.7 & 100 & 25.1 & 0 & 
22.6 & 0 \\ 
2 & 3.3 & 17 & 8.7 & 1 & 2.7 & 0 & 2.1 & 0 & 2.3 & 0 & 18.0 & 0 & 9.8 & 0 \\ 
3 & 2.3 & 0 & 7.0 & 99 & 2.6 & 0 & 2.0 & 0 & 2.2 & 0 & 8.7 & 100 & 7.9 & 100
\\ \hline
$T=400$ &  &  &  &  &  &  &  &  &  &  &  &  &  &  \\ 
0 & --- & 0 & --- & 5 & --- & 100 & --- & 0 & --- & 0 & --- & 0 & --- & 0 \\ 
1 & 40.5 & 52 & 6.3 & 33 & 2.2 & 0 & 48.8 & 100 & 46.5 & 100 & 23.8 & 0 & 
21.1 & 0 \\ 
2 & 3.4 & 47 & 5.4 & 45 & 2.1 & 0 & 1.8 & 0 & 1.8 & 0 & 17.9 & 0 & 6.4 & 0
\\ 
3 & 2.2 & 1 & 4.8 & 18 & 2.1 & 0 & 1.7 & 0 & 1.7 & 0 & 8.8 & 100 & 5.4 & 100
\\ \hline
$T=600$ &  &  &  &  &  &  &  &  &  &  &  &  &  &  \\ 
0 & --- & 0 & --- & 91 & --- & 100 & --- & 0 & --- & 0 & --- & 0 & --- & 0
\\ 
1 & 38.6 & 38 & 4.9 & 9 & 1.9 & 0 & 49.3 & 100 & 47.0 & 100 & 23.3 & 0 & 20.8
& 0 \\ 
2 & 3.5 & 58 & 4.3 & 0 & 1.9 & 0 & 1.6 & 0 & 1.6 & 0 & 17.9 & 0 & 4.9 & 4 \\ 
3 & 2.3 & 5 & 3.9 & 0 & 1.8 & 0 & 1.5 & 0 & 1.6 & 0 & 8.9 & 100 & 4.3 & 96
\\ \hline
$T=800$ &  &  &  &  &  &  &  &  &  &  &  &  &  &  \\ 
0 & --- & 0 & --- & 100 & --- & 100 & --- & 0 & --- & 0 & --- & 0 & --- & 0
\\ 
1 & 37.0 & 33 & 4.1 & 0 & 1.8 & 0 & 49.4 & 100 & 47.0 & 100 & 23.2 & 0 & 20.5
& 33 \\ 
2 & 3.5 & 60 & 3.7 & 0 & 1.7 & 0 & 1.5 & 0 & 1.5 & 0 & 17.7 & 0 & 4.1 & 43
\\ 
3 & 2.4 & 8 & 3.3 & 0 & 1.7 & 0 & 1.5 & 0 & 1.5 & 0 & 9.2 & 100 & 3.7 & 24
\\ \hline
$T=1000$ &  &  &  &  &  &  &  &  &  &  &  &  &  &  \\ 
0 & --- & 0 & --- & 100 & --- & 100 & --- & 0 & --- & 0 & --- & 0 & --- & 0
\\ 
1 & 35.5 & 33 & 3.6 & 0 & 1.7 & 0 & 49.5 & 100 & 47.1 & 100 & 22.8 & 0 & 20.4
& 86 \\ 
2 & 3.5 & 59 & 3.3 & 0 & 1.7 & 0 & 1.5 & 0 & 1.5 & 0 & 17.6 & 0 & 3.7 & 14
\\ 
3 & 2.5 & 8 & 3.0 & 0 & 1.6 & 0 & 1.4 & 0 & 1.4 & 0 & 9.2 & 100 & 3.3 & 0 \\ 
\hline\hline
\end{tabular}
\vspace{5pt} \newline
Notes to Table \ref{tab_stationary_coint}. $R^2$ indicates the percentage of
total variance accounted for by the $j$th principal component for $j=1,2$ or
3. $r^*$ indicates the percentage of samples for which the criterion (\ref%
{eqn_bic}) selects the number of factors to be $j=0,1,2,$ or $\ge 3$. In
columns (1)-(6), the true number of factors is $r=0$. In columns (7)-(14),
the true number of factors is $r=1$. In every case, the cross-section
dimension is $N=100$.
\end{table}

In columns (3)-(4) we apply PCA to the residuals from a 24-period-ahead
forecasting regression. Again we estimated the same regression for all
variables, whether persistent or not. And again using regression residuals
solves the problem pretty reliably in samples larger than $T=600.$ If we
look for a factor structure in the one-period-ahead regression residuals as
in columns (5)-(6), the problem is solved 100\% of the time even in a sample
of $T=100.$

\subsection{Results for other data-generating processes}

We also report simulations for a number of other data-generating processes.
For details of these and other simulations see online Appendix C. We first
considered a dataset in which half the variables are cointegrated with a
single common factor and the other are independent white noise. Columns (7)
and (8) of Table \ref{tab_stationary_coint}\ reproduce the well-known result
that PCA on raw nonstationary cointegrated data can correctly pick out the
single common factor; for more discussion see \cite{harris1997} and \cite%
{Onatski_cointegration_2018}. We find that this result is preserved when all 
$N$ series are prefiltered as proposed here, as seen in columns (9)-(10) of
Table \ref{tab_stationary_coint}.

We also considered an example where the common factor is stationary while
the idiosyncratic components are a mix of stationary and nonstationary
processes. When PCA is applied directly to the observed data $y_{it},$ the
familiar \cite{onatski2021spurious} result is observed in column (11) of
Table \ref{tab_stationary_coint}: higher-order factors spuriously appear to
explain a large amount of the variance of the data. When PCA is applied
instead to the $h=24$-period-ahead regression forecast residuals $\hat{c}%
_{it},$ the contribution of higher-order factors $\hat{F}_{jt}$ for $j\geq 1$
is substantially lower, though the selection criterion (\ref{eqn_bic}) would
still typically incorrectly conclude that $r>2$ for $T\leq 600.$ The correct
conclusion $(r=1)$ would be reached 86\% of the time when $T=1000.$
Moreover, we found that the average correlation between the true realization
of $F_{t}$ for a particular simulation and the first estimated factor $\hat{F%
}_{1t}$ of the OLS regression residuals from that simulation is $0.98$ for $%
T\geq 600$. In other words, the first principal component of the OLS
regression residuals accurately uncovers the true single common feature of
these data.

Table C1 in online Appendix C examines local-to-unity and fractionally
integrated processes. The results in these simulations turn out to be in
between the unit-root example in Table \ref{tab_random_walk} and the
stationary-but-persistent example in columns (1)-(6) of Table \ref%
{tab_stationary_coint}. Table C2 compares our approach with the PANIC method
of \cite{BaiNg2004}. They proposed to take first-differences of all the
original data, apply PCA to the changes, and then accumulate the resulting
principal components. We find PANIC performs very similarly to our approach
for these examples when we set $h=1.$

Overall, these results confirm the asymptotic theory that applying PCA to
OLS regression residuals is a promising approach to handling both
nonstationarity and stationary persistence of unknown form provided that the
time-series dimension $T$ is reasonably large.

\section{Characterizing a large macroeconomic data set}

\label{sec_fred_md}

The use of large macroeconomic data sets was pioneered by \cite%
{stock1999forecasting}. Their goal was to use the information of 168
different macroeconomic variables to produce better forecasts of inflation.
They found that the first principal component of indicators of real economic
activity produced the best inflation forecasts over the period 1959:1 to
1997:9. Their findings led to the development of the Chicago Fed National
Activity Index (CFNAI).

\cite{mccracken2016fred} developed the FRED-MD database whose 2015:4 vintage
covered 134 macroeconomic variables. \ These include monthly measures in
eight broad categories: (1) output and income; (2) labor market; (3)
housing; (4) consumption, orders, and inventories; (5) money and credit; (6)
interest and exchange rates; (7) prices; and (8) stock market. This data set
offers benefits of continuity and continuous updating and is the basis for
the analysis in this paper.

\begin{figure}[tbp]
\caption{Level, transformed value, and cyclical component of industrial
production, unemployment, and PMI Composite, 1962:3 to 2014:12}
\label{fig_fred_selected_series}\centering
\includegraphics[width=\textwidth, height = 9.5cm]{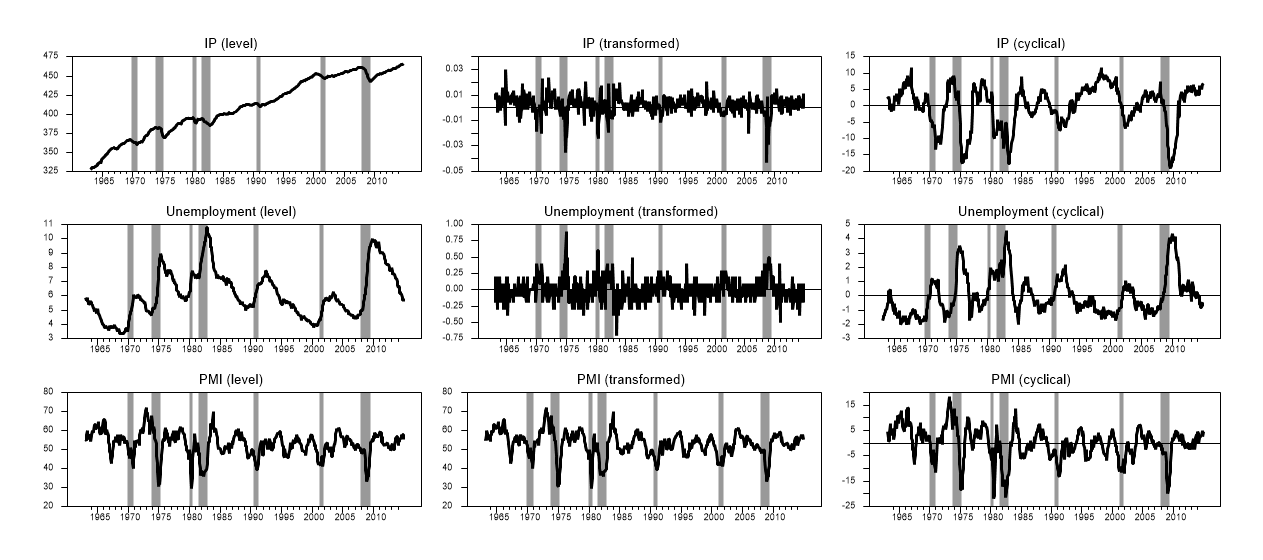}
\end{figure}

In previous applications of PCA to large macroeconomic data sets, each of
the variables needed to be transformed using a detrending method that was
selected individually for each series. For details of how this has been done
for the CFNAI see \cite{cfnai_background} and for FRED-MD see the data
appendix to \cite{mccracken2016fred}. Figure \ref{fig_fred_selected_series}
illustrates these transformations for three important macroeconomic
indicators. The first column plots the raw data, while the second column
plots the data as transformed by \cite{mccracken2016fred} using the same
data set as in their original paper. Everyone agrees that industrial
production (row 1) is nonstationary, and all previous researchers have used
first differences of the log of industrial production shown in panel (1,2).
While there is little doubt that this is a valid way to generate a
stationary series for this variable, monthly growth rates of industrial
production exhibit a lot of high-frequency fluctuations around the dominant
cyclical patterns. For the unemployment rate (row 2), it is less clear
whether the series should be regarded as stationary. \cite{mccracken2016fred}
used first differences of unemployment, which behave quite differently from
the level. The purchasing managers composite index from the Institute of
Supply Management (row 3) appears to be stationary, and \cite%
{mccracken2016fred} entered this series directly into PCA without any
transformation.

The third column of Figure \ref{fig_fred_selected_series} plots the cyclical
components of industrial production, unemployment, and PMI as estimated by
the residuals of the OLS regression (\ref{eqn_recommended_regression}) with $%
h=24$ and $p=12$.\footnote{%
For those series that \cite{mccracken2016fred} transformed using logs, first
differences of logs, or second differences of logs (their transformations
4-6), we simply took the log of the variable before performing the
regression. Thus for example the series plotted in the upper left panel of
Figure \ref{fig_fred_selected_series} is 100 times the natural logarithm of
the industrial production index. For those series that they used as is, as
first differences, or second differences (their transformations 1-3), we
simply used the variable as is. They employed a special transformation (7)
for nonborrowed reserves. One would have expected to take logs of a variable
like this, but the variable took on negative values in 2008. For this series
their transformation was $y_{it}=\Delta (x_{t}/x_{t-1}-1.0)$ and we used $%
y_{it}=$ $x_{t}/x_{t-1}.$} PMI is almost impossible to predict two years in
advance, and our cyclical component is almost identical to the original
series. Thus both our method and the traditional approach use this variable
essentially as is. There is some but not much predictability of the
unemployment rate at the two-year horizon, so for this variable our
transformation much more closely resembles the original series than it does
the first-difference transformation. For industrial production, our approach
takes out the broad trend while retaining the essential cyclical behavior
observed in the raw data. The three variables in the third column, unlike
those in the second column, all share a common characterization of what is
happening over the business cycle. Consistent with a long tradition in
business cycle research, when plotted this way PMI appears as a leading
indicator, industrial production as a coincident indicator, and unemployment
as a coincident or lagging indicator, with all three clearly following the
same cycle.

\begin{figure}[tbp]
\caption{First PC of FRED-MD variables as transformed by \protect\cite%
{mccracken2016fred}, the Chicago Fed National Activity Index, and first PC
of cyclical components of FRED-MD variables, 1962:3 to 2014:12}
\label{fig_macro_fred_2015}\centering
\includegraphics[width=\textwidth, height = 9.5cm]{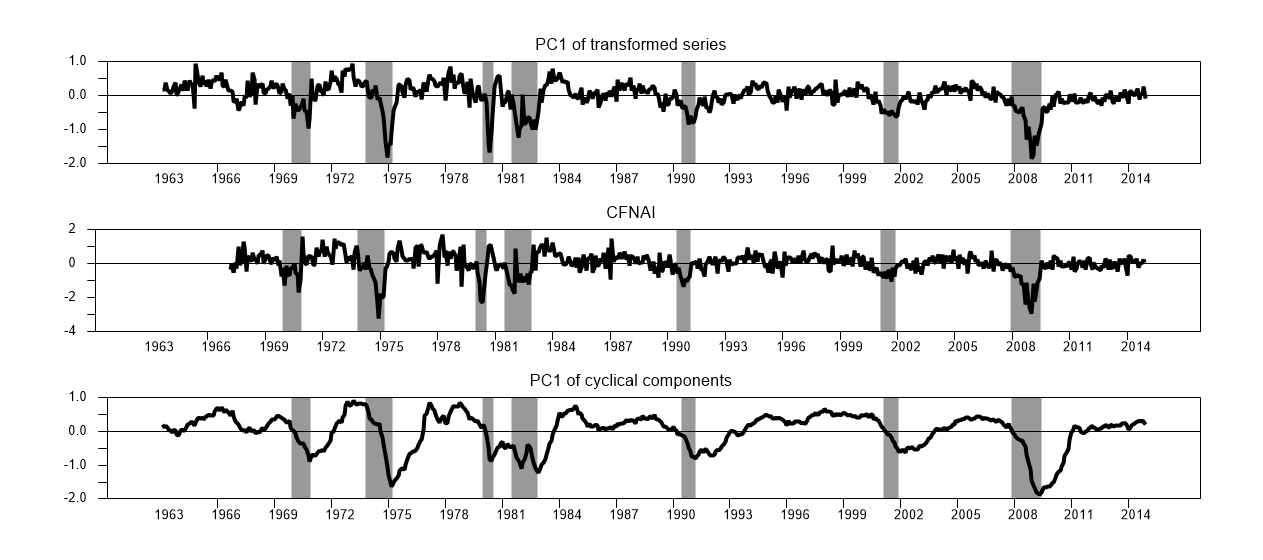}
\end{figure}

The top panel of Figure \ref{fig_macro_fred_2015} plots the first principal
component of the transformed series arrived at by \cite{mccracken2016fred}.%
\footnote{%
We generated this figure using the exact data and code posted at
https://research.stlouisfed.org/econ/mccracken/fred-databases/. Note that we
have multiplied the series by $-1$ in order to give it the property that the
factor declines in recessions, and that their Figure 3 plots the
accumulations $(s_{1t}=\tsum\nolimits_{j=1}^{t}\tilde{f}_{1t})$ whereas our
graph shows $\tilde{f}_{1t}$ itself.} This inherits some of the
high-frequency fluctuations seen in the (1,2) and (2,2) panels of Figure \ref%
{fig_fred_selected_series}. Indeed, \cite{mccracken2016fred} regarded this
series as too volatile to reliably identify business cycles and turning
points, and instead plotted in their Figure 3 the accumulation of this
series. The CFNAI (shown in panel 2 of Figure \ref{fig_macro_fred_2015}) is
very similar to the first principal component of the FRED-MD macro data set.

The first principal component of the estimated cyclical components of the
variables in the data set is plotted in the bottom panel of Figure \ref%
{fig_macro_fred_2015}. Unlike the CFNAI, this provides a very clean summary
of historical business cycles. There is another interesting difference
between the third panel and the first two. The NBER defines the
business-cycle trough (the end point of the shaded regions) as the low point
in the \textit{level} of overall economic activity. For example, in the
first month of a new expansion, the unemployment rate is still very high,
but it has started to come down. Our series in the bottom panel of Figure %
\ref{fig_macro_fred_2015} captures this feature very well, reaching a trough
at exactly the point identified by the NBER Business Cycle Dating Committee.
By contrast, the low point in the series plotted in the first two panels
typically comes more towards the middle of the recession. This is because
the \textit{rate of decline }of real output (a common raw input in the
variables as usually transformed) starts to ease well before the recession
has ended. Both in terms of the cleanness of the series and its timing, we
would suggest that our approach offers a better characterization of the
state of the U.S. business cycle over this sample period.

\subsection{Choice of $h$}

Our method identifies common components in $h$-period-ahead forecast errors.
Any value of $h\geq 1$ is sufficient to produce a stationary series. Which
value of $h$ to use depends on which common components the researcher is
interested in.

McCracken and Ng applied the \cite{BryBoschan} algorithm for dating
business-cycle turning points to the first principal component that resulted
from their analysis. They found that the resulting series was in agreement
with NBER recession dates only 65\% of the time, meaning that 65\% of NBER
recession periods fell within the peak-to-trough phases of the series. We
were able to reproduce this result by adapting the code in \cite%
{StockWatsonTurningPoints}.\footnote{%
As discussed in \citet[f.n. 2]{StockWatsonTurningPoints}, the first step in
applying the Bry and Boschan algorithm is to take a centered moving average
of the series for which turning points are to be assigned, which in this
case is a moving average of the first principal component. Specifying nma =
12 seems to come closest to reproducing the numbers reported in 
\citet[p.
583]{mccracken2016fred}. Our numbers, however, were not in all instances
identical. This appears to be in part because McCracken and Ng were
reporting results for a data set that has some slightly different series
from those in 2015-04.csv. In the case of the percent agreement with NBER
recession dates, our calculation (65\%) exactly reproduces theirs. Other
numbers reported here in the text are based on our reproduction using the
2015-04.csv data, which sometimes differ from those reported in 
\citet[p.
583]{mccracken2016fred}.} We found a similar result (67\% agreement) when
Bry and Boschan is applied to the CFNAI.

\begin{figure}[tb]
\caption{Agreement with NBER-dated recessions and number of series
exhibiting outliers as a function of forecast horizon $h$, 1967:2-2014:12}
\label{fig_Bry_Boschan}
\begin{center}
\includegraphics[width=\textwidth, height = 9.5cm]{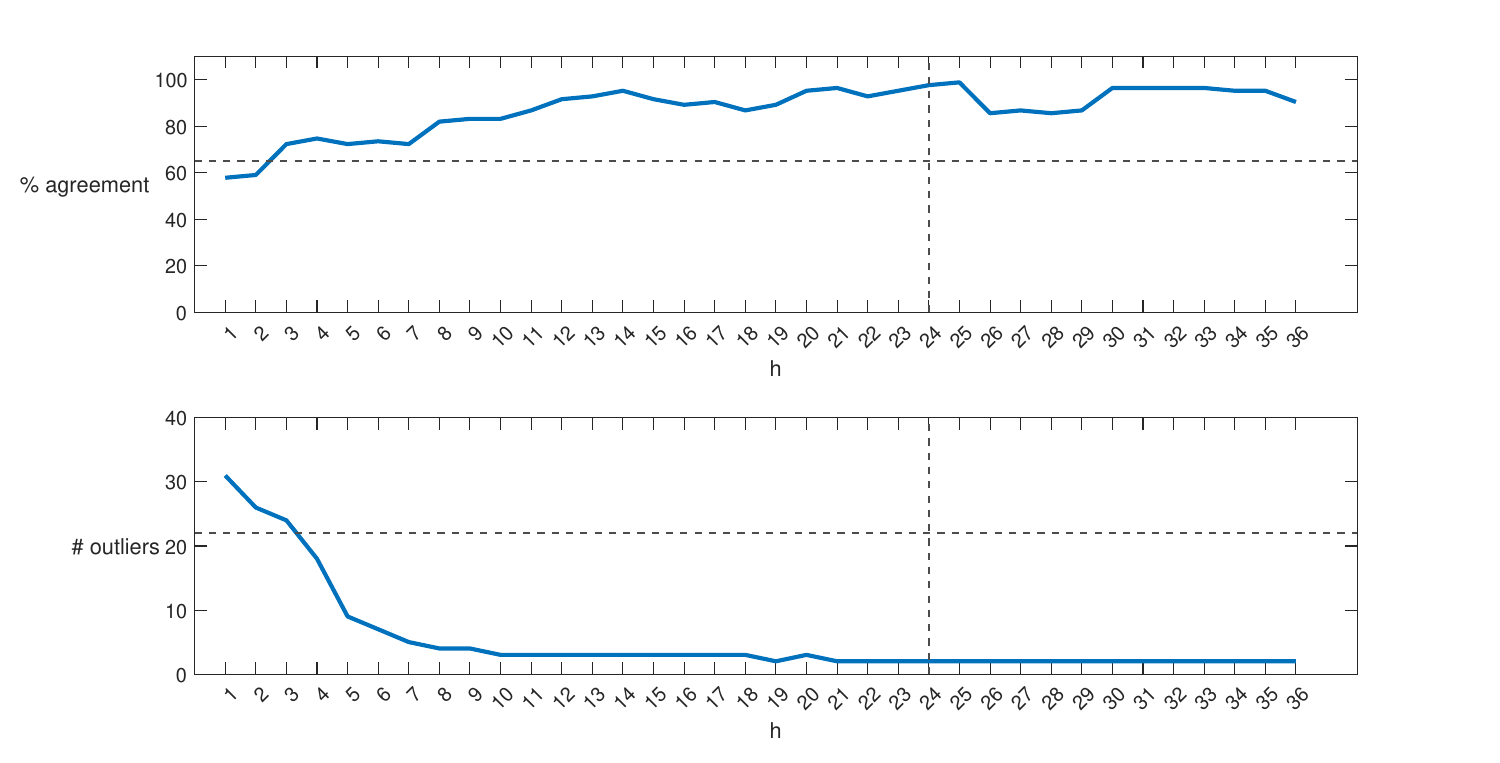}
\end{center}
\par
{\small {\ \emph{Notes to Figure \ref{fig_Bry_Boschan}.} Horizontal axis:
forecast horizon $h$. Top panel: percent of NBER-dated recession periods
that fall within Bry-Boschan-designated downturns. Bottom panel: number of
series exhibiting outliers. Horizontal lines denote results for the first
principal component found by \cite{mccracken2016fred}.} }
\end{figure}

The top panel of Figure \ref{fig_Bry_Boschan} plots the percent agreement
with NBER recessions that results from our method for every value of $h$
between 1 and 36 months. The common component of $h=1$- or 2-month-ahead
forecast errors has less correspondence with NBER recession dates than
either the McCracken and Ng series or the CFNAI. We attribute this to the
fact that replacing stationary series like the PMI in the bottom row of
Figure \ref{fig_fred_selected_series} with a 1-month-ahead forecast error
removes what we would normally think of as the business-cycle indicator
provided by the series. By contrast, for every $h\geq 3,$ our approach has a
closer correspondence to NBER recessions than either McCracken-Ng or the
CFNAI. The maximal agreement (99\%) is obtained using $h=25$. Choosing $h=24$
is very similar (98\%).

McCracken and Ng noted that agreement of the original $\hat{f}_{t}^{MN}$
with NBER expansion dates was even weaker. We found that only 51\% of the
expansion dates based on $\hat{f}_{t}^{MN}$ were also designated as
expansion by NBER and still only 61\% agreement using $\hat{F}_{t}^{MN}$.
Our series with $h=24$ does a little better than either of these with 65\%
agreement.

\subsection{Outliers}

Previous users of large macro data sets devoted a lot of attention to
outliers and implemented procedures to mitigate their influence. Prior to
the COVID-19 recession of 2020, the CFNAI discarded observations that were
more than six times the interquartile range, as did \cite%
{stock1999forecasting} in some of their analysis. \cite{mccracken2016fred}
discarded observations that were more than ten times the interquartile
range. This criterion identifies 79 different observations on 22 different
variables as outliers in the 1960:3 to 2014:12 data set.

We noted in footnote \ref{fn_CLT} that as a result of the Central Limit
Theorem, we should expect fewer outliers when we calculate principal
components of $h$-period-ahead forecasts for larger values of $h$. To
calculate outliers for our method, we constructed forecast errors from
leave-one-out regressions and designated values exceeding ten times the
interquartile range as outliers.\footnote{%
That is, we calculated $\tilde{c}_{it}=y_{it}-\tilde{\alpha}_{i,t}\tilde{z}%
_{i,t-h}$ with $\tilde{\alpha}_{i,t}=\left( \tsum\nolimits_{s=1,s\neq t}^{T}%
\tilde{z}_{i,s-h}\tilde{z}_{i,s-h}^{\prime }\right) ^{-1}\left(
\tsum\nolimits_{s=1,s\neq t}^{T}\tilde{z}_{i,s-h}y_{is}\right) $ estimated
separately for each $i$ and $t$ and then divided $\tilde{c}_{i.t}$ by its
observed interquartile range.} The number of series exhibiting outliers is
plotted as a function of $h$ in the bottom panel of Figure \ref%
{fig_Bry_Boschan}. Using $h=1$ results in 31 series exhibiting outliers,
substantially more than in calculating the \cite{mccracken2016fred}
principal components. Many of the additional outliers come from interest
rate spreads. \cite{mccracken2016fred} used these as is without filtering,
which turns out to be close to our definition of the cyclical component of
these variables. By contrast, using $h=1$-month-ahead forecast errors
highlights some of the unusual behavior of interest rates during the Volcker
monetary contraction in 1980-81. The number of outliers steadily decreases
as $h$ is increased. Only two series exhibit outliers for $h\geq 21.$%
\footnote{%
These are total bank reserves and nonborrowed reserves in the Fed's response
to the financial crisis.}

Our recommended procedure is to use two-year-ahead regression residuals and
make no corrections for outliers. The series that we have plotted in the
bottom panel of Figure \ref{fig_macro_fred_2015} is the unadjusted first
principal component of the full set of OLS residuals $\hat{c}_{it}.$

\begin{figure}[tbh]
\caption{First PC of FRED-MD variables as transformed by \protect\cite%
{mccracken2016fred}, the Chicago Fed National Activity Index, and first and
second PC of cyclical components of FRED-MD variables, 1962:3 to 2024:9}
\label{fig_macro_fred_2022}\centering
\includegraphics[width=\textwidth, height = 11cm]{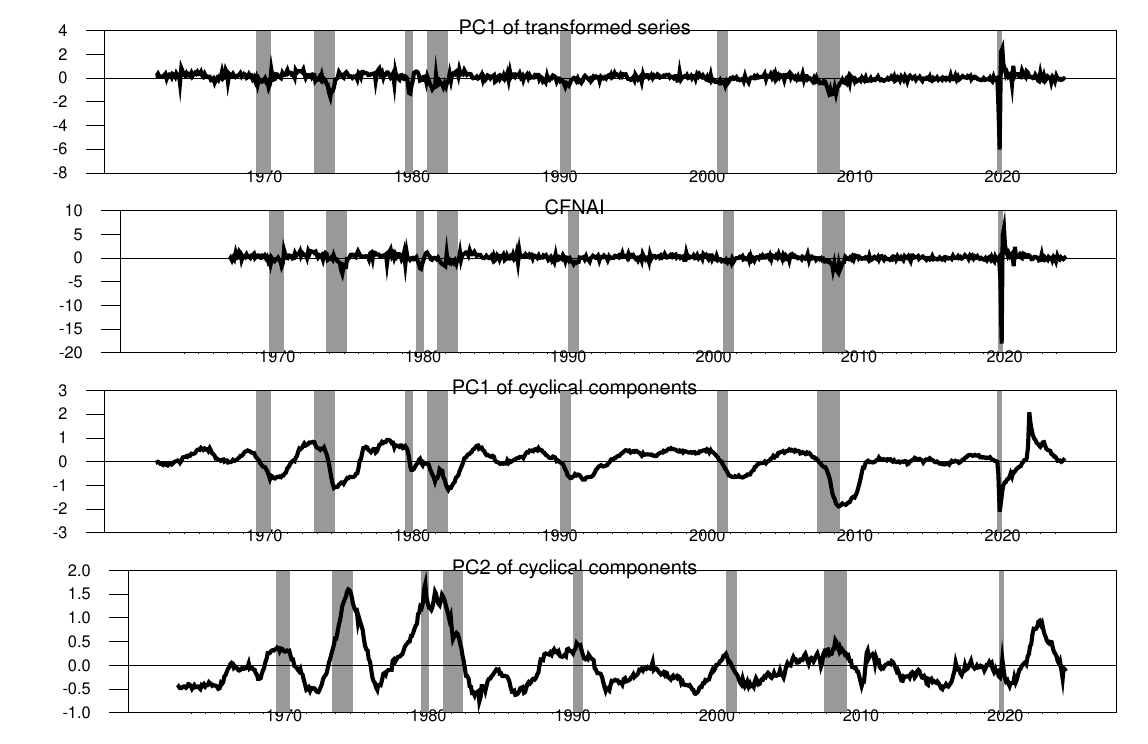}
\end{figure}

Outliers are an even bigger issue when data for 2020 are included. For the
2024:12 vintage of FRED-MD, the McCracken-Ng procedure would identify 40 of
the 126 variables as all being outliers in the single month of 2020:4.
Despite dropping all of these 40 observations, the first principal component
calculated using their algorithm shows an enormous decline in this month.
Indeed, in order to include the 2020 observations in the top panel of Figure %
\ref{fig_macro_fred_2022}, the scale must be so large that it makes all the
previous cyclical fluctuations barely noticeable. The CFNAI modified its
procedure for dealing with anomalous observations to handle these
observations. Even so the CFNAI still displays an unprecedented drop in
2020, as seen in the second panel.

By contrast, only two variables are identified as outliers for 2020:4 for
purposes of our approach, these being new claims for unemployment insurance
and the number unemployed for less than 5 weeks. The result of applying our
procedure to the FRED-MD database available as of December of 2024 with no
corrections for outliers is displayed in the third panel of Figure \ref%
{fig_macro_fred_2022}. Note that, unlike the top two panels, our series
describes the downturn in 2020 on a comparable scale as earlier recessions,
although our series indicates that the speed of the downturn was
unprecedented, as was the growth in the first two months of the recovery.
Also in contrast to the first two panels, our series indicates (correctly,
in our view) that the economy did not fully recover from the COVID-19 shock
until September of 2021. The sharp spike up in our series in April of 2022
reflects the fact that most macro variables were substantially higher in
April 2022 than one would have predicted based on observations in April
2020. This conclusion is also consistent with the aggressive actions of
policy makers in the spring of 2020. Our series further indicates that
economic activity remained unusually strong through the fall of 2023.

Another issue comes from discontinued, newly added, or missing variables. 
\cite{mccracken2016fred} adapted the \cite{stock2002forecasting} algorithm
for unbalanced panels, though they found in their original data set that the
results are essentially identical if one simply drops variables as needed to
create a balanced panel. For our application, we have simply calculated
principal components of $\hat{c}_{it}$ on a balanced panel, though there is
no obstacle to applying the \cite{stock2002forecasting} algorithm to an
unbalanced panel of $\hat{c}_{it}.$

\subsection{Uses of macroeconomic cyclical factors}

A key use of PCA is to summarize the statistical information in a large
cross section of indicators. The movement in variable $i$ that is captured
by the $j$th factor alone is given by $\hat{\lambda}_{ij}\hat{f}_{jt}.$
Since $\hat{c}_{it}$ is normalized to have unit variance, the fraction of
the variance of the stationary component of variable $i$ that is explained
by the $j$th macro factor is given by $\hat{\lambda}_{ij}^{2}T^{-1}\tsum%
\nolimits_{t=1}^{T}\hat{f}_{jt}^{2}.$

Table B1 in the online appendix reports the $R^{2}$ explained by the first
macroeconomic cyclical factor for each of the 119 variables used in our
analysis of the 2023 vintage data set. The first factor alone accounts for
almost 2/3 of the variance of typical indicators of real output or income
and more than half of the variance of typical indicators of labor-market
conditions. The first cyclical factor is far less successful at describing
financial indicators and nominal prices. It is interesting that when we add
the second cyclical factor, the $R^{2}$ for the median price indicator rises
to 66\%. The first factor thus seems mainly to capture real economic
conditions and the second characterizes nominal prices and interest rates.

The fourth panel in Figure \ref{fig_macro_fred_2022} plots the second
cyclical factor. This by construction is orthogonal to the first, and often
continues to fall even as the recovery in real economic activity is
beginning. This is consistent with the view that nominal variables may
respond sluggishly to business-cycle developments. It describes events
beginning in 2022 as a third big U.S. inflation wave, though less dramatic
than the big inflations of 1973-74 and 1979-81.

We next explore the use of the cyclical factors in forecasting. \cite%
{stock1999forecasting} demonstrated that the first principal component of a
large data set of real macroeconomic variables could be very helpful for
forecasting inflation. Their finding gave rise to the Chicago Fed National
Activity Index (CFNAI), a PCA-based indicator that is still widely used
today. We compare the usefulness for forecasting of the CFNAI (denoted $\hat{%
f}_{t}^{CF}),$ the first or second principal component calculated using the
algorithm and data set of \cite{mccracken2016fred} (denoted $\hat{f}%
_{t}^{MN1}$ and $\hat{f}_{t}^{MN2},$ respectively$),$ or the first or second
principal component of the forecasting residuals (denoted $\hat{f}%
_{t}^{HMX1} $ or $\hat{f}_{t}^{HMX2})$ calculated from the FRED-MD data set
using equation (\ref{eqn_HX_f1_definition}). Our approach to comparing
different forecasts is similar to that used by \cite{stock1999forecasting}
and \cite{mccracken2016fred}.

A particular model $m$ uses a set of variables $\mathbf{x}_{t}^{m}$ that are
observed at date $t$ to try to forecast the value of a variable of interest $%
y_{t+h}^{h}$ that will not be observed until $t+h$:%
\begin{equation}
y_{t+h}^{h}=\boldsymbol{\pi }^{m\prime }\mathbf{x}_{t}^{m}+u_{t+h}^{m,h}.
\label{eqn_forecast_regression}
\end{equation}%
We estimated the value of $\boldsymbol{\pi }^{m}$ by OLS regression on the
subsample $t=T_{0},T_{0}+1,...,T_{1}-h-1$ and used these coefficients to
forecast $y_{T_{1}}^{h}.$ We then augmented the sample by one observation,
estimating the regression over $t=T_{0},T_{0}+1,...,T_{1}-h$ and using those
coefficients to forecast $y_{T_{1}+1}^{h}.$ We repeated this for an
evaluation period $T_{1}$ to $T_{2}$ and calculated the average squared
forecast error over this evaluation period. We performed this analysis using
three different evaluation periods. The first evaluation period is specified
by $T_{1}=1970$:1 to $T_{2}=$ 1996:12, which was the evaluation period in
the original study by \cite{stock1999forecasting}. The second evaluation
period is $T_{1}=$ 1997:1 to $T_{2}=$ 2014:12, which corresponds to the new
data used by \cite{mccracken2016fred} that were not available to \cite%
{stock1999forecasting}. The third evaluation period is $T_{1}=$ 2015:1 to $%
T_{2}=$ 2024:9, which is the new data available since publication of \cite%
{mccracken2016fred}. The models we considered were a pure autoregressive
model, $\mathbf{x}_{t}^{AR}=(1,y_{t}^{1},y_{t-1}^{1},...,y_{t-5}^{1})^{%
\prime },$ and models that add to the autoregressive model six lags of one
of the principal components estimates. For example, $\mathbf{x}_{t}^{CF}=(%
\mathbf{x}_{t}^{AR\prime },\hat{f}_{t}^{CF},\hat{f}_{t-1}^{CF},...,\hat{f}%
_{t-5}^{CF})^{\prime }.$ This differs a little from the forecast evaluations
performed by \cite{stock1999forecasting} and \cite{mccracken2016fred} in
that these authors used $BIC$ to select different lag lengths for the
autoregressive and principal components and for each subsample, whereas we
set the lag length to six for every evaluation. Also, since $\hat{f}%
_{t}^{CF} $ is only available beginning in 1967:3, we used $T_{0}=$ 1967:9
as the first date for estimation of all models. In every case, for each $T$
we re-estimated the coefficients $\boldsymbol{\hat{\pi}}^{m}$ for the
forecasting regression (\ref{eqn_forecast_regression}) using an expanding
data set ending $h$ periods before the variable being forecast.

A separate question is the data set used to estimate the factors $\hat{f}%
_{t} $ themselves. Insofar as the factors are only identified up to sign,
the meaning of coefficients multiplying $\hat{f}_{t}$ could change across
expanding samples. We formed data sets to calculate the factors $\hat{f}_{t}$
for each of the three methods as follows. (i) For the explanatory variable $%
\hat{f}_{t}^{MN},$ for the first two evaluation samples we followed \cite%
{mccracken2016fred} in calculating the factor $\hat{f}_{t}^{MN}$ using the
full historical vintage of the FRED-MD database available as of 2015:4.%
\footnote{%
Our series for $\hat{f}_{t}^{MN1}$ for this subsample is almost (but not
quite) identical to the series analyzed by \cite{mccracken2016fred}. We have
been unable to identify the source of the small discrepancies.} For the
third evaluation sample, we estimated $\hat{f}_{t}^{MN}$ using the full
database available as of 2024:12. (ii) For the explanatory variable $\hat{f}%
_{t}^{CF}$ we used the value of the CFNAI as it was reported in 2024 for all
historical dates.\footnote{%
We downloaded $\hat{f}_{t}^{CF}$ on January 13, 2025 from the FRED database
at https://\allowbreak fred.stlouisfed.org/series/CFNAI.} (iii) For the
explanatory variable $\hat{f}_{t}^{HMX},$ we need to specify both (a) the
data set used to estimate the coefficients $\hat{\alpha}_{i}$ for the
detrending regression (\ref{eqn_recommended_regression}) and (b) the data
set used to calculate principal components of the forecast errors $\hat{c}%
_{it}=y_{it}-\hat{\alpha}_{i}^{\prime }z_{i,t-h}.$ For (a), for all three
evaluation samples, we estimated $\hat{\alpha}_{i}$ from the 2015:4 vintage
data set. For (b), for the first two evaluation samples, we calculated
principal components of $\hat{c}_{it}$ using the 2015:4 vintage data set.
For the last evaluation sample, we calculated principal components of $\hat{c%
}_{it}$ using the 2024:12 data set. Thus our forecasting exercise for the
third subsample is not affected by the potential concern that the detrending
regression used information that was not available at the time the forecast
was made. This is because the coefficients for the detrending regression
were all known as of 2015. We also repeated the exercise using instead for
(b) trend coefficients estimated using all the data through 2024. This
alternative specification produced similar results (not reported here).

\begin{table}[tbph]
\caption{Mean squared forecast errors for different models}
\label{tab_mse}%
\begin{tabular}{cccccccc}
\hline
&  & \multicolumn{6}{c}{Consumer Price Index} \\ \cline{3-8}
sample & horizon & AR & CF & MN1 & HMX1 & MN2 & HMX2 \\ \hline
&  &  &  &  &  &  &  \\ 
1970-1996 & h=1 & 7.91 & 1.00 & 0.99 & 1.03 & 0.98 & 0.90 \\ 
& h=6 & 4.26 & 0.81 & 0.77 & 0.80 & 0.95 & 0.88 \\ 
& h=12 & 5.32 & 0.70 & 0.62 & 0.74 & 1.02 & 1.33 \\ 
&  &  &  &  &  &  &  \\ 
1997-2014 & h=1 & 12.26 & 1.03 & 1.04 & 1.02 & 0.97 & 1.09 \\ 
& h=6 & 6.08 & 1.23 & 1.23 & 1.23 & 0.96 & 1.11 \\ 
& h=12 & 4.21 & 1.22 & 1.22 & 1.28 & 0.95 & 1.17 \\ 
&  &  &  &  &  &  &  \\ 
2015-2024 & h=1 & 8.07 & 1.51 & 1.27 & 1.14 & 0.94 & 1.04 \\ 
& h=6 & 3.70 & 1.95 & 1.48 & 1.76 & 1.03 & 1.01 \\ 
& h=12 & 3.64 & 1.71 & 1.30 & 1.49 & 1.09 & 0.96 \\ 
&  &  &  &  &  &  &  \\ \hline
&  & \multicolumn{6}{c}{Industrial Production} \\ \cline{3-8}
sample & horizon & AR & CF & MN1 & HMX1 & MN2 & HMX2 \\ \hline
&  &  &  &  &  &  &  \\ 
1970-1996 & h=1 & 76.73 & 0.97 & 0.94 & 0.96 & 1.01 & 1.02 \\ 
& h=6 & 38.66 & 0.92 & 0.93 & 0.83 & 0.71 & 0.79 \\ 
& h=12 & 27.19 & 1.01 & 1.06 & 1.21 & 0.49 & 0.87 \\ 
&  &  &  &  &  &  &  \\ 
1997-2014 & h=1 & 58.90 & 0.85 & 0.83 & 0.98 & 1.04 & 1.00 \\ 
& h=6 & 22.61 & 0.93 & 0.94 & 1.05 & 1.24 & 1.12 \\ 
& h=12 & 20.11 & 0.96 & 1.01 & 1.06 & 1.26 & 1.11 \\ 
&  &  &  &  &  &  &  \\ 
2015-2024 & h=1 & 506.67 & 1.73 & 0.93 & 1.10 & 1.02 & 1.00 \\ 
& h=6 & 87.37 & 2.29 & 1.16 & 1.03 & 1.05 & 0.94 \\ 
& h=12 & 38.10 & 2.31 & 1.14 & 0.85 & 1.03 & 0.93 \\ 
&  &  &  &  &  &  &  \\ \hline\hline
\end{tabular}
\vspace{5pt} \newline
Notes to Table \ref{tab_mse}. AR column reports simulated out-of-sample mean
squared forecast error for purely autoregressive model evaluated over three
different out-of-sample periods. CF column reports the relative MSE when
lags of the Chicago Fed National Activity Index are added to the
autoregression, with a value less than one indicating the variable is useful
for forecasting. MN1 column reports the relative MSE when lags of the first
principal component calculated using the procedures in \cite%
{mccracken2016fred} are used in place of the CFNAI. HMX1 reports relative
MSE when lags of the first principal component of the estimated cyclical
components are used in place of CFNAI. MN2 and HMX2 report results when the
second principal component is used instead of the first.
\end{table}

For our first set of evaluations we set $y_{t+h}^{h}$ to be the average
inflation rate between month $t$ and $t+h,$ quoted at an annual rate,%
\begin{equation*}
y_{t+h}^{h}=(1200/h)\log (CPI_{t+h}/CPI_{t}),
\end{equation*}%
where $CPI_{t}$ denotes the level of the consumer price index in month $t.$%
\footnote{%
Here again our evaluation design differs slightly from that in \cite%
{stock1999forecasting} and \cite{mccracken2016fred} in that those authors
took the object of interest to be to forecast the \textit{change }in the
inflation rate as a function of lagged changes:%
\begin{equation*}
y_{t+h}^{h}=(1200/h)\log (CPI_{t+h}/CPI_{t})-1200\log (CPI_{t}/CPI_{t-1}).
\end{equation*}%
} The column labeled $AR$ in Table \ref{tab_mse} reports the simulated
out-of-sample mean squared error of a purely autoregressive model for each
of the three evaluation samples and for forecast horizons of $h=1,$ 6, or 12
months.\footnote{\cite{stock1999forecasting} reported results for $h=12$
months whereas \cite{mccracken2016fred} reported results for $h=1,$ 6, and
12.} The first panel reproduces the finding of \cite{stock1999forecasting}
that an index like CFNAI significantly improves forecasts for longer
horizons over the 1970-1996 period. The alternative measures $\hat{f}%
_{t}^{MN1}$ or $\hat{f}_{t}^{HMX1}$ offer similar improvements. The indexes
offer little or no improvement for one-month-ahead forecasts over this
period, but again are similar to each other. All three indexes are
outperformed by simple autoregressive forecasts over either of the later two
evaluation periods. The observation that inflation has become much harder to
forecast in data since 1996 has been reported by a number of other
researchers, including \cite{Atkeson}, \cite{Fisher}, \cite{StockWatson2007}%
, and \cite{StockWatson2008}. Interestingly, the second cyclical factor $%
\hat{f}_{t}^{HMX2}$ does better than any of the other four indexes at
forecasting inflation at the one-month horizon for the 1970-1996 sample and
at the 6-12 month horizons for the 2015-2024 sample.

Table \ref{tab_mse} also reports forecasts of industrial production, setting%
\begin{equation*}
y_{t+h}^{h}=(1200/h)\log (IP_{t+h}/IP_{t}),
\end{equation*}%
for $IP_{t}$ the level of the industrial production index in month $t.$ The
indexes $\hat{f}_{t}^{CF},$ $\hat{f}_{t}^{MN1},$ and $\hat{f}_{t}^{HMX1}$
all help forecast industrial production over near horizons in the first two
evaluation periods. The CFNAI does particularly poorly at forecasting either
inflation or industrial production at any horizon for 2015-2024. Both $\hat{f%
}_{t}^{HMX1}$ and $\hat{f}_{t}^{HMX2}$ do significantly better than CFNAI in
every case over this period.

We conclude that our approach offers similar benefits to conventional PCA
when evaluated in terms of simulated out-of-sample forecasts, and does much
better than measures like the CFNAI for recent data. We share the conclusion
of the earlier literature that the usefulness of any
principal-component-based measure for purposes of forecasting depends on the
variable, evaluation period, and horizon of the forecast.{}

\section{Conclusion}

Calculating principal components of medium-horizon forecast errors is a
viable approach to identifying the common cyclical factors that drive a
large collection of potentially nonstationary economic indicators. This
avoids the need to decide how to detrend each individual series and is much
more promising than approaches such as the Chicago Fed National Activity
Index for handling data that include the large outliers of 2020.

\newpage 
\bibliographystyle{apalike}
\bibliography{HXrefs.bib}

\end{document}


\title{Supplementary Materials for ``Principal Component Analysis for a Mix of Stationary and
Nonstationary Variables''}
\author[]{James D. Hamilton\thanks{%
Email: \texttt{jhamilton@ucsd.edu}} \and Xinwei Ma\thanks{%
Email: \texttt{x1ma@ucsd.edu}} \and Jin Xi\thanks{%
Email: \texttt{xijin@amss.ac.cn}}}
\date{August 17, 2026}
\maketitle

\begin{abstract}
\doublespacing This supplementary material provides proofs of the results in the main paper and their extensions, discusses additional methodological and technical results, and reports additional simulation evidence.
\end{abstract}


\affil[1]{University of California at San Diego} 
\affil[2]{University of
California at San Diego} \affil[3]{Chinese Academy of Sciences}

\bigskip \thispagestyle{empty} \vspace{2em}

\clearpage
\pagenumbering{arabic} \renewcommand*{\thepage}{A-%
	\arabic{page}}

\appendix\doublespacing

\renewcommand{\theequation}{A-\arabic{equation}} \setcounter{equation}{0}

\section{Proofs}

\subsection{Proof of Theorem 1}

Recall that the operator norm of a symmetric matrix $S$ is defined by%
\begin{equation*}
\left\Vert S\right\Vert _{op}=\underset{\gamma \in \Gamma }{\sup }\frac{%
| \gamma ^{\prime }S\gamma | }{\gamma ^{\prime }\gamma }.
\end{equation*}%
When $S$ is positive semidefinite, this is the largest eigenvalue: $%
\left\Vert S\right\Vert _{op}=\varrho _{\max }(S).$ Since the trace of $S$
(the sum of the diagonal elements) is equal to the sum of the eigenvalues,
it follows immediately that for any symmetric positive semidefinite matrix $%
S,$

\begin{equation}
\underset{\gamma \in \Gamma }{\sup }\frac{\gamma ^{\prime }S\gamma }{\gamma
^{\prime }\gamma }\leq \tsum\nolimits_{i=1}^{N}s_{ii}.  \label{eqn_lemma1}
\end{equation}

\subsubsection*{Proof of Theorem 1%
(i)-(ii)}

Note first that 
\begin{eqnarray}
&&\big\Vert(NT)^{-1}\tsum\nolimits_{t=1}^{T}\hat{C}_{t}\hat{C}_{t}^{\prime
}-(NT)^{-1}\tsum\nolimits_{t=1}^{T}C_{t}C_{t}^{\prime }\big\Vert_{\mathrm{op}%
}  \notag \\
&=&\underset{\gamma \in \Gamma }{\sup }\frac{\Big|\gamma ^{\prime }\Big(%
(NT)^{-1}\tsum\nolimits_{t=1}^{T}\hat{C}_{t}\hat{C}_{t}^{\prime
}-(NT)^{-1}\tsum\nolimits_{t=1}^{T}C_{t}C_{t}^{\prime }\Big)\gamma \Big|}{%
\gamma ^{\prime }\gamma }  \notag \\
&=&\underset{\gamma \in \Gamma }{\sup }\big|(N^{2}T)^{-1}\gamma ^{\prime
}\tsum\nolimits_{t=1}^{T}\hat{C}_{t}\hat{C}_{t}^{\prime }\gamma
-(N^{2}T)^{-1}\gamma ^{\prime }\tsum\nolimits_{t=1}^{T}C_{t}C_{t}^{\prime
}\gamma \big|  \notag \\
&\leq &\underset{\gamma \in \Gamma }{\sup }(N^{2}T)^{-1}\gamma ^{\prime
}\tsum\nolimits_{t=1}^{T}\hat{V}_{t}\hat{V}_{t}^{\prime }\gamma +2\underset{%
\gamma \in \Gamma }{\sup }(N^{2}T)^{-1}\big|\gamma ^{\prime
}\tsum\nolimits_{t=1}^{T}C_{t}\hat{V}_{t}^{\prime }\gamma \big|,
\label{eqn_thm2_starting_eq}
\end{eqnarray}%
where the last line uses the relation $\hat{C}_{t}=C_{t}+\hat{V}_{t}$. We
first show that the two terms in the last line of (\ref{eqn_thm2_starting_eq}%
) converge in probability to 0. To show this for the first term, notice from
(\ref{eqn_lemma1}) that%
\begin{eqnarray}
\underset{\gamma \in \Gamma }{\sup }(N^{2}T)^{-1}\gamma ^{\prime
}\tsum\nolimits_{t=1}^{T}\hat{V}_{t}\hat{V}_{t}^{\prime }\gamma &=&\,%
\underset{\gamma \in \Gamma }{\sup }(NT)^{-1}\frac{\gamma ^{\prime
}\sum\nolimits_{t=1}^{T}\hat{V}_{t}\hat{V}_{t}^{\prime }\gamma }{\gamma
^{\prime }\gamma }  \notag \\
&\leq &(NT)^{-1}\tsum\nolimits_{i=1}^{N}\tsum\nolimits_{t=1}^{T}\hat{v}%
_{it}^{2}\ \leq \ \max_{1\leq i\leq N}T^{-1}\tsum\nolimits_{t=1}^{T}\hat{v}%
_{it}^{2},  \label{eqn_thm2_GNT}
\end{eqnarray}%
which tends to zero in probability by Assumption 3(ii).

For the last term in (\ref{eqn_thm2_starting_eq}),%
\begin{equation*}
\underset{\gamma \in \Gamma }{\sup }\left\vert (N^{2}T)^{-1}\gamma ^{\prime
}\tsum\nolimits_{t=1}^{T}C_{t}\hat{V}_{t}^{\prime }\gamma \right\vert \leq %
\left[ \underset{\gamma \in \Gamma }{\sup }(N^{2}T)^{-1}\gamma ^{\prime
}\tsum\nolimits_{t=1}^{T}C_{t}C_{t}^{\prime }\gamma \right] ^{1/2}\left[ 
\underset{\gamma \in \Gamma }{\sup }(N^{2}T)^{-1}\gamma ^{\prime
}\tsum\nolimits_{t=1}^{T}\hat{V}_{t}\hat{V}_{t}^{\prime }\gamma \right]
^{1/2}.
\end{equation*}%
The first term converges in probability to a number no larger than $\omega
_{11}^{1/2}$ from result (R8) in \cite{stock2002forecasting} and the second
converges in probability to 0 from (\ref{eqn_thm2_GNT}). Thus
$$\underset{%
\gamma \in \Gamma }{\sup }\left\vert (N^{2}T)^{-1}\gamma ^{\prime
}\tsum\nolimits_{t=1}^{T}C_{t}\hat{V}_{t}^{\prime }\gamma \right\vert 
\overset{p}{\rightarrow }0.$$ 
We thus conclude from (\ref%
{eqn_thm2_starting_eq}) that 
\begin{equation}
\underset{\gamma \in \Gamma }{\sup }\big|(N^{2}T)^{-1}\gamma ^{\prime
}\tsum\nolimits_{t=1}^{T}\hat{C}_{t}\hat{C}_{t}^{\prime }\gamma
-(N^{2}T)^{-1}\gamma ^{\prime }\tsum\nolimits_{t=1}^{T}C_{t}C_{t}^{\prime
}\gamma \big|\overset{p}{\rightarrow }0,  \label{eqn_thm 2ii final equation}
\end{equation}%
or equivalently 
\begin{equation*}
\big\Vert(NT)^{-1}\tsum\nolimits_{t=1}^{T}\hat{C}_{t}\hat{C}_{t}^{\prime
}-(NT)^{-1}\tsum\nolimits_{t=1}^{T}C_{t}C_{t}^{\prime }\big\Vert_{\mathrm{op}%
}\overset{p}{\rightarrow }0.
\end{equation*}%
Recall that $T^{-1}\tsum\nolimits_{t=1}^{T}\hat{f}_{jt}^{2}$ and $%
T^{-1}\tsum\nolimits_{t=1}^{T}\tilde{f}_{jt}^{2}$ denote the $j$th largest
eigenvalues of $(NT)^{-1}\tsum\nolimits_{t=1}^{T}\hat{C}_{t}\hat{C}%
_{t}^{\prime }$ and $(NT)^{-1}\tsum\nolimits_{t=1}^{T}C_{t}C_{t}^{\prime }$,
respectively, and that $T^{-1}\tsum\nolimits_{t=1}^{T}\tilde{f}_{jt}^{2}$
converges in probability to $\omega_{jj}$. Therefore, by Weyl's inequality on perturbation,
it is true for all $j$ that 
\begin{equation*}
|T^{-1}\tsum\nolimits_{t=1}^{T}\hat{f}_{jt}^{2}-T^{-1}\tsum%
\nolimits_{t=1}^{T}\tilde{f}_{jt}^{2}|\leq \big\Vert(NT)^{-1}\tsum%
\nolimits_{t=1}^{T}\hat{C}_{t}\hat{C}_{t}^{\prime
}-(NT)^{-1}\tsum\nolimits_{t=1}^{T}C_{t}C_{t}^{\prime }\big\Vert_{\mathrm{op}%
},
\end{equation*}%
which implies that $T^{-1}\tsum\nolimits_{t=1}^{T}\hat{f}_{jt}^{2}$
converges in probability to $\omega _{jj}$ for $j=1,2,\dots ,r$, and to $0$
for $j=r+1,\dots ,k$.

\subsubsection*{Proof of Theorem 1(iii)%
}

Notice that $(N^{2}T)^{-1}\hat{\Lambda}^{\prime }\tsum\nolimits_{t=1}^{T}%
\hat{C}_{t}\hat{C}_{t}^{\prime }\hat{\Lambda}$ is a diagonal matrix for all $%
N$ and $T$ by the definition of $\hat{\Lambda}$ with diagonal elements
converging in probability to $\omega _{jj}$ by result (\textit{i}):%
\begin{equation}
(N^{2}T)^{-1}\hat{\Lambda}^{\prime }\tsum\nolimits_{t=1}^{T}\hat{C}_{t}\hat{C%
}_{t}^{\prime }\hat{\Lambda}\overset{p}{\rightarrow }\Omega _{FF}.
\label{eqn_thm 2iii first equation}
\end{equation}%
Equation (\ref{eqn_thm 2ii final equation}) then establishes that $%
(N^{2}T)^{-1}\hat{\Lambda}^{\prime
}\tsum\nolimits_{t=1}^{T}C_{t}C_{t}^{\prime }\hat{\Lambda}\overset{p}{%
\rightarrow }\Omega _{FF}.$ We also know from results (R2)-(R6) in \cite%
{stock2002forecasting} that%
\begin{equation*}
(N^{2}T)^{-1}\gamma ^{\prime }\tsum\nolimits_{t=1}^{T}C_{t}C_{t}^{\prime
}\gamma -(N^{2}T)^{-1}\gamma ^{\prime }\tsum\nolimits_{t=1}^{T}\Lambda
^{\prime }F_{t}F_{t}^{\prime }\Lambda \gamma \overset{p}{\rightarrow }0
\end{equation*}%
where the convergence is uniform for $\gamma \in \Gamma ,$ meaning%
\begin{eqnarray}
\mathrm{plim}\Big\{(N^{2}T)^{-1}\hat{\Lambda}^{\prime
}\tsum\nolimits_{t=1}^{T}C_{t}C_{t}^{\prime }\Big\} = \mathrm{plim}\Big\{%
(N^{2}T)^{-1}\hat{\Lambda}^{\prime }\tsum\nolimits_{t=1}^{T}\Lambda
F_{t}F_{t}^{\prime }\Lambda ^{\prime }\hat{\Lambda}\Big\} = H\Omega
_{FF}H^{\prime }  \label{eqn_thm 2iii second equation}
\end{eqnarray}%
for $H=$ plim $(N^{-1}\hat{\Lambda}^{\prime }\Lambda ).$ Combining results (%
\ref{eqn_thm 2ii final equation})-(\ref{eqn_thm 2iii second equation}),%
\begin{equation}
\Omega _{FF}=H\Omega _{FF}H^{\prime }.  \label{eqn_thm 2iii third equation}
\end{equation}

Let $\hat{h}_{j}^{\prime }$ denote the $j$th row of $\hat{\Lambda}^{\prime
}\Lambda /N$, 
\begin{equation*}
\underset{(1\times r)}{\hat{h}_{j}^{\prime }}=\underset{(1\times N)}{\hat{%
\lambda}_{j}^{\prime }}\underset{(N\times r)}{\Lambda }/N,
\end{equation*}%
for $\hat{\lambda}_{j}^{\prime }$ the $j$th row of $\hat{\Lambda}^{\prime }.$
Then%
\begin{equation*}
\hat{h}_{j}^{\prime }\hat{h}_{j}=\frac{\hat{\lambda}_{j}^{\prime }}{\sqrt{N}}%
\frac{\Lambda \Lambda ^{\prime }}{N}\frac{\hat{\lambda}_{j}}{\sqrt{N}}.
\end{equation*}%
This is less than or equal to the largest eigenvalue of $\Lambda \Lambda
^{\prime }/N,$ which converges to 1. Letting $h_{j}^{\prime
}=(h_{j1},h_{j2},...,h_{jr})^{\prime }$ denote the $j$th row of $H,$ we thus
have 
\begin{equation*}
\hat{h}_{j}^{\prime }\hat{h}_{j}\overset{p}{\rightarrow }%
h_{j1}^{2}+h_{j2}^{2}+\cdots h_{jr}^{2}\leq 1.
\end{equation*}%
The (1,1) element of (\ref{eqn_thm 2iii third equation}) states%
\begin{equation*}
h_{1}^{\prime }\Omega _{FF}h_{1}=h_{11}^{2}\omega _{11}+h_{12}^{2}\omega
_{22}+\cdots +h_{1r}^{2}\omega _{rr}=\omega _{11}.
\end{equation*}%
Since $\omega _{11}>\omega _{22}>\cdots >\omega _{rr}>0,$ this requires $%
h_{11}^{2}=1$ and $h_{12}=\cdots =h_{1r}=0.$ Thus the (1,1) element of $\hat{%
\Lambda}^{\prime }\Lambda /N$ converges in probability to $\pm 1$ and other
elements of the first row converge to zero.

The (2,2) element of (\ref{eqn_thm 2iii third equation}) states%
\begin{equation}
h_{21}^{2}\omega _{11}+h_{22}^{2}\omega _{22}+\cdots +h_{2r}^{2}\omega
_{rr}=\omega _{22}  \label{eqn_thm 2iii second row}
\end{equation}%
where $h_{21}=$ plim $\hat{\lambda}_{2}\lambda _{1}/N.$ Regress $\lambda
_{1} $ on $\hat{\lambda}_{1}$ with residual $q_{1}$:%
\begin{equation}
\lambda _{1}=\hat{k}_{1}\hat{\lambda}_{1}+q_{1}
\label{eqn_thm 2iii lambda1 regression}
\end{equation}%
\begin{equation*}
\hat{k}_{1}=(\hat{\lambda}_{1}^{\prime }\hat{\lambda}_{1}/N)^{-1}(\hat{%
\lambda}_{1}^{\prime }\lambda _{1}/N)
\end{equation*}%
\begin{equation*}
q_{1}^{\prime }\hat{\lambda}_{1}=0
\end{equation*}%
\begin{equation*}
\lambda _{1}^{\prime }\lambda _{1}/N=\hat{k}_{1}^{2}(\hat{\lambda}%
_{1}^{\prime }\hat{\lambda}_{1}/N)+q_{1}^{\prime }q_{1}/N.
\end{equation*}%
We saw above that $\hat{k}_{1}^{2}\overset{p}{\rightarrow }1,$ which along
with $\lambda _{1}^{\prime }\lambda _{1}/N\rightarrow 1$ and $\hat{\lambda}%
_{1}^{\prime }\hat{\lambda}_{1}/N=1$ establishes $q_{1}^{\prime }q_{1}/N%
\overset{p}{\rightarrow }0.$ Premultiply (\ref{eqn_thm 2iii lambda1
regression}) by $\hat{\lambda}_{2}^{\prime }/N$:%
\begin{equation*}
\hat{\lambda}_{2}^{\prime }\lambda _{1}/N=\hat{k}_{1}\hat{\lambda}%
_{2}^{\prime }\hat{\lambda}_{1}/N+\hat{\lambda}_{2}^{\prime }q_{1}/N=\hat{%
\lambda}_{2}^{\prime }q_{1}/N.
\end{equation*}%
But from Cauchy-Schwarz%
\begin{equation*}
(\hat{\lambda}_{2}^{\prime }q_{1}/N)^{2}\leq (\hat{\lambda}_{2}^{\prime }%
\hat{\lambda}_{2}/N)(q_{1}^{\prime }q_{1}/N)\overset{p}{\rightarrow }0.
\end{equation*}%
Thus $\hat{\lambda}_{2}^{\prime }\lambda _{1}/N\overset{p}{\rightarrow }%
h_{21}=0$ and (\ref{eqn_thm 2iii second row}) becomes%
\begin{equation*}
h_{22}^{2}\omega _{22}+h_{23}^{2}\omega _{33}+\cdots +h_{2r}^{2}\omega
_{rr}=\omega _{22}.
\end{equation*}%
Since $\omega _{22}>\omega _{33}>\cdots >\omega _{rr}$ and $%
h_{22}^{2}+h_{23}^{2}+\cdots +h_{2r}^{2}\leq 1,$ this requires $h_{22}^{2}=1$
and all other elements of the second row of $H$ to be zero, establishing the
second row of the claim in Theorem 1(\textit{%
iii}). Proceeding iteratively through rows 3,4,..,$r$ establishes the rest
of the result in (\textit{iii}).

\subsubsection*{Proof of Theorem 1(iv)}

Write%
\begin{eqnarray}
\hat{\Xi}\hat{F}_{t}-F_{t} &=&N^{-1}\hat{\Xi}\hat{\Lambda}^{\prime }\hat{C}%
_{t}-F_{t}  \notag \\
&=&N^{-1}\hat{\Xi}\hat{\Lambda}^{\prime }(\Lambda F_{t}+e_{t}+\hat{V}%
_{t})-F_{t}  \notag \\
&=&(N^{-1}\hat{\Xi}\hat{\Lambda}^{\prime }\Lambda -I_{r})F_{t}+N^{-1}\hat{\Xi%
}\hat{\Lambda}^{\prime }e_{t}+N^{-1}\hat{\Xi}\hat{\Lambda}^{\prime }\hat{V}%
_{t}.  \label{eqn_thm2iv eq1}
\end{eqnarray}%
The task is to show that all three terms in (\ref{eqn_thm2iv eq1}) have plim
0. That $(N^{-1}\hat{\Xi}\hat{\Lambda}^{\prime }\Lambda -I_{r})F_{t}\overset{%
p}{\rightarrow }0$ follows immediately from result (\textit{iii}). For the
second term,%
\begin{equation}
N^{-1}\hat{\Xi}\hat{\Lambda}^{\prime }e_{t}=N^{-1}(\hat{\Xi}\hat{\Lambda}%
^{\prime }-\Lambda ^{\prime })e_{t}+N^{-1}\hat{\Xi}\Lambda ^{\prime }e_{t}.
\label{eqn_thm2iv eq2}
\end{equation}%
Consider the square of the $j$th element of the first term in (\ref%
{eqn_thm2iv eq2}):%
\begin{equation}
\left[ \frac{(\hat{\Xi}_{j}\hat{\lambda}_{j}^{\prime }-\lambda _{j}^{\prime
})e_{t}}{N}\right] ^{2}\leq \left[ \frac{(\hat{\Xi}_{j}\hat{\lambda}%
_{j}^{\prime }-\lambda _{j}^{\prime })(\hat{\Xi}_{j}\hat{\lambda}%
_{j}-\lambda _{j})}{N}\right] \left[ \frac{e_{t}^{\prime }e_{t}}{N}\right] .
\label{eqn_thm2iv eq3}
\end{equation}%
The first term in (\ref{eqn_thm2iv eq3}) is%
\begin{equation*}
\frac{(\hat{\Xi}_{j}\hat{\lambda}_{j}^{\prime }-\lambda _{j}^{\prime })(\hat{%
\Xi}_{j}\hat{\lambda}_{j}-\lambda _{j})}{N}=\frac{\hat{\Xi}_{j}^{2}\hat{%
\lambda}_{j}^{\prime }\hat{\lambda}_{j}}{N}-\frac{\lambda _{j}^{\prime }\hat{%
\Xi}_{j}\hat{\lambda}_{j}}{N}-\frac{\hat{\Xi}_{j}\hat{\lambda}_{j}^{\prime
}\lambda _{j}}{N}+\frac{\lambda _{j}^{\prime }\lambda _{j}}{N},
\end{equation*}%
which converges in probability to zero by Theorem 1(\textit{iii}). The second term in (\ref%
{eqn_thm2iv eq3}) is $O_{p}(1),$ by result (R1) in \cite%
{stock2002forecasting}, meaning the plim of (\ref{eqn_thm2iv eq3}) is zero.
The second term in (\ref{eqn_thm2iv eq2}) also converges in probability to
zero as in \cite{stock2002forecasting} Result (R15). Hence $N^{-1}\hat{\Xi}%
\hat{\Lambda}^{\prime }e_{t}\overset{p}{\rightarrow }0.$

For the third term in (\ref{eqn_thm2iv eq1}), $N^{-1}\hat{\Xi}\hat{\Lambda}%
^{\prime }\hat{V}_{t},$ note that the $j$th element is $N^{-1}\hat{\Xi}_{j}%
\hat{\lambda}_{j}^{\prime }\hat{V}_{t}$ whose square is 
\begin{equation}
N^{-2}\hat{\lambda}_{j}^{\prime }\hat{V}_{t}\hat{V}_{t}^{\prime }\hat{\lambda%
}_{j}\leq N^{-1}\tsum\nolimits_{i=1}^{N}\hat{v}_{it}^{2}\overset{p}{%
\rightarrow }0  \label{eqn_thm2 vit convergence}
\end{equation}%
due to our Assumption 3.

\subsection{Proof of Theorem 2}

We first verify Assumption 3(ii) that $\max_{1\leq i\leq N}T^{-1}\sum_{t=1}^{T}\hat{%
v}_{it}^{2}\overset{p}{\rightarrow }0$. Write (16) as 
$\tsum\nolimits_{t=1}^{T}\hat{v}_{it}^{2}=q_{iT}^{\prime }Q_{iT}^{-1}q_{iT}$
where 
\begin{equation*}
q_{iT}=\sum_{t=1}^{T}\Upsilon _{iT}^{-1}z_{i,t-h}c_{it}\text{ }\ \text{and }%
Q_{iT}=\sum_{t=1}^{T}\Upsilon _{iT}^{-1}z_{i,t-h}z_{i,t-h}^{\prime }\Upsilon
_{iT}^{-1}.
\end{equation*}%
Then 
\begin{equation}
\max_{1\leq i\leq N}\frac{1}{T}\sum_{t=1}^{T}\hat{v}_{it}^{2}\leq \big(\frac{%
1}{T}\max_{1\leq i\leq N}\Vert q_{iT}\Vert ^{2}\big)\cdot \big(\min_{1\leq
i\leq N}\varrho _{\min }(Q_{iT})\big)^{-1}.  \label{eqn_app_sum_vit2}
\end{equation}%
By Condition 4(i) and a standard
maximal inequality argument (see, for example, Lemma 2.2.2 of %
\citealt{van1996weak}), 
\begin{equation}
\frac{1}{T}\max_{1\leq i\leq N}\Vert q_{iT}\Vert ^{2}=O_{p}(N/T).
\label{eqn_small qit bound}
\end{equation}

To control the minimum eigenvalue in (\ref{eqn_app_sum_vit2}), we employ
Weyl's inequality on perturbation, which bounds the difference in minimum
eigenvalues by the norm of the matrix difference: 
\begin{equation*}
\left\vert \varrho _{\min }(Q_{iT})-\varrho _{\min }(Q_{i})\right\vert \leq
\Vert Q_{iT}-Q_{i}\Vert .
\end{equation*}%
An implication of this is that if $\min_{1\leq i\leq N}\varrho _{\min
}(Q_{iT})<\delta_N $ for some $\delta_N$ and $\max_{1\leq i\leq N}\left\Vert
Q_{iT}-Q_{i}\right\Vert \leq \mathfrak{c}_{6}(N),$ then $\min_{1\leq i\leq
N}\varrho _{\min }(Q_{i})<\delta_N +\mathfrak{c}_{6}(N)$. We can then use
Condition 4(iii) to conclude 
\begin{align*}
\mathnormal{Prob}\big(\min_{1\leq i\leq N}\varrho _{\min }(Q_{iT})<\delta_N %
\big)& =\mathnormal{Prob}\big(\min_{1\leq i\leq N}\varrho _{\min
}(Q_{iT})<\delta_N \text{ and }\max_{1\leq i\leq N}\left\Vert
Q_{iT}-Q_{i}\right\Vert >\mathfrak{c}_{6}(N)\big) \\
\text{ }\qquad \qquad \qquad & \qquad +\mathnormal{Prob}\big(\min_{1\leq
i\leq N}\varrho _{\min }(Q_{iT})<\delta_N \text{ and }\max_{1\leq i\leq
N}\left\Vert Q_{iT}-Q_{i}\right\Vert \leq \mathfrak{c}_{6}(N)\big) \\
& \leq \ \mathnormal{Prob}\big(\max_{1\leq i\leq N}\Vert Q_{iT}-Q_{i}\Vert >%
\mathfrak{c}_{6}(N)\big) \\
& \qquad +\mathnormal{Prob}\big(\min_{1\leq i\leq N}\varrho _{\min
}(Q_{i})<\delta_N +\mathfrak{c}_{6}(N)\big) \\
& =\ o(1)+\mathnormal{Prob}\big(\min_{1\leq i\leq N}\varrho _{\min
}(Q_{i})<\delta_N +\mathfrak{c}_{6}(N)\big) \\
& \leq \ o(1)+N\cdot \max_{1\leq i\leq N}\mathnormal{Prob}\big(\varrho
_{\min }(Q_{i})<\delta_N +\mathfrak{c}_{6}(N)\big).
\end{align*}%
Now set $\delta _{N}=C\log ^{-\mathfrak{c}_{4}}(N)$ for some small constant $%
C$ satisfying $\mathfrak{c}_{3}/(2C)^{1/\mathfrak{c}_{4}}>1$, where $%
\mathfrak{c}_{4}$ is defined in Condition 4(ii). Since $\mathfrak{c}_{6}(N)=o(\delta _{N})$, the above
is further bounded by 
\begin{align*}
\mathnormal{Prob}\big(\min_{1\leq i\leq N}\varrho _{\min }(Q_{iT})<C\log ^{-%
\mathfrak{c}_{4}}(N)\big)& \leq \ o(1)+N\cdot \max_{1\leq i\leq N}%
\mathnormal{Prob}\big(\varrho _{\min }(Q_{i})<2\delta_{N} \big) \\
& \leq o(1)+N\cdot \mathfrak{c}_{2}\exp \big\{-\frac{\mathfrak{c}_{3}}{%
(2C\log ^{-\mathfrak{c}_{4}}(N))^{1/\mathfrak{c}_{4}}}\big\} \\
& =o(1)+\mathfrak{c}_{2}\exp \big\{\big(-\frac{\mathfrak{c}_{3}}{(2C)^{1/%
\mathfrak{c}_{4}}}+1\big)\log (N)\big\}\rightarrow 0.
\end{align*}%
As a result, we have that 
\begin{equation}
\big(\min_{1\leq i\leq N}\varrho _{\min }(Q_{iT})\big)^{-1}=O_{p}\big(\log ^{%
\mathfrak{c}_{4}}(N)\big).  \label{eqn_big Qit bound}
\end{equation}%
Results (\ref{eqn_app_sum_vit2}), (\ref{eqn_small qit bound}), and (\ref%
{eqn_big Qit bound}) establish that Assumption 3(ii) follows from
Assumption 4 provided that $N\log ^{\mathfrak{c}%
_{4}}(N)/T=o(1)$.

We next verify Assumption 3(i) that $N^{-1}\sum_{i=1}^{N}\hat{v}_{it}^{2}\overset{p%
}{\rightarrow }0$ for each $t$. To start, we have 
\begin{align*}
\frac{1}{N}\sum_{i=1}^{N}\hat{v}_{it}^{2}& \leq \frac{1}{N}\sum_{i=1}^{N}%
\Big\{\big\Vert\sqrt{T}\Upsilon _{iT}^{-1}z_{i,t-h}\big\Vert^{2}\cdot %
\big\Vert\big(\sum_{t=1}^{T}\Upsilon _{iT}^{-1}z_{i,t-h}z_{i,t-h}^{\prime
}\Upsilon _{iT}^{-1}\big)^{-1}\big(\frac{1}{\sqrt{T}}\sum_{t=1}^{T}\Upsilon
_{iT}^{-1}z_{i,t-h}c_{it}\big)\big\Vert^{2}\Big\} \\
& \leq \big(\max_{1\leq i\leq N}\big\Vert Q_{iT}^{-1}\big(\frac{1}{\sqrt{T}}%
q_{iT}\big)\big\Vert^{2}\big)\big(\frac{1}{N}\sum_{i=1}^{N}\Vert \sqrt{T}%
\Upsilon _{iT}^{-1}z_{i,t-h}\Vert ^{2}\big) \\
& \leq \big(\min_{1\leq i\leq N}\varrho _{\min }(Q_{iT})\big)^{-2}\big(\frac{%
1}{T}\max_{1\leq i\leq N}\Vert q_{iT}\Vert ^{2}\big)\big(\frac{1}{N}%
\sum_{i=1}^{N}\Vert \sqrt{T}\Upsilon _{iT}^{-1}z_{i,t-h}\Vert ^{2}\big).
\end{align*}%
From \eqref{eqn_small qit bound} and \eqref{eqn_big Qit bound}, we have 
\begin{equation*}
\big(\min_{1\leq i\leq N}\varrho _{\min }(Q_{iT})\big)^{-2}\big(\frac{1}{T}%
\max_{1\leq i\leq N}\Vert q_{iT}\Vert ^{2}\big)=O_{p}\big(\frac{N\log ^{2%
\mathfrak{c}_{4}}(N)}{T}\big).
\end{equation*}%
Finally, Markov's inequality and Condition 4(i) together imply that 
\begin{equation*}
\frac{1}{N}\sum_{i=1}^{N}\Vert \sqrt{T}\Upsilon _{iT}^{-1}z_{i,t-h}\Vert
^{2}=O_{p}\big(\frac{1}{N}\sum_{i=1}^{N}E(\Vert \sqrt{T}\Upsilon
_{iT}^{-1}z_{i,t-h}\Vert ^{2})\big)=O_{p}(1),
\end{equation*}%
which concludes the proof.

\subsection{Proof of Theorem 3}

Recall that $u_{it}=(1-L)^{d_{i}}y_{it}-\mu _{i}=\sum_{\ell =0}^{\infty
}\psi _{i\ell }\eta _{i,t-\ell }$. The asymptotic distribution is different
depending on whether: (1) $y_{it}$ is stationary ($d_{i}=0$); (2) $y_{it}$
has a unit root but no time trend ($d_{i}=1$ and $\mu _{i}=0$); (3) $y_{it}$
has a unit root with time trend bounded away from zero ($d_{i}=1$ and $%
\exists \mathfrak{c}_{12}>0$ such that $\left\vert \mu _{i}\right\vert \geq 
\mathfrak{c}_{12}$ for all $i$). If there is a subset of observations for
which $\mu _{i}$ is nonzero but extremely close to zero, as long as the
number of such observations does not increase with $N$ and $T,$ the
existence of such a $\mathfrak{c}_{12}$ is immediate. For completeness we
also discuss the case when the number of observations with $\mu _{i}$
arbitrarily close to zero grows with $T$: (4) $\mu _{i}=\tilde{\mu}_{i}/%
\sqrt{T}$ for $\tilde{\mu}_{i}\neq 0.$ For ease of exposition, we will
partition $\{1,2,\dots ,N\}$ into $\mathcal{N}_{1}$, $\mathcal{N}_{2}$, $%
\mathcal{N}_{3}$, and $\mathcal{N}_{4}$ corresponding to the four settings.

\subsubsection*{Case 1 (stationary $y_{it})$}

In the stationary case, there is no need to introduce any additional scaling
or rotation of the regressors, so $u_{it}=y_{it}-\mu _{i}$ and $\Upsilon
_{iT}^{-1}=T^{-1/2}I_{p+1}$ where $I_{p+1}$ is the identity matrix. For this
case Assumption 5 guarantees that%
\begin{equation*}
Q_{iT}=\tsum\nolimits_{t=1}^{T}\Upsilon _{iT}^{-1}z_{i,t-h}z_{i,t-h}^{\prime
}\Upsilon
_{iT}^{-1}=T^{-1}\tsum\nolimits_{t=1}^{T}z_{i,t-h}z_{i,t-h}^{\prime }\overset%
{p}{\rightarrow }E(z_{i,t-h}z_{i,t-h}^{\prime })\coloneqq Q_{i}.
\end{equation*}%
Below we will verify Conditions 4(i--%
iii).

For the first part of 4(i), our
Assumption 5 directly implies that $E(y_{it}^{2})$ is
uniformly bounded. To show the second part of 4(i), it suffices to bound the following terms individually: 
\begin{align*}
&E\big(\big|\frac{1}{\sqrt{T}}\sum_{t=1}^{T}c_{it}\big|^2\big) = {var}\big(%
\frac{1}{\sqrt{T}}\sum_{t=1}^{T}c_{it}\big), \\
\text{and }&E\big(\big|\frac{1}{\sqrt{T}}\sum_{t=1}^{T}u_{i,t-l}c_{it}\big|^2%
\big) = {var}\big(\frac{1}{\sqrt{T}}\sum_{t=1}^{T}u_{i,t-l}c_{it}\big), 
\text{ for }l=h,\dots,h+p-1.
\end{align*}
As $c_{it}$ is a linear combination of $u_{it},u_{i,t-h},\dots,u_{i,t-h-p+1}$%
, the above can further be bounded by the following variances: 
\begin{align*}
&{var}\big(\frac{1}{\sqrt{T}}\sum_{t=1}^{T}u_{it}\big), \text{ and } {var}%
\big(\frac{1}{\sqrt{T}}\sum_{t=1}^{T}u_{i,t-l}u_{i,t-l^{\prime }}\big),
\end{align*}
for $l=h,\dots,h+p-1$ and $l^{\prime }=0, h,\dots,h+p-1$. Given our linear
process specification, the above is bounded. See, for example, Appendix A of 
\cite{shumway2000time} for an exact calculation. The same argument can be
applied to show that $E(\Vert Q_{iT} - Q_i \Vert^2) = O(1/T)$.

Condition 4(ii) holds automatically
given our assumption that the minimum eigenvalue of $\tilde{Q}_{i}$ is
bounded away from zero (i.e., it is satisfied for any $\mathfrak{c}_{4}>0$).

Finally we verify 4(iii). We adopt
Markov's inequality with a union bound, which implies that 
\begin{align*}
\mathnormal{Prob}\Big(\max_{i\in \mathcal{N}_{0}}\big\Vert Q_{iT}-Q_{i}%
\big\Vert>\sqrt{\frac{N\log ^{1/2}(N)}{T}}\Big)& \leq \sum_{i\in \mathcal{N}%
_{0}}\mathnormal{Prob}\Big(\big\Vert Q_{iT}-Q_{i}\big\Vert>\sqrt{\frac{N\log
^{1/2}(N)}{T}}\Big) \\
& \leq C\frac{|\mathcal{N}_{0}|}{N}\frac{1}{\log ^{1/2}(N)}\leq C\frac{1}{%
\log ^{1/2}(N)},
\end{align*}%
where $C$ is some constant and $|\mathcal{N}_{0}|$ is the cardinality of $%
\mathcal{N}_{0}.$ 

\subsubsection*{Case 2 (unit-root $y_{it}$ with no time trend)}

For these observations, we use $\Upsilon _{iT}^{-1}=\tilde{\Upsilon}%
_{iT}^{-1}R_{i}^{-1}$, where $\tilde{\Upsilon}_{iT}$ is a diagonal scaling
matrix 
\begin{equation}
\tilde{\Upsilon}_{iT}=%
\begin{bmatrix}
\sqrt{T}I_{p} & 0 \\ 
0 & T%
\end{bmatrix}
\label{eq_appendix_scaling_case_2}
\end{equation}%
and $R_{i}$ is a nonsingular rotation matrix such that $\Upsilon
_{iT}^{-1}z_{i,t-h}=\tilde{\Upsilon}_{iT}^{-1}\tilde{z}_{i,t-h}$ with 
\begin{equation*}
\tilde{z}_{i,t-h}=(u_{i,t-h},\ u_{i,t-h-1},\ \dots ,\ u_{i,t-h-p+2},\ 1,\
y_{i,t-h}).
\end{equation*}%
Notice that if $p=1$, we simply set $\tilde{z}_{i,t-h}={z}_{i,t-h}=(1,\
y_{i,t-h})$ so no rotation of the regressors is needed. Assumption 5 guarantees that as in 
\citet[eq.
(17.7.18)]{hamilton1994},%
\begin{equation}
Q_{iT}=\tsum\nolimits_{t=1}^{T}\Upsilon _{iT}^{-1}z_{i,t-h}z_{i,t-h}^{\prime
}\Upsilon _{iT}^{-1}\overset{d}{\rightarrow }%
\begin{bmatrix}
\tilde{Q}_{i} & 0 & 0 \\[5pt] 
0 & 1 & \psi _{i}(1)\int_{0}^{1}W_{i}(r)dr \\[5pt] 
0 & \psi _{i}(1)\int_{0}^{1}W_{i}(r)dr & \psi
_{i}(1)^{2}\int_{0}^{1}W_{i}(r)^{2}dr%
\end{bmatrix}%
\coloneqq Q_{i}  \label{eqn_thm5_Qi_unit_root}
\end{equation}%
where $\psi _{i}(1)=\sum_{l=0}^{\infty }\psi _{il}$, and $W_{i}(r)$ is a
standard Brownian motion for each $i$.

For Condition 4(i), we first note
that for any stationary component of $\tilde{z}_{i,t-h}$, the condition
holds by standard variance calculation (see the proof of the stationary
case). For the nonstationary component of $\tilde{z}_{i,t-h}$, namely $%
y_{i,t-h},$ we rewrite it as 
\begin{equation}
y_{i,t-h}=\psi _{i}(1)\sum_{s=1-h}^{t-h}\eta _{is}-\tilde{\epsilon}_{i,t-h}+%
\tilde{\epsilon}_{i,-h},\text{ where }\tilde{\epsilon}_{i,t-h}=\sum_{l=0}^{%
\infty }\tilde{\psi}_{il}\eta _{i,t-h-l}\text{ and }\tilde{\psi}%
_{il}=\sum_{l^{\prime }=l+1}^{\infty }{\psi }_{il^{\prime }};
\label{eqn_beveridge_nelson}
\end{equation}%
see for example \citet[eq. (17.5.3)]{hamilton1994}. Thus to bound moments of 
$T^{-1}\tsum\nolimits_{t=1}^{T}y_{i,t-h}u_{it}$ it is sufficient to note that%
\begin{equation*}
E\Big(\Big|\frac{1}{T}\sum_{t=1}^{T}\Big(\sum_{s=1-h}^{t-h}\eta _{is}\Big)%
\Big(\sum_{l=0}^{\infty }\psi _{il}\eta _{i,t-l}\Big)\Big|^{2}\Big)=O(1),
\end{equation*}%
where the bound is uniform for $i\in \mathcal{N}_{1}$.

We next verify Condition 4(ii).
Assumption 5(iv) bounds the
smallest eigenvalue of the upper-left block of (\ref{eqn_thm5_Qi_unit_root}%
), while Assumption 5(ii)
allows us to ignore the additional scaling $\psi _{i}(1)$ for the Brownian
motion. Thus it is sufficient to establish a bound on the smallest
eigenvalue of the $(2\times 2)$ matrix

\begin{align*}
Q_i = 
\begin{bmatrix}
1 & \int_{0}^1 {W}_i(r)dr \\ 
\int_{0}^1 {W}_i(r)dr & \int_{0}^1 {W}_i(r)^2dr%
\end{bmatrix}%
\end{align*}
as a slight abuse of notation.

To provide a probabilistic bound on its minimum eigenvalue, notice that the
determinant of $Q_{i}$ is given by 
\begin{equation*}
\left\vert Q_{i}\right\vert =\int_{0}^{1}\Big({W}_{i}(r)-\int_{0}^{1}{W}%
_{i}(s)ds\Big)^{2}dr=\varrho _{\min }(Q_{i})\varrho _{\max }(Q_{i})
\end{equation*}%
while the norm $\left\Vert Q_{i}\right\Vert \geq \varrho _{\max }(Q_{i}).$
This means 
\begin{equation*}
\varrho _{\min }(Q_{i})\geq \Vert Q_{i}\Vert ^{-1}\int_{0}^{1}\Big({W}%
_{i}(r)-\int_{0}^{1}{W}_{i}(s)ds\Big)^{2}dr.
\end{equation*}%
\cite{beghin2005exact} showed in their Section 3 that 
\begin{equation*}
\mathnormal{Prob}\Big(\int_{0}^{1}\Big({W}_{i}(r)-\int_{0}^{1}{W}_{i}(s)ds%
\Big)^{2}dr<\epsilon \Big)=O(\exp \{-C\epsilon ^{-1}\}).
\end{equation*}%
To bound the matrix norm $\Vert Q_{i}\Vert $, we notice that the supremum of
a Brownian motion is sub-Gaussian by the reflection principle (see, for
example, \citealt[Chapter VIII, 1.21]{cinlar2011probability}), and therefore
the entries of $Q_{i}$ are at most sub-exponential, which leads to the
following: 
\begin{equation*}
\mathnormal{Prob}\Big(\Vert Q_{i}\Vert >\frac{1}{\epsilon }\Big)=O(\exp
\{-C\epsilon ^{-1}\}).
\end{equation*}%
As a result, Condition 4(ii) holds
with $\mathfrak{c}_{4}=2$.

Finally, we verify Condition 4(iii) by
establishing a uniform distributional approximation result. We can write 
\begin{align*}
Q_{iT} &= \sum_{t=1}^T \Upsilon_{iT}^{-1}z_{i,t-h}z_{i,t-h}^{\prime
}\Upsilon_{iT}^{-1} = 
\begin{bmatrix}
\frac{1}{T}\sum_{t=1}^T x_{i,t-h}x_{i,t-h}^{\prime } & \frac{1}{T}%
\sum_{t=1}^T x_{i,t-h} & \frac{1}{T^{3/2}}\sum_{t=1}^T y_{i,t-h}x_{i,t-h} \\ 
\frac{1}{T}\sum_{t=1}^T x_{i,t-h}^{\prime } & 1 & \frac{1}{T^{3/2}}%
\sum_{t=1}^T y_{it} \\ 
\frac{1}{T^{3/2}}\sum_{t=1}^T y_{i,t-h}x_{i,t-h}^{\prime } & \frac{1}{T^{3/2}%
}\sum_{t=1}^T y_{i,t-h} & \frac{1}{T^{2}}\sum_{t=1}^T y_{i,t-h}^2%
\end{bmatrix}%
,
\end{align*}
where $x_{i,t-h} = (u_{i,t-h},\ u_{i,t-h-1},\ \dots,\ u_{i,t-h-p+2})$. We
will only illustrate convergence of the bottom-right $2\times 2$ block of $%
Q_{iT}$, as the other entries converge in probability, which can be
demonstrated by standard variance calculation.

To start, we recall that the decomposition in \eqref{eqn_beveridge_nelson} expresses $y_{it}$ as a sum of independent components and two remainder terms. Our assumption implies that $\tilde{\psi}_{il} $ are absolutely summable, from which we conclude that $\tilde{\epsilon}_{it}$ have bounded fourth moments. As a result, 
\begin{equation*}
\max_{i \in \mathcal{N}_1}\max_{1\leq t\leq T }\left|\tilde{\epsilon}%
_{i,t-h}\right| = O_p((NT)^{1/4}) = o_p\Big(\frac{\sqrt{T}}{\log^{3}(N)}\Big).
\end{equation*}
In the above, the last step follows from the condition $N\log
^{12}(N)/T\rightarrow 0$. We next employ Theorem 4 of \cite{komlos1976approximation} with a union bound: setting $x = \sqrt{T} /\log^3(N)$ and $H(x) = x^4$ in their theorem, we have 
\begin{align}  \label{eqn_strong approx}
\mathnormal{Prob}\Big( \max_{i \in \mathcal{N}_1}\max_{1\leq t\leq T} \Big|%
y_{i,t-h}^\star- \tilde{W}_{i}(t)\Big| > \frac{\sqrt{T}}{\log^{3}(N)} \Big) = O%
\Big(\frac{N\log^{12}(N)}{T}\Big) + o(1) = o(1).
\end{align}
Here, $\tilde{W}_{i}(t)$ and $y_{it}^\star$ are defined on a possibly new probability space, such that $\tilde{W}_{i}(t)$ is a standard Brownian motion scaled by $\psi_{i}(1)$, and that $y_{it}^\star$ and the original $y_{it}$ have the same distribution. To improve readability, however, we slightly abuse notation
and do not distinguish $y_{it}^\star$ and $y_{it}$. Given this theoretical device, we will show below that under the condition $N\log
^{12}(N)/T\rightarrow 0$, the following hold: 
\begin{align}
& \mathnormal{Prob}\Big(\max_{i\in \mathcal{N}_{1}}\Big|\frac{1}{T}%
\sum_{t=1}^{T}\frac{y_{i,t-h}}{\sqrt{T}}-\int_{0}^{1}\tilde{W}_{i}(t)dt\Big|>%
\frac{1}{\log ^{3}(N)}\Big)=o(1)  \label{eqn_convergence rate 1} \\
& \mathnormal{Prob}\Big(\max_{i\in \mathcal{N}_{1}}\Big|\frac{1}{T}%
\sum_{t=1}^{T}\Big(\frac{y_{i,t-h}}{\sqrt{T}}\Big)^{2}-\int_{0}^{1}\tilde{W}%
_{i}(t)^{2}dt\Big|>\frac{1}{\log ^{9/4}(N)}\Big)=o(1).
\label{eqn_convergence rate 2}
\end{align}%
Since $\log ^{3}(N)>\log ^{9/4}(N),$ we conclude that if we set 
\begin{equation*}
\mathfrak{c}_{6}(N)\propto \frac{1}{\log ^{9/4}(N)}=o\Big(\frac{1}{\log ^{%
\mathfrak{c}_{4}}(N)}\Big)\text{ with }\mathfrak{c}_{4}=2,
\end{equation*}%
then Condition 4(iii) would hold.

We first prove \eqref{eqn_convergence rate 1}. Write 
\begin{align*}
\max_{i \in \mathcal{N}_1}\Big|\frac{1}{T}\sum_{t=1}^T\frac{y_{i,t-h}}{\sqrt{%
T}}- \int_{0}^1 \tilde{W}_i(t)dt\Big| \leq \underbrace{\max_{i \in \mathcal{N%
}_1}\Big|\frac{1}{T}\sum_{t=1}^T\frac{y_{i,t-h}}{\sqrt{T}}- \frac{1}{T}%
\sum_{t=1}^T\tilde{W}_i(t/T)\Big|}_{\textstyle \text{(I)}} + \underbrace{%
\max_{i \in \mathcal{N}_1}\Big|\frac{1}{T}\sum_{t=1}^T\tilde{W}_i(t/T)-
\int_{0}^1 \tilde{W}_i(t)dt\Big|}_{\textstyle \text{(II)}}.
\end{align*}
A bound for (I) directly follows from our earlier discussion, which implies 
\begin{align*}
\mathnormal{Prob}\Big( \text{\text(I)} > \frac{1}{\log^{3}(N)} \Big) = o(1).
\end{align*} 

The term (II) is simply the approximation error of
the Riemann sum of the Brownian path, which satisfies 
\begin{align*}
\text{(II)} \leq \max_{i \in \mathcal{N}_1}\frac{1}{T}\sum_{t=1}^T \sup_{%
\frac{t}{T}\leq s< \frac{t+1}{T}}|\tilde{W}_i(t/T) - \tilde{W}_i(s)|.
\end{align*}
To provide a probabilistic bound, we notice that the supremum $\sqrt{T}%
\sup_{ \frac{t}{T}\leq s< \frac{t+1}{T}}|\tilde{W}_i(t/T) - \tilde{W}_i(s)|$
is sub-Gaussian by the reflection principle. Therefore, by a maximal
inequality argument, one has 
\begin{align*}
\mathnormal{Prob}\Big( \text{(II)} > \frac{\log(N)}{\sqrt{T}} \Big) = o(1).
\end{align*}
The conclusion \eqref{eqn_convergence rate 1} then follows from the rates we
established for (I) and (II) as $\log(N) / \sqrt{T}$ is negligible compared
to $\log^{-3}(N)$.

We now show \eqref{eqn_convergence rate 2}. We first write 
\begin{align*}
\Big|\frac{1}{T}\sum_{t=1}^T\Big(\frac{y_{i,t-h}}{\sqrt{T}}\Big)^2-
\int_{0}^1 \tilde{W}_i(t)^2dt\Big| \leq \Big|\frac{1}{T}\sum_{t=1}^T\Big(%
\frac{y_{i,t-h}}{\sqrt{T}}\Big)^2- \frac{1}{T}\sum_{t=1}^T\tilde{W}_i(t/T)^2%
\Big| + \Big|\frac{1}{T}\sum_{t=1}^T\tilde{W}_i(t/T)^2- \int_{0}^1 \tilde{W}%
_i(t)^2dt\Big|.
\end{align*}
Next, we further bound the first term on the right-hand side by 
\begin{align*}
&\Big|\frac{1}{T}\sum_{t=1}^T\Big(\frac{y_{i,t-h}}{\sqrt{T}}\Big)^2- \frac{1%
}{T}\sum_{t=1}^T\tilde{W}_i(t/T)^2\Big| \leq \frac{2}{T}\sum_{t=1}^T\Big|%
\frac{y_{i,t-h}}{\sqrt{T}}- \tilde{W}_i(t/T)\Big|\tilde{W}_i(t/T) +\frac{1}{T%
}\sum_{t=1}^T\Big|\frac{y_{i,t-h}}{\sqrt{T}}- \tilde{W}_i(t/T)\Big|^2 \\
&\qquad\qquad\leq 2\sqrt{\frac{1}{T}\sum_{t=1}^T\Big|\frac{y_{i,t-h}}{\sqrt{T%
}}- \tilde{W}_i(t/T)\Big|^2} \sqrt{\frac{1}{T}\sum_{t=1}^T\tilde{W}_i(t/T)^2}
+\frac{1}{T}\sum_{t=1}^T\Big|\frac{y_{i,t-h}}{\sqrt{T}}- \tilde{W}_i(t/T)%
\Big|^2.
\end{align*}
Therefore, 
\begin{align*}
&\max_{i \in \mathcal{N}_1}\Big|\frac{1}{T}\sum_{t=1}^T\Big(\frac{y_{i,t-h}}{%
\sqrt{T}}\Big)^2- \int_{0}^1 \tilde{W}_i(t)^2dt\Big| \leq 2\underbrace{\sqrt{%
\max_{i \in \mathcal{N}_1}\frac{1}{T}\sum_{t=1}^T\Big|\frac{y_{i,t-h}}{\sqrt{%
T}}- \tilde{W}_i(t/T)\Big|^2}}_{\textstyle \text{(III)}} \underbrace{\sqrt{%
\max_{i \in \mathcal{N}_1}\frac{1}{T}\sum_{t=1}^T\tilde{W}_i(t/T)^2}}_{%
\textstyle \text{(IV)}} \\
&\qquad +\underbrace{\max_{i \in \mathcal{N}_1}\frac{1}{T}\sum_{t=1}^T\Big|%
\frac{y_{i,t-h}}{\sqrt{T}}- \tilde{W}_i(t/T)\Big|^2 }_{\textstyle \text{(V)}%
} + \underbrace{\max_{i \in \mathcal{N}_1}\Big|\frac{1}{T}\sum_{t=1}^T\tilde{%
W}_i(t/T)^2- \int_{0}^1 \tilde{W}_i(t)^2dt\Big|}_{\textstyle \text{(VI)}}.
\end{align*}

Clearly from \eqref{eqn_strong approx}, 
\begin{align*}
\mathnormal{Prob}\Big(\text{(III)} > \frac{1}{\log^{3}(N)}\Big) = o(1),\quad 
\mathnormal{Prob}\Big(\text{(V)} > \frac{1}{\log^{6}(N)}\Big) = o(1).
\end{align*}
By a maximal inequality for sub-Gaussian random variables, one has 
\begin{align*}
\mathnormal{Prob}\Big( \text{(IV)} > \log^{3/4}(N) \Big) = o(1).
\end{align*}
Term (VI) is again the approximation error of a Riemann sum, which is
bounded by 
\begin{align*}
\text{ (VI)} \leq \max_{i \in \mathcal{N}_1}\frac{1}{T}\sum_{t=1}^T \sup_{%
\frac{t}{T}\leq s< \frac{t+1}{T}}|\tilde{W}_i(t/T)^2 - \tilde{W}_i(s)^2|
\leq \max_{i \in \mathcal{N}_1}\Big(\frac{1}{T}\sum_{t=1}^T \sup_{\frac{t}{T}%
\leq s< \frac{t+1}{T}}|\tilde{W}_i(t/T) - \tilde{W}_i(s)|\Big)\Big(%
2\sup_{1\leq s\leq T}|\tilde{W}_i(s)|\Big),
\end{align*}
and therefore it satisfies 
\begin{align*}
\mathnormal{Prob}\Big( \text{ (VI)} > \frac{\log^{2}(N)}{\sqrt{T}} \Big) =
o(1).
\end{align*}
Finally, we note that the rate for the product (III)$\cdot$(IV) is $%
\log^{-9/4}(N)$, which dominates the rates for (V) and (VI).

\subsubsection*{Case 3 (unit-root $y_{it}$ with linear time trend)}

With a time trend, $\Upsilon _{iT}^{-1}=\tilde{\Upsilon}_{iT}^{-1}R_{i}^{-1}$
with 
\begin{equation*}
\tilde{\Upsilon}_{iT}=%
\begin{bmatrix}
\sqrt{T}I_{p} & 0 \\ 
0 & T^{3/2}%
\end{bmatrix}%
\end{equation*}%
and $\Upsilon _{iT}^{-1}z_{i,t-h}=\tilde{\Upsilon}_{iT}^{-1}\tilde{z}%
_{i,t-h} $ with $\tilde{z}_{i,t-h}=(u_{i,t-h},\ u_{i,t-h-1},\ \dots ,\
u_{i,t-h-p+2},\ 1,\ y_{i,t-h})$. Here 
\begin{equation*}
\tsum\nolimits_{t=1}^{T}\Upsilon _{iT}^{-1}z_{i,t-h}z_{i,t-h}^{\prime
}\Upsilon _{iT}^{-1}\overset{p}{\rightarrow }%
\begin{bmatrix}
\tilde{Q}_{i} & 0 & 0 \\[5pt] 
0 & 1 & \mu _{i}/2 \\[5pt] 
0 & \mu _{i}/2 & \mu _{i}^{2}/3%
\end{bmatrix}%
\coloneqq Q_{i}
\end{equation*}

To verify Condition 4(i), it
suffices to compute $E((\sqrt{T}T^{-3/2})^{2}y_{i,t-h}^{2})$, which is
uniformly bounded in $i$ and $t$, and 
\begin{align*}
E\Big(\Big|\frac{1}{T^{3/2}}\sum_{t=1}^{T}t\Big(\sum_{l=1}^{\infty }\psi
_{il}\eta _{i,t-l}\Big)\Big|^{2}\Big)& =\frac{1}{T^{3}}\sum_{t=1}^{T}%
\sum_{t^{\prime }=1}^{T}tt^{\prime }cov(u_{it},u_{it^{\prime }}) \\
& \leq \frac{1}{T^{2}}\sum_{t=1}^{T}t\sum_{t^{\prime
}=1}^{T}cov(u_{it},u_{it^{\prime }})\leq C\frac{1}{T^{2}}\sum_{t=1}^{T}t
\end{align*}%
for some constant $C$, making the above bounded.

Condition 4(ii) is satisfied by
Assumption 5, and we take $\mathfrak{c}_{4}=1/2$.

Finally we check Condition 4(iii). The
upper diagonal block of $Q_{iT}$ converges to $\tilde{Q}_i$, which has been
discussed in Case 1 in this proof. Take any stationary component $u_{i,t-l}$
of $x_{i,t-h}$, then we can consider 
\begin{align*}
E\Big(\Big|\frac{1}{T^2}\sum_{t=1}^T y_{i,t-h}u_{i,t-l}\Big|^2\Big) \leq 2E%
\Big(\Big|\frac{1}{T^2}\sum_{t=1}^T (y_{i,t-h}-\mu_it)u_{i,t-l}\Big|^2\Big) %
+ 2\mu_i^2E\Big(\Big|\frac{1}{T^2}\sum_{t=1}^T tu_{i,t-l}\Big|^2\Big).
\end{align*}
The first term on the right-hand side is of order $T^{-2}$ by our discussion
in Case 2, and the second term is of order $T^{-1}$ by Condition 4(i), which we verified earlier.
Therefore, $T^{-2}\sum_{t=1}^T y_{i,t-h}u_{i,t-l} = O_p(T^{-1/2})$, where
the bound is uniform for $i \in \mathcal{N}_2$. This establishes the
convergence of top-right and bottom-left entries of $Q_{iT}$.

Now consider $T^{-2}\sum_{t=1}^T y_{i,t-h}$. We write it as the sum $%
T^{-2}\sum_{t=1}^T (y_{i,t-h}-\mu_{i}t)$ and $T^{-2}\sum_{t=1}^T \mu_{i}t$.
Clearly the former is of order $T^{-1/2}$ by standard variance calculation.

Finally, we consider 
\begin{equation*}
\frac{1}{T^{3}}\sum_{t=1}^{T}y_{i,t-h}^{2}=\frac{1}{T^{3}}\sum_{t=1}^{T}\Big(%
(y_{i,t-h}-\mu _{i}t)+\mu _{i}t\Big)^{2}=\frac{1}{T^{3}}%
\sum_{t=1}^{T}(y_{i,t-h}-\mu _{i}t)^{2}+\frac{2\mu _{i}}{T^{3}}%
\sum_{t=1}^{T}(y_{i,t-h}-\mu _{i}t)t+\frac{\mu _{i}^{2}}{T^{3}}%
\sum_{t=1}^{T}t^{2}.
\end{equation*}%
The first term on the right-hand side is of order $T^{-1}$ by mean
calculation. The second term has order $T^{-1/2}$ by variance calculation.

\subsubsection*{Case 4 (unit root with time trend arbitrarily close to zero)}

Finally we consider observations for which $\mu _{i}=\tilde{\mu}_{i}/\sqrt{T}
$ for $\tilde{\mu}_{i}\neq 0.$ For this case (\ref{eqn_beveridge_nelson})
becomes%
\begin{equation*}
y_{i,t-h}=\xi _{i,t-h}+\tilde{\mu}_{i}(t-h)/\sqrt{T}
\end{equation*}%
for $\xi _{i,t-h}=\psi _{i}(1)\sum_{s=1-h}^{t-h}\eta _{is}-\tilde{\epsilon}%
_{i,t-h}+\tilde{\epsilon}_{i,-h}.$ It turns out for this case we can use the
same scaling matrix as in (\ref{eq_appendix_scaling_case_2}). The (3,2)
element in (\ref{eqn_thm5_Qi_unit_root}) would then be 
\begin{equation*}
T^{-3/2}\tsum\nolimits_{t=1}^{T}\left[ \xi _{i,t-h}+\tilde{\mu}_{i}(t-h)/%
\sqrt{T}\right] \overset{d}{\rightarrow }\psi _{i}(1)\int_{0}^{1}W_{i}(r)dr+%
\tilde{\mu}_{i}/2.
\end{equation*}%
The (3,3) element is%
\begin{eqnarray*}
&&T^{-2}\tsum\nolimits_{t=1}^{T}\left[ \xi _{i,t-h}+\tilde{\mu}_{i}(t-h)/%
\sqrt{T}\right] ^{2} \\
&=&T^{-2}\tsum\nolimits_{t=1}^{T}\xi
_{i,t-h}^{2}+2T^{-5/2}\tsum\nolimits_{t=1}^{T}\xi _{i,t-h}\tilde{\mu}%
_{i}(t-h)+T^{-3}\tsum\nolimits_{t=1}^{T}\tilde{\mu}_{i}^{2}(t-h)^{2} \\
&&\overset{d}{\rightarrow }\left[ \psi _{i}(1)\right] ^{2}\int_{0}^{1}\left[
W_{i}(r)dr\right] ^{2}+2\tilde{\mu}_{i}\psi _{i}(1)\int_{0}^{1}rW_{i}(r)dr+%
\tilde{\mu}_{i}^{2}/3
\end{eqnarray*}%
using results (h), (i), and (k) in Proposition 17.3 in \cite{hamilton1994}.
Thus for Case 4, the bottom-right block of the limiting matrix $Q_{i}$ takes
the form 
\begin{align*}
& 
\begin{bmatrix}
1 & \psi _{i}(1)\int_{0}^{1}{W}_{i}(r)dr+\tilde{\mu}_{i}/2 \\ 
\psi _{i}(1)\int_{0}^{1}{W}_{i}(r)dr+\tilde{\mu}_{i}/2 & \psi
_{i}(1)^{2}\int_{0}^{1}{W}_{i}(r)^{2}dr+2\tilde{\mu}_{i}\psi
_{i}(1)\int_{0}^{1}rW_{i}(r)dr+\tilde{\mu}_{i}^{2}/3%
\end{bmatrix}
\\
& =%
\begin{bmatrix}
1 & \int_{0}^{1}\left[ \psi _{i}(1){W}_{i}(r)+\tilde{\mu}_{i}r\right] dr \\ 
\int_{0}^{1}\left[ \psi _{i}(1){W}_{i}(r)+\tilde{\mu}_{i}r\right] dr & 
\int_{0}^{1}\left[ \psi _{i}(1){W}_{i}(r)+\tilde{\mu}_{i}r\right] ^{2}dr%
\end{bmatrix}%
.
\end{align*}%
The determinant of this $(2\times 2)$ matrix is 
\begin{equation*}
\int_{0}^{1}\Big[\psi _{i}(1){W}_{i}(r)+\tilde{\mu}_{i}r-\int_{0}^{1}(\psi
_{i}(1){W}_{i}(s)+\tilde{\mu}_{i}s)ds\Big]^{2}dr.
\end{equation*}%
Its left-tail behavior follows from $L_{2}$ small-ball probabilities of
Brownian motions with drifts. Finally, note that all moment bounds
established for earlier cases continue to hold.

\subsection{Extension to the local-to-unity case}

In this section we extend our uniform distributional approximation result to series that are local to unity. To be specific, let $\mathcal{I}_{\mathrm{LTU}}$ be a collection of series where $i\in\mathcal I_{\mathrm{LTU}}$ takes the form
\[
y_{it}=\rho_{iT}y_{i,t-1}+u_{it},
\qquad\text{with}\ 
\rho_{iT}=1-\frac{c_i}{T},
\]
and $c_i\in[0,\mathfrak{c}_{13}]$ for some constant $0<\mathfrak{c}_{13}<\infty$, and $u_{it}$ satisfies Assumption 5. As before, we define
\[
Q_{iT}
=
\sum_{t=1}^T \Upsilon_{iT}^{-1} z_{i,t-h} z_{i,t-h}' \Upsilon_{iT}^{-1}
\]
Our goal is to show that
\[
\mathnormal{Prob}\Big(
\max_{i\in\mathcal I_{\mathrm{LTU}}}\bigl\|Q_{iT}-Q_i(c_i)\bigr\|\le c_6(N)
\Big)\to1,
\]
and hence Assumption 4(iii) holds for the local-to-unity case. Throughout the proof we impose the same growth condition as in Appendix~A.3, namely
\begin{equation*}
\frac{N\log^{12}(N)}{T}\to 0.
\end{equation*}
We will set $c_6(N)\asymp \log^{-9/4}N$.

We first discuss the limiting matrix $Q_i(c_i)$. Define the continuous-time Brownian motion
\[
\mathbb W_i(r)=\frac{\widetilde W_i(\lfloor Tr\rfloor)}{\sqrt T},\qquad r\in[0,1],
\]
where $\lfloor \cdot \rfloor$ denotes the floor function, and $\widetilde W_i(\cdot)$ is the Gaussian process (scaled Brownian motion) coupled to the partial sums
as in \eqref{eqn_strong approx} of Appendix~A.3. For each $c\in[0,\mathfrak{c}_{13}]$ consider the Ornstein--Uhlenbeck (OU) functional
\begin{equation*}
J_{i,c}(r)=\int_0^r e^{-c(r-s)}\,d\mathbb W_i(s),\qquad r\in[0,1].
\end{equation*}
We then let $Q_i(c)$ be the same limiting matrix as in Appendix~A.3, Case~2, with the modification that
every appearance of the Brownian path $\widetilde W_i(\cdot)$ is replaced by the OU path $J_{i,c}(\cdot)$.
Concretely, in the bottom-right $2\times 2$ block, the entries $\int_0^1\widetilde W_i(r)\,dr$ and
$\int_0^1\widetilde W_i(r)^2\,dr$ are replaced by $\int_0^1 J_{i,c}(r)\,dr$ and $\int_0^1 J_{i,c}(r)^2\,dr$.

Let $\Upsilon_{iT}$ be the same scaling matrix as in the $d_i=1$ case
(the one used in Appendix~A.3, Case~2). Under this normalization, Appendix~A.3, Case~2 shows
that it suffices to verify convergence of the bottom-right $2\times 2$ block of $Q_{iT}$, because all
other entries converge in probability by standard variance calculations (the argument is unchanged here
since it only uses moment bounds implied by Assumption~5 and the same scaling).
Hence it remains to establish uniform approximation of
\begin{equation*}
\frac1T\sum_{t=1}^T \frac{y_{i,t-h}}{\sqrt T}
\quad\text{and}\quad
\frac1T\sum_{t=1}^T \Big(\frac{y_{i,t-h}}{\sqrt T}\Big)^2
\end{equation*}
by the corresponding OU functionals $\int_0^1J_{i,c_i}(r)\,dr$ and $\int_0^1J_{i,c_i}(r)^2\,dr$.

Fix $i\in\mathcal I_{\mathrm{LTU}}$ and write $\rho=\rho_{iT}=1-c_i/T$ to save notation.
Iterating $y_{it}=\rho y_{i,t-1}+u_{it}$ gives
\begin{equation*}
y_{it}=\rho^t y_{i0}+\sum_{s=1}^t\rho^{t-s}u_{is}.
\end{equation*}
Let $S_{it}=\sum_{s=1}^t u_{is}$ and set $S_{i0}=0$. Since $u_{is}=S_{is}-S_{i,s-1}$,
we can rewrite the convolution via summation by parts:
\begin{align*}
\sum_{s=1}^t\rho^{t-s}u_{is}
&=\sum_{s=1}^t\rho^{t-s}(S_{is}-S_{i,s-1})
=\sum_{s=1}^t\rho^{t-s}S_{is}-\sum_{s=1}^t\rho^{t-s}S_{i,s-1}\notag\\
&=\sum_{s=1}^t\rho^{t-s}S_{is}-\sum_{r=0}^{t-1}\rho^{t-1-r}S_{ir} =S_{it}+\sum_{s=1}^{t-1}\big(\rho^{t-s}-\rho^{t-1-s}\big)S_{is}\notag\\
&=S_{it}-(1-\rho)\sum_{s=1}^{t-1}\rho^{t-1-s}S_{is}
=S_{it}-\frac{c_i}{T}\sum_{s=1}^{t-1}\rho^{t-1-s}S_{is}.
\end{align*}
Combining results in the two displays above, one has that, for every $t\le T$,
\begin{equation*}
\frac{y_{it}}{\sqrt T}
=\rho^t\frac{y_{i0}}{\sqrt T}
+\frac{S_{it}}{\sqrt T}
-\frac{c_i}{T}\sum_{s=1}^{t-1}\rho^{t-1-s}\frac{S_{is}}{\sqrt T}.
\end{equation*}
Let $Y_{iT}(r)=y_{i,\lfloor Tr\rfloor}/\sqrt T$. Substituting $\mathbb W_{iT}$ for the scaled partial sums, we obtain the decomposition
\begin{align*}
Y_{iT}(r)
&=
\rho^t\frac{y_{i0}}{\sqrt T}
+\mathbb W_{iT}(r)
-\frac{c_i}{T}\sum_{s=1}^{t-1}\rho^{t-1-s}\mathbb W_{iT}(s/T)
+R_{iT}(r),
\end{align*}
where the remainder is
\[
R_{iT}(r)
=
\Big(\frac{S_{it}}{\sqrt T}-\mathbb W_{iT}(r)\Big)
-\frac{c_i}{T}\sum_{s=1}^{t-1}\rho^{t-1-s}\Big(\frac{S_{is}}{\sqrt T}-\mathbb W_{iT}(s/T)\Big).
\]

For each $i\in\mathcal I_{\mathrm{LTU}}$, define the partial sums
\[
S_{it}=\sum_{s=1}^t u_{is},\qquad S_{i0}=0.
\]
Appendix~A.3 establishes a uniform strong approximation (their display \eqref{eqn_strong approx}).
Applying that result to the partial sums $S_{i,t-h}$,
there exist Gaussian processes $\{\widetilde W_i(t)\}_{t=0}^T$ such that
\begin{equation*}
\mathnormal{Prob}\Bigg(
\max_{i\in\mathcal I_{\mathrm{LTU}}}\max_{1\le t\le T}
\big|S_{i,t-h}-\widetilde W_i(t)\big|
>\frac{\sqrt T}{\log^3 (N)}
\Bigg)=o(1).
\end{equation*}
Define the event
\[
\mathcal E_T
=
\Bigg\{
\max_{i\in\mathcal I_{\mathrm{LTU}}}\max_{1\le t\le T}
\big|S_{i,t-h}-\widetilde W_i(t)\big|
\le\frac{\sqrt T}{\log^3 (N)}
\Bigg\},
\]
then $\mathnormal{Prob}(\mathcal E_T)\to 1$. On $\mathcal E_T$, 
\begin{equation*}
\sup_{i\in\mathcal I_{\mathrm{LTU}}}\sup_{0\le r\le 1}
\Big|\frac{S_{i,\lfloor Tr\rfloor}}{\sqrt T}-\mathbb W_{iT}(r)\Big|
\le \frac{1}{\log^3 (N)}.
\end{equation*}
This further implies that for all $i$ and all $r\in[0,1]$,
\begin{align*}
|R_{iT}(r)|
&\le
\sup_{0\le r\le 1}\Big|\frac{S_{i,\lfloor Tr\rfloor}}{\sqrt T}-\mathbb W_{iT}(r)\Big|
+
\frac{c_i}{T}\sum_{s=1}^{t-1}\rho^{t-1-s}
\sup_{1\le s\le T}\Big|\frac{S_{is}}{\sqrt T}-\mathbb W_{iT}(s/T)\Big|\notag\\
&\le
\frac{1}{\log^3 (N)}
+\Big(\frac{c_i}{T}\cdot T\Big)\frac{1}{\log^3 (N)}
\leq\frac{1 + \mathfrak{c}_{13}}{\log^3 (N)}.
\end{align*}
In other words, on the event $\mathcal E_T$,
\begin{equation*}
\max_{i\in\mathcal I_{\mathrm{LTU}}}\sup_{0\le r\le 1}
\Big|
Y_{iT}(r)-
\Big\{
\rho^{\lfloor Tr\rfloor}\frac{y_{i0}}{\sqrt T}
+\mathbb W_{iT}(r)
-\frac{c_i}{T}\sum_{s=1}^{\lfloor Tr\rfloor-1}\rho^{\lfloor Tr\rfloor-1-s}\mathbb W_{iT}(s/T)
\Big\}
\Big|
\le \frac{1 + \mathfrak{c}_{13}}{\log^3 (N)}.
\end{equation*}

To connect the discrete representation to an OU functional, we now embed the Gaussian increments
into a continuous-time Brownian motion. Let $B_i(\cdot)$ be a standard Brownian motion on $[0,1]$
such that
\begin{equation*}
B_i(s/T)=\frac{\widetilde W_i(s)}{\sqrt T},\qquad s=0,1,\dots,T.
\end{equation*}
(For example, one may take $B_i$ to be the continuous-time version of the Gaussian partial sums
$\widetilde W_i(\cdot)/\sqrt T$ with independent increments on the grid, and extend it to $[0,1]$
by adding independent Brownian bridges on each interval $[s/T,(s+1)/T]$.)

Define the modulus term
\[
\omega_{iT}=\sup_{0\le r\le 1}\big|B_i(r)-B_i(\lfloor Tr\rfloor/T)\big|.
\]
By the proof in Appendix A.3, we have 
\begin{equation}\label{eq:step-to-cont}
        \mathnormal{Prob}\left(\omega_{iT} > \frac{\log (N)}{\sqrt{T}}\right) = o(1).
\end{equation}

For each $c\in[0,\mathfrak{c}_{13}]$, define the OU process driven by $B_i$:
\begin{equation*}
J_{i,c}(r)=\int_0^r e^{-c(r-s)}\,dB_i(s),\qquad r\in[0,1].
\end{equation*}
By It\^o integration by parts,
\begin{equation*}
J_{i,c}(r)=B_i(r)-c\int_0^r e^{-c(r-s)}B_i(s)\,ds.
\end{equation*}
Then the discrete OU term 
can be compared to $J_{i,c_i}(r)$ by adding and subtracting $B_i$:
\begin{align*}
&\sup_{0\le r\le 1}
\Big|
\mathbb W_{iT}(r)
-\frac{c_i}{T}\sum_{s=1}^{t-1}\rho^{t-1-s}\mathbb W_{iT}(s/T)
-
\Big\{B_i(r)-c_i\int_0^r e^{-c_i(r-u)}B_i(u)\,du\Big\}
\Big|
\notag\\
&\le
\sup_{0\le r\le 1}|\mathbb W_{iT}(r)-B_i(r)|
+
c_i\sup_{0\le r\le 1}\Big|
\frac{1}{T}\sum_{s=1}^{t-1}\rho^{t-1-s}\mathbb W_{iT}(s/T)
-
\int_0^r e^{-c_i(r-u)}B_i(u)\,du
\Big|
\end{align*}
The first term is $O_{p}(\sqrt{\log(N)/T})$ by \eqref{eq:step-to-cont}. For the second term on the right-hand side of the above display, fix $i\in\mathcal I_{\mathrm{LTU}}$ and
$r\in[0,1]$, we obtain
\begin{align}
&\Bigg|
\frac{1}{T}\sum_{s=1}^{t-1}\rho^{t-1-s}\mathbb W_{iT}(s/T)
-
\int_0^r e^{-c_i(r-u)}B_i(u)\,du
\Bigg|
\notag\\
&\le
\underbrace{\Bigg|
\frac{1}{T}\sum_{s=1}^{t-1}\rho^{t-1-s}\big(\mathbb W_{iT}(s/T)-B_i(s/T)\big)
\Bigg|}_{=:A_{iT}(r)}
+
\underbrace{\Bigg|
\frac{1}{T}\sum_{s=1}^{t-1}\rho^{t-1-s}B_i(s/T)
-
\frac{1}{T}\sum_{s=1}^{t-1}e^{-c_i(r-s/T)}B_i(s/T)
\Bigg|}_{=:B_{iT}(r)}
\notag\\
&\quad+
\underbrace{\Bigg|
\frac{1}{T}\sum_{s=1}^{t-1}e^{-c_i(r-u_s)}B_i(s/T)
-
\int_0^r e^{-c_i(r-u)}B_i(u)\,du
\Bigg|}_{=:C_{iT}(r)}.
\label{eq:ABC-split}
\end{align}
We next bound the three terms uniformly in $r\in[0,1]$.

\textbf{Bound for $A_{iT}(r)$}:
Since $0<\rho<1$, we have
\[
\sup_{0\le r\le 1} A_{iT}(r)
\le
\Big(\frac{1}{T}\sum_{s=1}^{t-1}\rho^{t-1-s}\Big)\cdot
\sup_{0\le u\le 1}\big|\mathbb W_{iT}(u)-B_i(u)\big|
\le
\omega_{iT},
\]
where $\omega_{iT}=\sup_{0\le u\le 1}|\mathbb W_{iT}(u)-B_i(u)|$. Hence, by
\eqref{eq:step-to-cont},
\begin{equation}\label{eq:A-bound}
\mathnormal{Prob}\Big(\sup_{0\le r\le 1}A_{iT}(r)>\frac{\log (N)}{\sqrt T}\Big)=o(1).
\end{equation}

\textbf{Bound for $B_{iT}(r)$}:
For each $r\in[0,1]$ and $s\le t-1$, write $k=t-1-s\ge 0$. Then
\[
\rho^{t-1-s}=(1-c_i/T)^k
=\exp\!\Big(k\log(1-c_i/T)\Big)
=\exp\!\Big(-\frac{c_i k}{T}+R_{kT}\Big),
\]
where $|R_{kT}|\le C\,k/T^2$ for a constant $C$ depending only on $\mathfrak{c}_{13}$ (by the Taylor expansion of
$\log(1-x)$ and $c_i\in[0,\mathfrak{c}_{13}]$). Since $k\le T$, this implies
\begin{equation}\label{eq:kernel-unif}
\sup_{0\le r\le 1}\max_{1\le s\le t-1}\Big|\rho^{t-1-s}-e^{-c_i(r-s/T)}\Big|
\le \frac{C'}{T},
\end{equation}
with a constant $C'$ depending only on $\mathfrak{c}_{13}$.
Therefore,
\begin{equation}\label{eq:B-bound}
\sup_{0\le r\le 1}B_{iT}(r)
\le
\frac{C'}{T}\cdot \frac{1}{T}\sum_{s=1}^{T}\big|B_i(s/T)\big|
\le
\frac{C'}{T}\sup_{0\le u\le 1}|B_i(u)|.
\end{equation}
In particular, $\sup_{0\le u\le 1}|B_i(u)|=O_{p}(1)$, so $\sup_{r}B_{iT}(r)=O_{p}(1/T)=o_{P}(1)$
uniformly over $i$.

\textbf{Bound for $C_{iT}(r)$}:
The term $C_{iT}(r)$ is the Riemann-sum approximation error for the integral
$\int_0^r e^{-c_i(r-u)}B_i(u)\,du$. Since $c_i\in[0,\mathfrak{c}_{13}]$, the weight
$u\mapsto e^{-c_i(r-u)}$ is uniformly bounded and uniformly Lipschitz on $[0,1]$.
Thus, the proof that bounds the Brownian Riemann-sum error applies here after inserting the bounded weight
$e^{-c_i(r-u)}$.
Consequently,
\begin{equation}\label{eq:C-bound}
\mathnormal{Prob}\Big(
\max_{i\in\mathcal I_{\mathrm{LTU}}}\sup_{0\le r\le 1} C_{iT}(r)
>\frac{\log (N)}{\sqrt T}
\Big)=o(1).
\end{equation}

Combining \eqref{eq:ABC-split}--\eqref{eq:C-bound}, we obtain
\[
\mathnormal{Prob}\Bigg(
\max_{i\in\mathcal I_{\mathrm{LTU}}}\sup_{0\le r\le 1}
\Bigg|
\frac{1}{T}\sum_{s=1}^{\lfloor Tr\rfloor-1}\rho^{\lfloor Tr\rfloor-1-s}\mathbb W_{iT}(s/T)
-
\int_0^r e^{-c_i(r-u)}B_i(u)\,du
\Bigg|
>\frac{C'\log (N)}{\sqrt T}
\Bigg)=o(1)
\]
for a constant $C'$ depending only on $\mathfrak{c}_{13}$. Finally, combining the bounds of the three parts,
\begin{align*}
\max_{i\in\mathcal I_{\mathrm{LTU}}}\sup_{0\le r\le 1}|Y_{iT}(r)-J_{i,c_i}(r)|
=&
O_{p}\!\Big(\frac{1}{\log^3(N)}+\frac{\log (N)}{\sqrt T}+\sqrt{\frac{\log(NT)}{T}}\Big)
+
\max_{i\in\mathcal I_{\mathrm{LTU}}}\Big|\frac{y_{i0}}{\sqrt T}\Big| \\
=&O_{p}(1/\log^3(N)),
\end{align*}
which concludes the proof.

\clearpage

\setcounter{table}{0} \setcounter{page}{1} \renewcommand{\theequation}{B-%
\arabic{equation}} \renewcommand{\thetable}{B\arabic{table}}%
\renewcommand{\thefigure}{B\arabic{figure}}\renewcommand*{\thepage}{B-%
\arabic{page}}

\section{Data appendix}

A balanced panel was created from the 126 variables in the 2024:12 dataset
by: using only data over 1960:1-2024:9; dropping the Michigan Survey of
Consumer Sentiment (UMCSENT), trade-weighted exchange rate (TWEXAFEGSMTH),
and new orders for consumer goods (ACOGNO) and nondefense capital goods
(ANDENO), which are the same four series dropped by McCracken and Ng to
create a balanced panel from the 2015:4 dataset; dropping the VIX (VIXCLS),
which was not included in the 2015:4 dataset and whose first value is July
1962; and dropping the financial commercial paper rate (CP3M) and the
commercial paper-fed funds spread (COMPAPFF) which were not reported for
April 2020. The particular variables used in our analysis of the 2023
vintage dataset are described in Table \ref{table_R2}.

\begin{table}[!ht]
\caption{$R^2$ for each variable explained by first and second principal
components, 1962:3 to 2023:6}
\label{table_R2}\centering
\begin{tabular}{lllll}
\multicolumn{5}{c}{Group 1. Output and income} \\ \hline\hline
Index & FRED & Description & PC1 & PC1\&2 \\ \hline
1 & RPI & Real Personal Income & 0.23 & 0.40 \\ 
2 & W875RX1 & Real personal income ex transfer receipts & 0.61 & 0.75 \\ 
6 & INDPRO & IP Index & 0.77 & 0.85 \\ 
7 & IPFPNSS & IP: Final Products and Nonindustrial Supplies & 0.81 & 0.88 \\ 
8 & IPFINAL & IP: Final Products (Market Group) & 0.78 & 0.82 \\ 
9 & IPCONGD & IP: Consumer Goods & 0.52 & 0.83 \\ 
10 & IPDCONGD & IP: Durable Consumer Goods & 0.49 & 0.82 \\ 
11 & IPNCONGD & IP: Nondurable Consumer Goods & 0.41 & 0.55 \\ 
12 & IPBUSEQ & IP: Business Equipment & 0.71 & 0.71 \\ 
13 & IPMAT & IP: Materials & 0.65 & 0.73 \\ 
14 & IPDMAT & IP: Durable Materials & 0.64 & 0.75 \\ 
15 & IPNMAT & IP: Nondurable Materials & 0.62 & 0.69 \\ 
16 & IPMANSICS & IP: Manufacturing (SIC) & 0.77 & 0.87 \\ 
17 & IPB51222S & IP: Residential Utilities & 0.02 & 0.03 \\ 
18 & IPFUELS & IP: Fuels & 0.07 & 0.07 \\ 
19 & CUMFNS & Capacity Utilization: Manufacturing & 0.68 & 0.73 \\ \hline
&  & Median & 0.63 & 0.74 \\ \hline
&  &  &  &  \\ 
&  &  &  & 
\end{tabular}%
\end{table}

\clearpage \pagenumbering{gobble} 
\begin{table}[h!t]
\centering
Table B1 (continued) \newline
\begin{tabular}{lllll}
&  &  &  &  \\ 
\multicolumn{5}{c}{Group 2. Labor market} \\ \hline\hline
Index & FRED & Description & PC1 & PC1\&2 \\ \hline
20 & HWI & Help-Wanted Index for United States & 0.55 & 0.56 \\ 
21 & HWIURATIO & Ratio of Help Wanted/No. Unemployed & 0.55 & 0.55 \\ 
22 & CLF16OV & Civilian Labor Force & 0.25 & 0.34 \\ 
23 & CE16OV & Civilian Employment & 0.75 & 0.75 \\ 
24 & UNRATE & Civilian Unemployment Rate & 0.69 & 0.71 \\ 
25 & UEMPMEAN & Average Duration of Unemployment (Weeks) & 0.24 & 0.24 \\ 
26 & UEMPLT5 & Civilians Unemployed - Less Than 5 Weeks & 0.38 & 0.46 \\ 
27 & UEMP5TO14 & Civilians Unemployed for 5-14 Weeks & 0.65 & 0.68 \\ 
28 & UEMP15OV & Civilians Unemployed - 15 Weeks and Over & 0.66 & 0.66 \\ 
29 & UEMP15T26 & Civilians Unemployed for 15-26 Weeks & 0.65 & 0.66 \\ 
30 & UEMP27OV & Civilians Unemployed for 27 Weeks and Over & 0.59 & 0.59 \\ 
31 & CLAIMSx & Initial Claims & 0.42 & 0.45 \\ 
32 & PAYEMS & All Employees: Total nonfarm & 0.81 & 0.81 \\ 
33 & USGOOD & All Employees: Goods-Producing Industries & 0.85 & 0.85 \\ 
34 & CES1021000001 & All Employees: Mining and Logging: Mining & 0.03 & 0.41
\\ 
35 & USCONS & All Employees: Construction & 0.67 & 0.74 \\ 
36 & MANEMP & All Employees: Manufacturing & 0.74 & 0.74 \\ 
37 & DMANEMP & All Employees: Durable goods & 0.77 & 0.77 \\ 
38 & NDMANEMP & All Employees: Nondurable goods & 0.52 & 0.53 \\ 
39 & SRVPRD & All Employees: Service-Providing Industries & 0.67 & 0.68 \\ 
40 & USTPU & All Employees: Trade, Transportation and Utilities & 0.80 & 0.80
\\ 
41 & USWTRADE & All Employees: Wholesale Trade & 0.74 & 0.80 \\ 
42 & USTRADE & All Employees: Retail Trade & 0.67 & 0.68 \\ 
43 & USFIRE & All Employees: Financial Activities & 0.42 & 0.43 \\ 
44 & USGOVT & All Employees: Government & 0.08 & 0.09 \\ 
45 & CES0600000007 & Avg Weekly Hours : Goods-Producing & 0.33 & 0.48 \\ 
46 & AWOTMAN & Avg Weekly Overtime Hours : Manufacturing & 0.38 & 0.62 \\ 
47 & AWHMAN & Avg Weekly Hours : Manufacturing & 0.31 & 0.53 \\ 
115 & CES0600000008 & Avg Hourly Earnings : Goods-Producing & 0.04 & 0.42 \\ 
116 & CES2000000008 & Avg Hourly Earnings : Construction & 0.00 & 0.34 \\ 
117 & CES3000000008 & Avg Hourly Earnings : Manufacturing & 0.02 & 0.33 \\ 
\hline
&  & Median & 0.55 & 0.59 \\ \hline
&  &  &  &  \\ 
\multicolumn{5}{c}{Group 3. Housing} \\ \hline\hline
Index & FRED & Description & PC1 & PC1\&2 \\ \hline
48 & HOUST & Housing Starts: Total New Privately Owned & 0.14 & 0.37 \\ 
49 & HOUSTNE & Housing Starts, Northeast & 0.17 & 0.39 \\ 
50 & HOUSTMW & Housing Starts, Midwest & 0.11 & 0.42 \\ 
51 & HOUSTS & Housing Starts, South & 0.12 & 0.28 \\ 
52 & HOUSTW & Housing Starts, West & 0.11 & 0.26 \\ 
53 & PERMIT & New Private Housing Permits (SAAR) & 0.11 & 0.35 \\ 
54 & PERMITNE & New Private Housing Permits, Northeast (SAAR) & 0.13 & 0.43
\\ 
55 & PERMITMW & New Private Housing Permits, Midwest (SAAR) & 0.10 & 0.47 \\ 
56 & PERMITS & New Private Housing Permits, South (SAAR) & 0.08 & 0.30 \\ 
57 & PERMITW & New Private Housing Permits, West (SAAR) & 0.09 & 0.24 \\ 
\hline
&  & Median & 0.11 & 0.36 \\ \hline
\end{tabular}%
\end{table}

\clearpage 
\begin{table}[th!]
\centering
Table B1 (continued) \vspace{8pt} 
\begin{tabular}{lllll}
&  &  &  &  \\ 
\multicolumn{5}{c}{Group 4. Consumption, orders, and inventories} \\ 
\hline\hline
Index & FRED & Description & PC1 & PC1\&2 \\ \hline
3 & DPCERA3M086SBEA & Real personal consumption expenditures & 0.54 & 0.77
\\ 
4 & CMRMTSPLx & Real Manu. and Trade Industries Sales & 0.73 & 0.89 \\ 
5 & RETAILx & Retail and Food Services Sales & 0.46 & 0.50 \\ 
58 & AMDMNOx & New Orders for Durable Goods & 0.69 & 0.69 \\ 
59 & AMDMUOx & Unfilled Orders for Durable Goods & 0.26 & 0.40 \\ 
60 & BUSINVx & Total Business Inventories & 0.27 & 0.75 \\ 
61 & ISRATIOx & Total Business: Inventories to Sales Ratio & 0.22 & 0.27 \\ 
\hline
&  & Median & 0.46 & 0.69 \\ \hline
&  &  &  &  \\ 
\multicolumn{5}{c}{Group 5. Money and credit} \\ \hline\hline
Index & FRED & Description & PC1 & PC1\&2 \\ \hline
62 & M1SL & M1 Money Stock & 0.02 & 0.02 \\ 
63 & M2SL & M2 Money Stock & 0.03 & 0.05 \\ 
64 & M2REAL & Real M2 Money Stock & 0.03 & 0.45 \\ 
65 & BOGMBASE & Monetary Base & 0.20 & 0.20 \\ 
66 & TOTRESNS & Total Reserves of Depository Institutions & 0.28 & 0.28 \\ 
67 & NONBORRES & Reserves Of Depository Institutions & 0.00 & 0.00 \\ 
68 & BUSLOANS & Commercial and Industrial Loans & 0.10 & 0.17 \\ 
69 & REALLN & Real Estate Loans at All Commercial Banks & 0.23 & 0.23 \\ 
70 & NONREVSL & Total Nonrevolving Credit & 0.29 & 0.30 \\ 
71 & CONSPI & Nonrevolving consumer credit to Personal Income & 0.09 & 0.17
\\ 
118 & DTCOLNVHFNM & Consumer Motor Vehicle Loans Outstanding & 0.03 & 0.03
\\ 
119 & DTCTHFNM & Total Consumer Loans and Leases Outstanding & 0.21 & 0.22
\\ 
120 & INVEST & Securities in Bank Credit at All Commercial Banks & 0.02 & 
0.04 \\ \hline
&  & Median & 0.09 & 0.17 \\ \hline
&  &  &  &  \\ 
\multicolumn{5}{c}{Group 6. Interest and exchange rates} \\ \hline\hline
Index & FRED & Description & PC1 & PC1\&2 \\ \hline
76 & FEDFUNDS & Effective Federal Funds Rate & 0.34 & 0.68 \\ 
77 & TB3MS & 3-Month Treasury Bill: & 0.37 & 0.69 \\ 
78 & TB6MS & 6-Month Treasury Bill: & 0.38 & 0.71 \\ 
79 & GS1 & 1-Year Treasury Rate & 0.36 & 0.71 \\ 
80 & GS5 & 5-Year Treasury Rate & 0.16 & 0.63 \\ 
81 & GS10 & 10-Year Treasury Rate & 0.08 & 0.59 \\ 
82 & AAA & Moody's Seasoned Aaa Corporate Bond Yield & 0.02 & 0.60 \\ 
83 & BAA & Moody's Seasoned Baa Corporate Bond Yield & 0.00 & 0.61 \\ 
84 & TB3SMFFM & 3-Month Treasury C Minus FEDFUNDS & 0.06 & 0.31 \\ 
85 & TB6SMFFM & 6-Month Treasury C Minus FEDFUNDS & 0.04 & 0.27 \\ 
86 & T1YFFM & 1-Year Treasury C Minus FEDFUNDS & 0.00 & 0.16 \\ 
87 & T5YFFM & 5-Year Treasury C Minus FEDFUNDS & 0.12 & 0.34 \\ 
88 & T10YFFM & 10-Year Treasury C Minus FEDFUNDS & 0.21 & 0.41 \\ 
89 & AAAFFM & Moody's Aaa Corporate Bond Minus FEDFUNDS & 0.33 & 0.49 \\ 
90 & BAAFFM & Moody's Baa Corporate Bond Minus FEDFUNDS & 0.40 & 0.49 \\ 
91 & EXSZUSx & Switzerland / U.S. Foreign Exchange Rate & 0.00 & 0.01 \\ 
92 & EXJPUSx & Japan / U.S. Foreign Exchange Rate & 0.00 & 0.03 \\ 
93 & EXUSUKx & U.S. / U.K. Foreign Exchange Rate & 0.07 & 0.08 \\ 
94 & EXCAUSx & Canada / U.S. Foreign Exchange Rate & 0.00 & 0.04 \\ \hline
&  & Median & 0.08 & 0.49 \\ \hline
\end{tabular}%
\end{table}
\clearpage \renewcommand*{\thepage}{B-\arabic{page}} \setcounter{page}{4} 
\begin{table}[th!]
\begin{center}
Table B1 (concluded) \vspace{8pt} 
\begin{tabular}{lllll}
&  &  &  &  \\ 
\multicolumn{5}{c}{Group 7. Prices} \\ \hline\hline
Index & FRED & Description & PC1 & PC1\&2 \\ \hline
95 & WPSFD49207 & PPI: Finished Goods & 0.07 & 0.74 \\ 
96 & WPSFD49502 & PPI: Finished Consumer Goods & 0.07 & 0.72 \\ 
97 & WPSID61 & PPI: Intermediate Materials & 0.06 & 0.65 \\ 
98 & WPSID62 & PPI: Crude Materials & 0.11 & 0.44 \\ 
99 & OILPRICEx & Crude Oil, spliced WTI and Cushing & 0.02 & 0.50 \\ 
100 & PPICMM & PPI: Metals and metal products: & 0.17 & 0.36 \\ 
101 & CPIAUCSL & CPI : All Items & 0.09 & 0.82 \\ 
102 & CPIAPPSL & CPI : Apparel & 0.04 & 0.40 \\ 
103 & CPITRNSL & CPI : Transportation & 0.04 & 0.56 \\ 
104 & CPIMEDSL & CPI : Medical Care & 0.11 & 0.41 \\ 
105 & CUSR0000SAC & CPI : Commodities & 0.08 & 0.73 \\ 
106 & CUSR0000SAD & CPI : Durables & 0.00 & 0.33 \\ 
107 & CUSR0000SAS & CPI : Services & 0.01 & 0.67 \\ 
108 & CPIULFSL & CPI : All Items Less Food & 0.04 & 0.77 \\ 
109 & CUSR0000SA0L2 & CPI : All items less shelter & 0.07 & 0.78 \\ 
110 & CUSR0000SA0L5 & CPI : All items less medical care & 0.10 & 0.82 \\ 
111 & PCEPI & Personal Cons. Expend.: Chain Index & 0.08 & 0.76 \\ 
112 & DDURRG3M086SBEA & Personal Cons. Exp: Durable goods & 0.00 & 0.37 \\ 
113 & DNDGRG3M086SBEA & Personal Cons. Exp: Nondurable goods & 0.06 & 0.78
\\ 
114 & DSERRG3M086SBEA & Personal Cons. Exp: Services & 0.04 & 0.66 \\ \hline
&  & Median & 0.06 & 0.66 \\ \hline
&  &  &  &  \\ 
\multicolumn{5}{c}{Group 8. Stock market} \\ \hline\hline
Index & FRED & Description & PC1 & PC1\&2 \\ \hline
72 & S\&P 500 & S\&P's Common Stock Price Index: Composite & 0.22 & 0.33 \\ 
73 & S\&P: indust & S\&P's Common Stock Price Index: Industrials & 0.18 & 
0.28 \\ 
74 & S\&P div yield & S\&P's Composite Common Stock: Dividend Yield & 0.05 & 
0.40 \\ 
75 & S\&P PE ratio & S\&P's Composite Common Stock: Price-Earnings Ratio & 
0.11 & 0.44 \\ \hline
&  & Median & 0.15 & 0.36 \\ \hline\hline
&  &  &  &  \\ 
&  & Overall median & 0.19 & 0.50 \\ \hline\hline
\end{tabular}%
\end{center}
\par
\vspace{5pt} Notes to Table \ref{table_R2}. Index refers to the index number
of the variable in our database. FRED refers to variable name in the FRED
database. PC1 is the fraction of the variance of the cyclical component of
that variable that is explained by the first principal component. PC1\&2 is
the fraction of the variance of the cyclical component of that variable that
is explained by the first and second principal components combined.
\end{table}

\clearpage

\setcounter{table}{0} \setcounter{page}{1} \renewcommand{\theequation}{C-	%
\arabic{equation}} \renewcommand{\thetable}{C\arabic{table}}%
\renewcommand{\thefigure}{C\arabic{figure}}\renewcommand*{\thepage}{C-3%
\arabic{page}}

\section{Additional simulations}

Here we provide additional details on the simulations.

\textit{Cointegration. }For this example, the nonstationary variables are
cointegrated and there is no factor structure for the stationary variables.
The single common factor for the nonstationary variables follows a random
walk:%
\begin{equation*}
F_{t}=F_{t-1}+v_{t}\text{ \ \ \ }t=1,....,T;\text{ }F_{0}=0
\end{equation*}%
\begin{equation*}
y_{it}=F_{t}+\varepsilon _{it}\text{ \ \ \ }i=1,2,...,N/2
\end{equation*}%
\begin{equation*}
y_{it}=\varepsilon _{it}\text{ \ \ \ }i=(N/2)+1,...,N
\end{equation*}%
with $v_{t}\sim N(0,1)$ and $\varepsilon _{it}\sim N(0,1)$ independent for
all $i$ and $t.$ Notice that the first $(N/2)$ variables are characterized by%
\begin{equation*}
y_{it}=F_{t-h}+v_{t-h+1}+v_{t-h+2}+\cdots +v_{t}+\varepsilon _{it}
\end{equation*}%
\begin{equation*}
{\mathbb{P}(y_{it}|1,y_{i,t-h},y_{i,t-h-1},...,y_{i,t-h-p+1})\simeq F_{t-h}}
\end{equation*}%
\begin{equation*}
c_{it}\simeq v_{t-h+1}+v_{t-h+2}+\cdots +v_{t}+\varepsilon _{it}.
\end{equation*}%
Thus there is a single factor (namely $v_{t-h+1}+v_{t-h+2}+\cdots +v_{t})$
that is common to the cyclical component of the first $(N/2)$ variables,%
\footnote{%
Another way to express this is that the variables are cointegrated with $%
(N/2)-1$ linearly independent cointegrating relations given by $%
y_{it}-y_{1t}\sim I(0)$ for $i=2,3,...,(N/2).$} and the true number of
common factors in the sample of $N$ variables is $r=1.$

Columns (7) and (8) of Table 2\ report the results
from applying PCA to the raw data. Note that even though half the variables
are nonstationary, PCA always correctly concludes that there is a single
common factor in these data. Columns (9) and (10) report results from
applying PCA to the residuals from 24-period-ahead forecasting regressions.
Again the same regression is estimated in the same way for stationary and
nonstationary observations. And again PCA on the regression residuals
results in the correct answer 100\% of the time, even if the sample size is
as small as $T=100.$

\textit{Stationary factor with a mix of stationary and nonstationary
indicators.} In this example the data-generating process features a common
stationary factor: $F_{t}=\rho _{F}F_{t-1}+v_{t}$ with $\rho _{F}=0.8.$ The $%
i$th observed variable $y_{it}$ is related to $F_{t}$ with a weight $\omega
_{i},$ $y_{it}=\omega _{i}F_{t}+g_{it},$ where the idiosyncratic components $%
g_{it}$ are a mix of stationary and nonstationary processes: $g_{it}=\rho
_{i}g_{i,t-1}+e_{it}$ where $v_{t}$ and $e_{it}$ ($i=1,...,N)$ are mutually
independent $N(0,1)$. Columns (11)-(14) of Table 2
report results for the following example:%
\begin{equation*}
\begin{array}{ll}
\rho _{i}=1,\omega _{i}=1 & \text{for }i=1,...,N/4 \\ 
\rho _{i}=0.5,\omega _{i}=1 & \text{for }i=1+N/4,...,N/2 \\ 
\rho _{i}=1,\omega _{i}=0 & \text{for }i=1+N/2,...,3N/4 \\ 
\rho _{i}=0.5,\omega _{i}=0 & \text{for }i=1+3N/4,...,N%
\end{array}
.
\end{equation*}%
Thus for this example half of the $N$ observed variables are nonstationary.
The variables are independent of each other apart from their potential
common dependence on the $r=1$-dimensional factor $F_{t},$ and this single
factor affects some of the observed variables but not others.

\textit{Local-to-unit roots.} For these simulations we replaced $\rho $ in (%
21) for half the variables with $\rho _{T}=1-c/T$ with $%
c=1$ and $T$ the number of time-series observations in a generated sample.
The other half of the generated variables were white noise.

\textit{Fractional integration.} Here we replaced equation (21) for half the variables with%
\begin{equation*}
(1-L)^{d}y_{it}=\varepsilon _{it}
\end{equation*}%
for $i=1,2,...,N/2$ and $d=0.6.$ The other half of the generated variables
were again white noise. We generated draws for this process using the MA($%
\infty )$ representation $y_{it}=\tsum\nolimits_{j=0}^{\infty }\psi
_{j}\varepsilon _{i,t-j}$ and the weights calculated recursively from%
\begin{equation*}
\psi _{j}=\frac{d+j-1}{j}\psi _{j-1}
\end{equation*}%
starting from $\psi _{0}=0.$ In practice we did this by truncating the sum
as $\tsum\nolimits_{j=0}^{M}\psi _{j}\varepsilon _{i,t-j}$ for $M=100,000$
using 100,000 pre-sample draws for $\varepsilon _{i,-1},\varepsilon
_{i,-2},...,\varepsilon _{i-M}.$

\textit{Comparison with the \cite{BaiNg2004} PANIC approach.} \cite%
{BaiNg2004} recommended taking first-differences of all the observations and
then applying PCA to the changes. Columns (4) and (8) in Table \ref%
{tab_PANIC} show that this correctly concludes that there are no common
factors when applied to the data-generating processes described in Sections
5.1 and 5.2. Columns (3) and (7) show the fraction of the variance of $%
\Delta y_{it}$ explained by the first three principal components. These
results are very similar to those for our method with $h=1,$ shown in
columns (1)-(2) (which reproduce columns (11) and (12) in Table 1) and columns (5)-(6) (which reproduce columns (5) and (6)
in Table 2). Note that the $R^{2}$ for the $h=1$
cases refer to the sample variance of the estimated cyclical factors, as it
did in the earlier tables.

\begin{table}[tbph]
\caption{Local to unity and fractionally integrated processes}
\label{tab_local_to_unity}%
\begin{tabular}{c|cc|cc|cc|cc|cc|cc}
\hline
& \multicolumn{6}{c|}{Local to unity} & \multicolumn{6}{c|}{Fractionally
integrated} \\ \hline
& \multicolumn{2}{c|}{Raw data} & \multicolumn{2}{c|}{$\hat{c}_t$ ($h$=24)}
& \multicolumn{2}{c|}{$\hat{c}_t$ ($h$=1)} & \multicolumn{2}{c|}{Raw data} & 
\multicolumn{2}{c|}{$\hat{c}_t$ ($h$=24)} & \multicolumn{2}{c|}{$\hat{c}_t$ (%
$h$=1)} \\ 
\cmidrule{2-3} \cmidrule{4-5} \cmidrule{6-7} \cmidrule{8-9} \cmidrule{10-11} %
\cmidrule{12-13} $j$ & $R^2$ & $r^*$ & $R^2$ & $r^*$ & $R^2$ & $r^*$ & $R^2$
& $r^*$ & $R^2$ & $r^*$ & $R^2$ & $r^*$ \\ 
& (1) & (2) & (3) & (4) & (5) & (6) & (7) & (8) & (9) & (10) & (11) & (12)
\\ \hline
$T=100$ &  &  &  &  &  &  &  &  &  &  &  &  \\ 
0 & --- & 0 & --- & 0 & --- & 100 & --- & 8 & --- & 99 & --- & 100 \\ 
1 & 43.0 & 98 & 15.8 & 3 & 4.1 & 0 & 10.5 & 88 & 8.0 & 1 & 4.1 & 0 \\ 
2 & 3.7 & 2 & 10.6 & 97 & 3.8 & 0 & 6.7 & 4 & 6.5 & 0 & 3.8 & 0 \\ 
3 & 2.8 & 0 & 5.5 & 0 & 3.6 & 0 & 4.9 & 0 & 5.4 & 0 & 3.6 & 0 \\ \hline
$T=200$ &  &  &  &  &  &  &  &  &  &  &  &  \\ 
0 & --- & 0 & --- & 0 & --- & 100 & --- & 0 & --- & 43 & --- & 100 \\ 
1 & 44.8 & 96 & 11.0 & 0 & 2.9 & 0 & 9.9 & 59 & 7.2 & 53 & 2.9 & 0 \\ 
2 & 2.7 & 4 & 8.9 & 0 & 2.7 & 0 & 5.9 & 41 & 5.8 & 4 & 2.7 & 0 \\ 
3 & 2.2 & 0 & 7.2 & 100 & 2.6 & 0 & 4.4 & 0 & 4.5 & 0 & 2.6 & 0 \\ \hline
$T=400$ &  &  &  &  &  &  &  &  &  &  &  &  \\ 
0 & --- & 0 & --- & 3 & --- & 100 & --- & 0 & --- & 99 & --- & 100 \\ 
1 & 46.1 & 98 & 6.4 & 25 & 2.2 & 0 & 9.3 & 26 & 4.8 & 1 & 2.2 & 0 \\ 
2 & 2.1 & 2 & 5.5 & 48 & 2.1 & 0 & 5.6 & 72 & 4.1 & 0 & 2.1 & 0 \\ 
3 & 1.8 & 0 & 4.9 & 24 & 2.1 & 0 & 4.0 & 2 & 3.6 & 0 & 2.1 & 0 \\ \hline
$T=600$ &  &  &  &  &  &  &  &  &  &  &  &  \\ 
0 & --- & 0 & --- & 89 & --- & 100 & --- & 0 & --- & 100 & --- & 100 \\ 
1 & 46.8 & 98 & 4.9 & 11 & 1.9 & 0 & 9.0 & 18 & 3.9 & 0 & 1.9 & 0 \\ 
2 & 1.9 & 2 & 4.3 & 0 & 1.9 & 0 & 5.4 & 77 & 3.4 & 0 & 1.9 & 0 \\ 
3 & 1.6 & 0 & 3.9 & 0 & 1.8 & 0 & 3.8 & 5 & 3.1 & 0 & 1.8 & 0 \\ \hline
$T=800$ &  &  &  &  &  &  &  &  &  &  &  &  \\ 
0 & --- & 0 & --- & 100 & --- & 100 & --- & 0 & --- & 100 & --- & 100 \\ 
1 & 47.2 & 99 & 4.1 & 0 & 1.8 & 0 & 8.7 & 8 & 3.4 & 0 & 1.8 & 0 \\ 
2 & 1.7 & 1 & 3.7 & 0 & 1.7 & 0 & 5.4 & 78 & 3.0 & 0 & 1.8 & 0 \\ 
3 & 1.5 & 0 & 3.4 & 0 & 1.7 & 0 & 3.8 & 14 & 2.7 & 0 & 1.7 & 0 \\ \hline
$T=1000$ &  &  &  &  &  &  &  &  &  &  &  &  \\ 
0 & --- & 0 & --- & 100 & --- & 100 & --- & 0 & --- & 100 & --- & 100 \\ 
1 & 47.5 & 99 & 3.7 & 0 & 1.7 & 0 & 8.7 & 8 & 3.1 & 0 & 1.7 & 0 \\ 
2 & 1.6 & 1 & 3.3 & 0 & 1.7 & 0 & 5.3 & 89 & 2.7 & 0 & 1.7 & 0 \\ 
3 & 1.4 & 0 & 3.0 & 0 & 1.6 & 0 & 3.6 & 3 & 2.5 & 0 & 1.6 & 0 \\ \hline\hline
\end{tabular}
\vspace{5pt} \newline
Notes to Table \ref{tab_local_to_unity}. $R^2$ indicates the percentage of
total variance accounted for by the $j$th principal component for $j=1,2$ or
3. $r^*$ indicates the percentage of samples for which the criterion (19) selects the number of factors to be $j=0,1,2,$ or $\ge 3$. In
every case, the true number of factors is $r=0$ and the cross-section
dimension is $N=100$.
\end{table}

\begin{table}[tbph]
\caption{Comparison of PANIC method with one-period-ahead forecast errors}
\label{tab_PANIC}%
\begin{tabular}{c|cc|cc|cc|cc}
\hline
& \multicolumn{4}{c|}{Random walk} & \multicolumn{4}{c}{Stationary} \\ \hline
& \multicolumn{2}{c|}{$h=1$} & \multicolumn{2}{c|}{PANIC} & 
\multicolumn{2}{c|}{$h=1$} & \multicolumn{2}{c}{PANIC} \\ 
\cmidrule{2-3} \cmidrule{4-5} \cmidrule{6-7} \cmidrule{8-9} $j$ & $R^2$ & $%
r^*$ & $R^2$ & $r^*$ & $R^2$ & $r^*$ & $R^2$ & $r^*$ \\ 
& (1) & (2) & (3) & (4) & (5) & (6) & (7) & (8) \\ \hline
$T=100$ &  &  &  &  &  &  &  &  \\ 
0 & --- & 100 & --- & 100 & --- & 100 & --- & 100 \\ 
1 & 4.1 & 0 & 4.2 & 0 & 4.1 & 0 & 4.2 & 0 \\ 
2 & 3.8 & 0 & 3.9 & 0 & 3.8 & 0 & 3.9 & 0 \\ 
3 & 3.6 & 0 & 3.7 & 0 & 3.6 & 0 & 3.7 & 0 \\ \hline
$T=200$ &  &  &  &  &  &  &  &  \\ 
0 & --- & 100 & --- & 100 & --- & 100 & --- & 100 \\ 
1 & 2.9 & 0 & 3.0 & 0 & 2.9 & 0 & 3.8 & 0 \\ 
2 & 2.7 & 0 & 2.8 & 0 & 2.7 & 0 & 2.9 & 0 \\ 
3 & 2.6 & 0 & 2.7 & 0 & 2.6 & 0 & 2.8 & 0 \\ \hline
$T=400$ &  &  &  &  &  &  &  &  \\ 
0 & --- & 100 & --- & 100 & --- & 100 & --- & 100 \\ 
1 & 2.2 & 0 & 2.3 & 0 & 2.2 & 0 & 3.8 & 0 \\ 
2 & 2.1 & 0 & 2.2 & 0 & 2.1 & 0 & 2.3 & 0 \\ 
3 & 2.1 & 0 & 2.1 & 0 & 2.1 & 0 & 2.2 & 0 \\ \hline
$T=600$ &  &  &  &  &  &  &  &  \\ 
0 & --- & 100 & --- & 100 & --- & 100 & --- & 100 \\ 
1 & 1.9 & 0 & 2.0 & 0 & 1.9 & 0 & 3.4 & 0 \\ 
2 & 1.9 & 0 & 1.9 & 0 & 1.9 & 0 & 2.0 & 0 \\ 
3 & 1.8 & 0 & 1.9 & 0 & 1.8 & 0 & 1.9 & 0 \\ \hline
$T=800$ &  &  &  &  &  &  &  &  \\ 
0 & --- & 100 & --- & 100 & --- & 100 & --- & 100 \\ 
1 & 1.8 & 0 & 1.9 & 0 & 1.8 & 0 & 3.1 & 0 \\ 
2 & 1.7 & 0 & 1.8 & 0 & 1.7 & 0 & 1.8 & 0 \\ 
3 & 1.7 & 0 & 1.8 & 0 & 1.7 & 0 & 1.8 & 0 \\ \hline
$T=1000$ &  &  &  &  &  &  &  &  \\ 
0 & --- & 100 & --- & 100 & --- & 100 & --- & 100 \\ 
1 & 1.7 & 0 & 1.8 & 0 & 1.7 & 0 & 2.8 & 0 \\ 
2 & 1.7 & 0 & 1.7 & 0 & 1.7 & 0 & 1.7 & 0 \\ 
3 & 1.6 & 0 & 1.7 & 0 & 1.6 & 0 & 1.7 & 0 \\ \hline\hline
\end{tabular}%
\par
Notes to Table \ref{tab_PANIC}. $R^2$ indicates the percentage of total
variance of the estimated cyclical component (for $h=1$) or of the observed
changes (for PANIC) accounted for by the $j$th principal component for $%
j=1,2 $ or 3. $r^*$ indicates the percentage of samples for which the
criterion (19) selects the number of factors to be $j=0,1,2,$ or $%
\ge 3$. In every case, the true number of factors is $r=0$ and the
cross-section dimension is $N=100$.
\end{table}

\newpage 
\bibliographystyle{apalike}
\bibliography{HXrefs.bib}